\documentclass[a4paper,fleqn]{cas-dc}

\usepackage[authoryear]{natbib}

\usepackage{amssymb}
\usepackage{amsmath}
\usepackage{bm}
\usepackage{booktabs}
\usepackage{subcaption}
\usepackage{hyperref} % keep this very late
\usepackage{cleveref}
\usepackage{tikz}
\usetikzlibrary{decorations.pathmorphing,patterns,calc,arrows.meta}
\usepackage{amsthm}

\theoremstyle{definition}

\theoremstyle{remark}

\usepackage{enumitem}
\usepackage{placeins}
\theoremstyle{definition}

\makeatletter
\global\sctrue
\global\dcfalse
\ExplSyntaxOn
\RenewDocumentCommand \maketitle { }
{
  \ifbool { usecasgrabsbox }
    {
      \setcounter{page}{0}
      \thispagestyle{empty}
      \unvbox\casgrabsbox
    } { }
  \pagebreak
  \ifbool { usecashlsbox }
    {
      \setcounter{page}{0}
      \thispagestyle{empty}
      \unvbox\casauhlbox
    } { }
  \pagebreak
  \thispagestyle{first}
  \ifbool{longmktitle}
    {
      \ifnum\theblind>0\relax
        \LongMaketitleBox[Blind]
      \else
        \LongMaketitleBox
      \fi
      \ProcessLongTitleBox
    }
    {
      \ifnum\theblind>0\relax
        \MaketitleBox[blind]
      \else
        \MaketitleBox
      \fi
      \printFirstPageNotes
    }
  \setcounter{footnote}{\int_use:N \g_stm_fnote_int}
  \renewcommand\thefootnote{\arabic{footnote}}
  \gdef\@pdfauthor{\infoauthors}
  \gdef\@pdfsubject{Complex ~STM ~Content}
  \ifbool{casreviewlayout}{\doublespacing}{}
}
\ExplSyntaxOff
\makeatother

\ExplSyntaxOn
\cs_set:Npn \__first_footerline:
{
  \group_begin:
  \small
  \sffamily
  \ifnum\theblind>0\relax
  \else
    \__short_authors:
  \fi
  \group_end:
}
\ExplSyntaxOff

\newcommand{\avg}[1]{\mkern-1mu\{\mkern-6mu\{\mkern-2mu #1 \mkern-2mu\}\mkern-6mu\}\mkern-1mu}

\newcommand{\jmp}[1]{[\![#1]\!]}

\newcommand{\pagefiguregraphics}[2][]{%
	\includegraphics[width=\textwidth,height=0.72\textheight,keepaspectratio,#1]{#2}%
}

\theoremstyle{definition}

\begin{document}
\let\WriteBookmarks\relax
\def\floatpagepagefraction{1}
\def\textpagefraction{.001}

\shorttitle{Shifted interface for poroelastic discontinuities}

\shortauthors{D.\,M.~Riley, G.~Scovazzi, I.~Stefanou}

\title[mode = title]{A shifted interface approach for internal discontinuities in poroelastic media}

\author[1]{David Michael Riley}
\credit{Formal analysis, Validation, Visualization, Software, Writing - Original Draft}

\affiliation[1]{organization={IMSIA (UMR 9219), CNRS, EDF, ENSTA Paris, Institut Polytechnique de Paris},
            city={Palaiseau},
            postcode={91120},
            country={France}}

\author[2]{Guglielmo Scovazzi}
\credit{Methodology, Supervision, Writing - Review and Editing}

\affiliation[2]{organization={Department of Civil and Environmental Engineering, Duke University},
            city={Durham},
            postcode={27708},
            state={North Carolina},
            country={USA}}

% Third author (corresponding)
\author[1]{Ioannis Stefanou}
\cormark[1]
\ead{ioannis.stefanou@ensta.fr}
\credit{Conceptualization, Methodology, Supervision, Writing - Review and Editing, Funding acquisition}

\cortext[1]{Corresponding author}

\begin{abstract}
Porous media containing cracks, fractures, or internal discontinuities
	arise throughout subsurface geomechanics, biomechanics, and materials
	science. Numerical simulation of the coupled hydromechanical response
	is inherently challenging because the pressure and displacement
	fields are tightly coupled through the Biot equations, requiring
	stable mixed formulations. These difficulties are compounded
	when cracks are present, because standard mesh-conforming approaches
	require costly, labor-intensive, body-fitted meshing, while unfitted methods often
	require cut-cell integration, enrichment functions, or additional
	stabilization. In this work, we use an alternative approach, we
	adapt the shifted interface method to coupled transient poroelasticity
	with embedded interfaces. The method replaces the true crack by a surrogate approximation where interface conditions are transferred through
	local expansions. A unified derivation yields shifted forms
	for both hydraulic transmission and mechanical traction coupling.
	Two enforcement strategies are extensively compared: a weak
	(integral) enforcement and a strong (pointwise) enforcement. Four test cases of
	increasing geometric complexity (offset mesh-aligned,
	boundary-intersecting angled, embedded angled, and multi-crack configurations) validate
	the formulation. Away from crack tips, interface residuals converge as
	$\mathcal{O}(h)$; near tips, localized post-processing artifacts degrade the global rate,
	but first-order convergence is recovered when a small tip region is excluded.
	A multi-crack demonstration with four simultaneously embedded cracks of
	distinct geometry and interface properties confirms the practical
	applicability of the framework. These results support the shifted
	interface method as a practical framework for poroelastic crack
	modeling on non-body-fitted meshes with geometrically complex
	embedded interfaces.
\end{abstract}

\begin{graphicalabstract}
\includegraphics{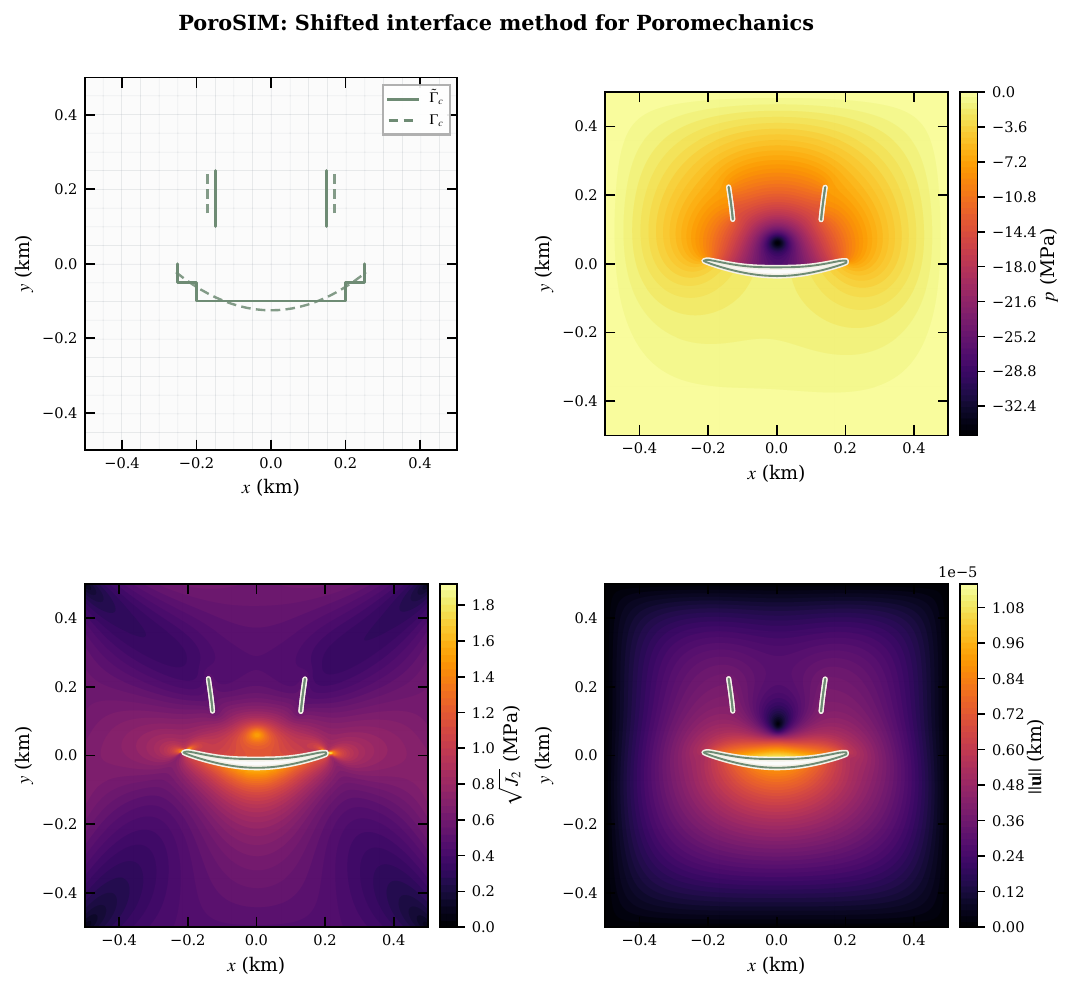}
\end{graphicalabstract}

\begin{highlights}
\item Shifted interface method is extended to transient poroelasticity with embedded cracks.
\item Formulation considers constitutive laws for both fluid and mechanics on the crack.
\item Two enforcement strategies of the interface laws are systematically compared.
\item Convergence benchmarks are conducted.
\end{highlights}
% Keywords
% Each keyword is separated by \sep

\maketitle


\section{Introduction}
\label{intro}
Porous materials permeated by cracks, fractures, and other internal discontinuities arise throughout engineering and the natural sciences. In subsurface geomechanics, faults and fractured rock masses play a critical role in the hydromechanical response of reservoirs and caprocks~\citep{Caine1996,Bense2013,Berre2019}; in biomechanics, cartilage, bone, and soft tissues are fluid-saturated porous media in which surface cracks and interfacial delaminations critically affect load bearing and repair~\citep{cowin1999bone,ateshian2009role}; and in materials science, the durability of concrete~\citep{ulm2004concrete} and the fracture of hydrogels~\citep{bouklas2015effect} and metallic foams~\citep{iaccarino2026investigating} are likewise controlled by the interplay between fluid transport and fractures. In all of these settings, internal discontinuities exert a controlling influence on both fluid flow and mechanical stress transfer, and may simultaneously accommodate displacement discontinuities through slip,
opening, or elastic compliance.

Capturing this coupled hydromechanical behavior is especially pressing for applications of broad societal importance, including CO$_2$ sequestration~\citep{rutqvist2012geomechanics,pan2016geomechanical}, geothermal energy extraction and induced-seismicity hazard assessment~\citep{ellsworth2013injection,foulger2018global,kivi2023global},
and the long-term safety of nuclear waste repositories~\citep{tsang2005geohydromechanical}. Efficient numerical methods that can represent the interaction between bulk poroelastic response and embedded interface physics---without requiring the mesh to conform to potentially complex interface geometries---are therefore of considerable practical value.

From a numerical simulation standpoint, modeling discontinuities is challenging. There are two common strategies. The first is to resolve the discontinuities as a thin 2D/3D region of very low (or highly anisotropic) material properties. While conceptually straightforward and compatible with standard poroelastic solvers, this approach degrades numerically as the barrier thickness decreases: extreme aspect ratios and small elements lead to element distortion, ill-conditioning, and restrictive time-step constraints in transient simulations. Moreover, mesh generation becomes fragile for complex geometries, if not impossible~\citep{Starnoni2021}. The second strategy is to reduce the barrier's dimensionality and model it as an \textit{internal interface} (i.e., a $(n-1)$D surface) endowed with an appropriate constitutive law on a fitted mesh that conforms to the interface.

This approach also enables the modeling of fractures without directly meshing thin interfacial layers or incurring large element distortions in complex geometries. In fact, when internal interfaces are geometrically complex, unfitted (or immersed or embedded) methods become attractive, since the interface geometry is independent of the computational mesh: the mesh need not conform to the interface, and interface conditions are instead imposed on a background grid that may cut through or lie near the true interface. XFEM introduces enrichment functions to capture field jumps across embedded interfaces without mesh alignment \citep{moes1999finite,Flemisch2016}. Its flexibility comes at the cost of specialized quadrature on elements cut by the interface, blending corrections, linear-algebra conditioning concerns (especially for intersecting or terminating features), and additional implementation complexity. CutFEM (and related unfitted Nitsche methods) instead keeps standard finite element spaces on a background mesh and imposes interface conditions weakly with stabilization (e.g., ghost penalties) to regularize small cut cells \citep{Hansbo2002,Burman2010,Burman2015}. Recent work has extended CutFEM to quasistatic poroelasticity for complex external boundaries~\citep{berre2025cut}, the coupling of free fluid flow with poroelastic media~\citep{ager2019nitsche}, and heterogeneous poroelastic domains with material interfaces (weak discontinuities where fields remain continuous but properties jump)~\citep{liu2025stabilized}. Recently, the tempered finite element method (TFEM) was proposed for meshes containing nearly-degenerate or zero-measure elements in the exactly-collapsed limit; it corresponds to mortaring \citep{quiriny2025tempered}. However, its coupling is geometry-induced rather than introduced through explicit fracture constitutive laws, so it does not directly provide the general hydraulic and mechanical interface relations considered here. Nonetheless, none of these approaches have focused on internal fractures with displacement and/or pressure jumps, which is the focus of the present contribution. Phase-field models offer yet another route, regularizing the discrete fracture as a continuous damage field and coupling it with poroelasticity~\citep{mikelic2015phase,wilson2016phase,chukwudozie2019poro}. While phase-field methods handle complex crack topologies and propagation naturally, they introduce a global auxiliary field that must be resolved on a mesh fine enough to capture the regularization length, leading to significant computational cost.

We adopt an alternative approach, the \emph{shifted interface method} (SIM), which emanates from the shifted boundary method (SBM) \citep{MainScovazzi2018,Scovazzi2018}. The key idea of this method is to replace the true internal interface by a nearby \emph{surrogate} interface that is fitted to the background mesh (e.g., a union of element faces), and to \emph{shift} the internal interface conditions in both location and value via local expansions to the true interface conditions~\citep{Li2020,Atallah2022}. Recent developments include the shifted fracture method (SFM) for cohesive crack simulations~\citep{Li2020,li2023simple}, shifted formulations for mechanical contact \citep{li2025shifted}, and extensions to coupled transient thermoelasticity \citep{li2024complex}. Together, these developments indicate that SIM provides a promising framework for complex multiphysics problems with intricate-geometry embedded interfaces, while keeping the numerical infrastructure close to standard mesh-conforming finite elements and avoiding the cut-cell integration, enrichment, and stabilization machinery typical of CutFEM and XFEM. Moreover, this approach is nonintrusive, in the sense that it only requires some additional terms in the variational form, without introducing new degrees of freedom, or changing the standard FEM data structure or requiring ad-hoc stabilization terms. The method has reached theoretical maturity in its approximation capabilities and convergence, providing the foundation upon which we build and extend herein.

We extend the SIM to coupled transient poroelasticity with embedded interfaces. The formulation accounts for both hydraulic transmission (via Robin-type conditions that relate the normal flux to the pressure jump) and mechanical response (via traction relations that relate the traction to the displacement jump in the normal and tangential directions). A unified derivation, based on decomposing a generic interface flux into true-normal and tangential-mismatch components, yields the shifted forms for both fields. The derived formulation is general to primal field methods (e.g., discontinuous and continuous Galerkin), but herein we apply continuous Galerkin on a mesh whose connectivity is split along the surrogate interface, combining the efficiency of conforming elements away from the interface with the flexibility to represent jumps across it.

The paper has the following structure. In Section~\ref{sec:problem_formulation}, we introduce the governing partial differential equations, the relevant external and internal boundary conditions, and the variational formulation on the computational domain. In Section~\ref{sec:sbm}, the derivation of the shifted interface forms is provided for the given PDE, including the interface constitutive closure and two enforcement strategies (weak and strong). Section~\ref{sec:spaces_and_split} describes the numerical implementation: the finite element spaces, the resulting semidiscrete problems for both enforcement strategies, and the post-processing pipeline used to evaluate interface residuals on the true crack geometry. Section~\ref{sec:numerical_results} presents numerical results on four test cases of increasing complexity: an offset mesh-aligned crack, a boundary-intersecting angled crack, an embedded angled crack, and a multi-crack configuration with four simultaneously embedded cracks of distinct geometry and interface properties.

\section{Problem formulation}
\label{sec:problem_formulation}
Let $\Omega\subset\mathbb{R}^d$ ($d=1,2,3$) be the porous body and let $\Gamma_c\subset \Omega$ be an embedded Lipschitz interface with unit normal $n_i$ (oriented from ``$-$'' to ``$+$''). The interface may terminate within $\Omega$ or extend to $\partial \Omega$, so that $\Omega\setminus\Gamma_c$ is either connected or consists of disjoint subdomains $\Omega^\pm$. In either case, for fields $\phi$ on $\Omega\setminus\Gamma_c$, we define one-sided traces 
$\phi^\pm$ on $\Gamma_c$ and the jump/average operators
\begin{equation}
  [\![\phi]\!] := \phi^+ - \phi^-,
  \qquad
  \avg{\phi} := \tfrac12\big(\phi^+ + \phi^-\big).
  \label{eq:avgjump_scalar}
\end{equation}

\subsection{Bulk equations}
On $\Omega\setminus\Gamma_c$, quasi-static Biot poroelasticity with Darcy flow is governed by ~\citep{cheng2016poroelasticity}
\begin{equation}
\begin{aligned}
  \beta\dot p + \alpha\dot u_{i,i} &= -q_{i,i} +\sum_{a=1}^{n}\mathcal{B}_a\,Q_a(t), \\[2pt]
  \sigma_{ij,j}&=0, \\[2pt]
\end{aligned}
\label{eq:bulk}
\end{equation}
where $u_i$ is the displacement, $p$ is the pore pressure, and
$\dot{\Box}=\partial(\Box)/\partial t$ denotes the time derivative.
Summation over repeated indices $i,j=1,\dots,d$ is assumed. The system is
closed by the constitutive relations, which, under linear small-strain
isotropic elasticity and Darcy's law, are given by
\begin{equation}
\begin{aligned}
  \sigma_{ij} &= \lambda\varepsilon_{kk}\delta_{ij} + 2G\varepsilon_{ij} 
                - \alpha p\delta_{ij},
  \qquad 
  \varepsilon_{ij} := \tfrac12(u_{i,j}+u_{j,i}), \\[2pt]
  q_i &= -\gamma p_{,i},\quad \gamma=\tfrac{k}{\eta}
\end{aligned}
\label{eq:bulk_constitutive}
\end{equation}
where $k$ is the permeability of the porous medium, $\eta$ is the dynamic viscosity of the fluid, and $\beta$ is the mixture compressibility, \textit{i.e.}, the combined compressibility of the fluid and solid. $\lambda$ and $G$ are the Lamé elastic parameters, and $\alpha$ is the Biot-Willis coefficient representing the ratio of changes in the fluid volume to the total bulk volume for deformation at constant pore pressure. All parameters may vary in space; the formulation remains consistent provided their spatial variation is $C^1$. When material properties are discontinuous (e.g., at a material interface), the resulting jumps can be accommodated by appropriately accounting for the jump and average contributions in the crack-face integrals. $Q(t)=[Q_{1}(t),\ldots,Q_{n}(t)]^T$ are flux sources applied through the weights $\mathcal{B}_i$, defined as
\begin{equation}
  \mathcal{B}_i(x) = \begin{cases}
  \frac{1}{|V_i^*|} \quad \textup{if} \quad x \in V_i^* \\ 
  \hspace{4pt} 0 \hspace{14pt} \textup{if} \quad x \notin V_i^*,
  \end{cases}
\end{equation}
where $V_i^*\subset \Omega$ is a prescribed subregion (control volume) over which the
$i$th source acts, and $|V_i^*|$ denotes its measure (length in 1D, area in 2D, volume in 3D),
so that $\int_{V_i^*} \mathcal B_i\,d\Omega = 1$ and $Q_i(t)$ represents the injection
(or production) flux of source $i$.

\subsection{Boundary and interface conditions}
\label{sec:bc_ic}
The external boundary $\partial \Omega$ carries two independent partitions,
one for the mechanical and one for the hydraulic subproblem:
\begin{alignat}{2}
	\partial \Omega &= \Gamma_u \cup \Gamma_t, &\quad \Gamma_u \cap \Gamma_t &= \emptyset
	\qquad\text{(mechanics)},
	\label{eq:bc_partition_mech}
	\\
	\partial \Omega &= \Gamma_p \cup \Gamma_q, &\quad \Gamma_p \cap \Gamma_q &= \emptyset
	\qquad\text{(flow)}.
	\label{eq:bc_partition_flow}
\end{alignat}

For the time-dependent problem, an initial condition for pressure is given by
\begin{equation}\label{eq:p_ic}
	p(x,0) = p_0(x) \quad \text{in } \Omega\setminus\Gamma_c.
\end{equation}
In quasi-static Biot poroelasticity, the momentum 
balance~\cref{eq:bulk} contains no time derivative of $u_i$, so it acts
as an elliptic constraint at each instant. Given the initial 
condition~\cref{eq:p_ic}, the displacement $u_i(\cdot,0)$ is therefore
determined by solving the equilibrium equation with the initial pressure
and boundary conditions.

% ------ Figure: Domain overview ------
\begin{figure*}[t]
	\centering
	\pagefiguregraphics[width=0.8\textwidth]{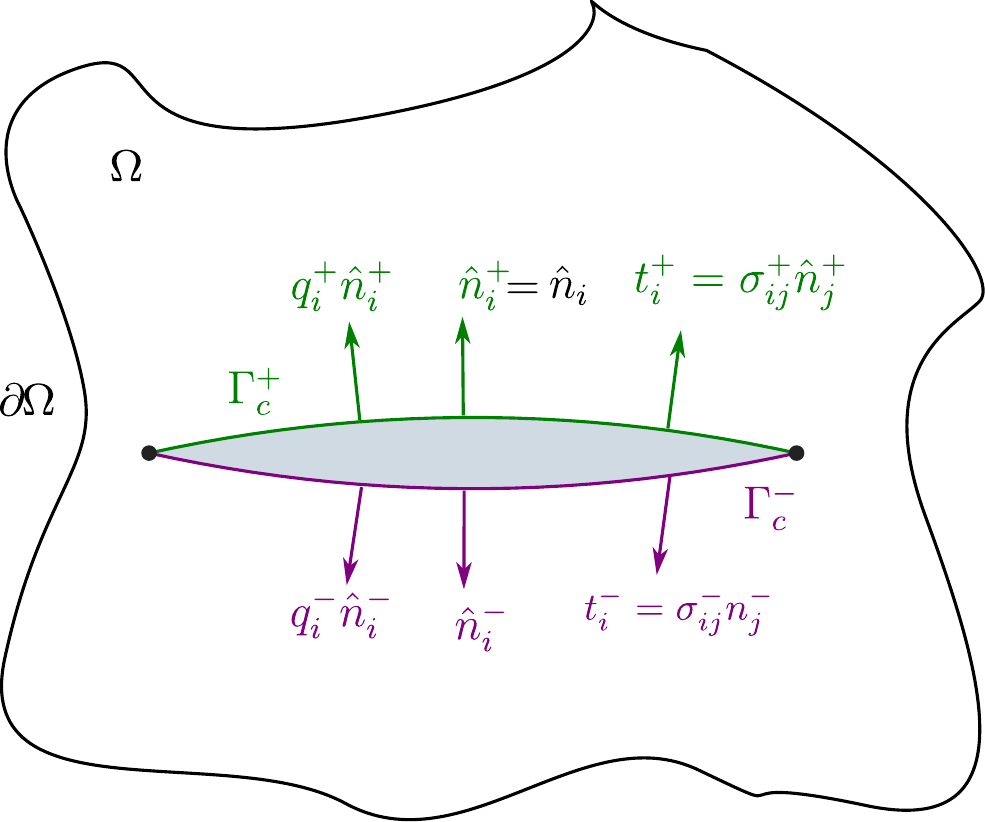}
		\caption{Two-sided notation on the embedded interface $\Gamma_c$. The interface is viewed from its two faces, $\Gamma_c^\pm$, with unit normals $\hat{n}_i^\pm$. The one-sided normal fluxes are $\hat{q}_i^\pm \hat{n}_i^\pm$ and the one-sided tractions are $\hat{t}_i^\pm=\hat{\sigma}_{ij}^\pm \hat{n}_j^\pm$; quantities on the $(+)$ side are shown in green and those on the $(-)$ side in purple.}
	\label{fig:interface_diagram}
\end{figure*}

On the internal interface $\Gamma_c$ we adopt a two-sided description(see~\cref{fig:interface_diagram}): bulk fields admit distinct traces $(\cdot)^+$ and $(\cdot)^-$ from the two adjacent subdomains $\Omega^+$ and $\Omega^-$, and interface relations are expressed through jumps $\jmp{\cdot}$ and averages $\avg{\cdot}$ as defined in~\cref{eq:avgjump_scalar}. We fix a unique reference normal $n_i$ on $\Gamma_c$ oriented from $\Gamma_c^-$ to $\Gamma_c^+$, so that
\begin{equation}\label{eq:normal_convention}
	\hat{n}_i^+ = +\hat{n}_i, \qquad \hat{n}_i^- = -\hat{n}_i,
\end{equation}
where $\hat{n}_i^\pm$ are the unit normals on $\Gamma_c^\pm$, respectively. The Cauchy traction on each face is $\hat{t}_i^\pm = \hat{\sigma}_{ij}^\pm\,\hat{n}_j^\pm$, and rewriting each in terms of
the reference normal gives
\begin{equation}\label{eq:traction_ref_normal}
	\hat{\sigma}_{ij}^+\hat{n}_j = +\hat{t}_i^+, \qquad
	\hat{\sigma}_{ij}^-\hat{n}_j = -\hat{t}_i^-.
\end{equation}
Hence, the jump and average of the traction vector become
\begin{align}
	\jmp{\hat{\sigma}_{ij}}\hat{n}_j
	&= \hat{\sigma}_{ij}^+\hat{n}_j - \hat{\sigma}_{ij}^-\hat{n}_j
	= \hat{t}_i^+ + \hat{t}_i^-=2\avg{\hat{t}_i},
	\label{eq:jump_traction}
	\\
	\avg{\hat{\sigma}_{ij}}\hat{n}_j\,
	&= \tfrac{1}{2}\!\bigl(\hat{\sigma}_{ij}^+\hat{n}_j + \hat{\sigma}_{ij}^-\hat{n}_j\bigr)
	= \tfrac{1}{2}\!\bigl(\hat{t}_i^+ - \hat{t}_i^-\bigr)=\tfrac{1}{2}\jmp{\hat{t}_i}.
	\label{eq:avg_traction}
\end{align}

An analogous decomposition holds for the normal flux. Defining the one-sided normal fluxes through the face-specific normals, $\hat{q}^{n,\pm} := \hat{q}_j^\pm\hat{n}_j^\pm$, and rewriting each in terms of the reference normal gives
\begin{equation}\label{eq:flux_ref_normal}
	\hat{q}_j^+\hat{n}_j = +\hat{q}^{n,+}, \qquad
	\hat{q}_j^-\hat{n}_j = -\hat{q}^{n,-},
\end{equation}
so that
\begin{align}
	\jmp{\hat{q}_j}\hat{n}_j
	&= \hat{q}^{n,+} + \hat{q}^{n,-} = 2\avg{\hat{q}^n},
	\label{eq:jump_flux}
	\\
	\avg{\hat{q}_j}\hat{n}_j
	&= \tfrac{1}{2}\bigl(\hat{q}^{n,+} - \hat{q}^{n,-}\bigr)
	= \tfrac{1}{2}\jmp{\hat{q}^n}.
	\label{eq:avg_flux}
\end{align}

In the present work we consider a zero-thickness interface without additional surface physics, for which conservation of momentum and mass imply
\begin{align}
	\jmp{\hat{\sigma}_{ij}}\hat{n}_j &= 0 \qquad \text{~on~} \Gamma_c,
	\label{eq:ic_traction_balance}
	\\
	\jmp{\hat{q}_i}\hat{n}_i &= 0 \qquad \text{~on~} \Gamma_c.
	\label{eq:ic_flux_balance}
\end{align}
By~\cref{eq:jump_traction,eq:avg_traction}, the traction balance~\cref{eq:ic_traction_balance} is equivalent to the action--reaction condition $\hat{t}_i^+ + \hat{t}_i^- = 0$, and the average reduces to
\begin{equation}\label{eq:avg_equals_transmitted}
	\avg{\hat{\sigma}_{ij}}\hat{n}_j = \hat{t}_i^+=\hat{t}_i,
\end{equation}
i.e., the average traction equals the traction from either side. An analogous reduction holds for the normal flux. More general interface models (e.g., reduced-dimensional fracture flow with interfacial storage, or mechanics with surface stresses) can be accommodated within the same jump/average
framework~\citep{chambat2014jump,dugstad2022dimensional,zemlyanova2023numerical}.

The system is closed by constitutive relations prescribing the traction $\avg{\hat{\sigma}_{ij}}\hat{n}_j$ and the average normal flux $\avg{\hat{q}_i}\hat{n}_i$ as functions of the two-sided primary fields. In jump/average form, we write
\begin{align}
	\avg{\hat{q}_i}\hat{n}_i
	&= \mathsf{J}_\Gamma\!\bigl(\jmp{\hat{p}},\avg{\hat{p}},\jmp{\hat{u}_i},\avg{\hat{u}_i},
	\jmp{\dot{\hat{p}}},\jmp{\dot{\hat{u}}_i},\hat{\theta};\hat{x},t\bigr)
	\quad &&  \text{~on~} \Gamma_c,
	\label{eq:interface_closure_flux}
	\\
	\avg{\hat{\sigma}_{ij}}\hat{n}_j
	&= \mathsf{T}_{\Gamma,i}\!\bigl(\jmp{\hat{u}_j},\avg{\hat{u}_j},\jmp{\hat{p}},\avg{\hat{p}},
	\jmp{\dot{\hat{u}}_j},\jmp{\dot{\hat{p}}},\hat{\theta};\hat{x},t\bigr)
	\quad && \text{~on~} \Gamma_c,
	\label{eq:interface_closure_traction}
\end{align}
where $\mathsf{J}_\Gamma$ and $\mathsf{T}_{\Gamma,i}$ encode the chosen hydraulic and mechanical interface laws, and $\hat{\theta}(\hat{x},t)$ collects any internal state variables residing on $\Gamma_c$ (e.g.\ the state variable in rate-and-state friction, accumulated plastic slip, or interfacial damage), governed by their own evolution equations on $\Gamma_c$. Since these constitutive choices are problem-dependent and largely orthogonal to the shifted-interface formulation developed below, we defer their explicit specification to \cref{sec:constitutive_closure}.

\subsection{Variational formulation on the computational domain}
\label{sec:discretization_prelim}

\begin{figure*}[t]
	\centering
	\pagefiguregraphics{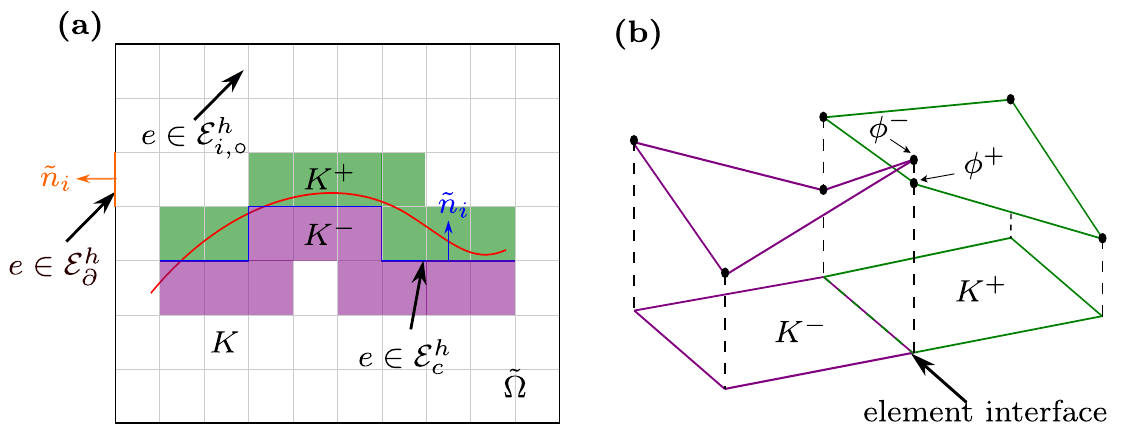}
	\caption{Computational mesh and two-sided notation. 
		(a) The physical interface $\Gamma_c$ (red) is approximated by a mesh-aligned surrogate $\tilde\Gamma_c$ (blue) composed of faces $\mathcal{E}^h_c \subset \mathcal{E}^h_i$; the remaining interior faces are $\mathcal{E}^h_{i,\circ}$ and boundary faces are $\mathcal{E}^h_\partial$.
		(b) On an interior face $e$ shared by elements $K^-$ and $K^+$, a unit normal $\tilde n_i$ is fixed from $K^-$ to $K^+$, defining one-sided traces $\phi^\pm=\phi|_{K^\pm}$ and the jump/average operators \eqref{eq:avgjump_scalar}. On boundary faces, $\tilde n_i$ denotes the outward unit normal.}
	\label{fig:mesh_notation}
\end{figure*}

Let $\tilde\Omega \approx \Omega$ be the computational domain and let $\mathcal{T}^h$ be a conforming partition of $\tilde\Omega$ into open elements $K \in \mathcal{T}^h$. The set of all (open) $(d-1)$-dimensional faces induced by $\mathcal{T}^h$ is denoted $\mathcal{E}^h$. Boundary faces form $\mathcal{E}^h_\partial$ and $\mathcal{E}^h_i$ contains the set of interior faces (see~\cref{fig:mesh_notation}).

The true interface $\Gamma_c$ is approximated by a \emph{surrogate} $\tilde\Gamma_c \subset \mathrm{skel}(\mathcal{T}^h)$, consisting of a subset of mesh faces. We denote by
$\mathcal{E}^h_c \subset \mathcal{E}^h_i$ the surrogate crack faces; the remaining interior faces are
$\mathcal{E}^h_{i,\circ} := \mathcal{E}^h_i \setminus \mathcal{E}^h_c$.

For any face $e \in \mathcal{E}^h_i$ shared by elements $K^-$ and $K^+$, we fix a unit normal $\tilde{n}_i$ oriented from $K^-$ to $K^+$, consistent with the convention adopted for the physical interface in~\cref{sec:bc_ic} (reference normal from the ``$-$'' to the ``$+$'' side). One-sided traces are $\phi^\pm := \phi|_{K^\pm}$, with jump and average as in~\cref{eq:avgjump_scalar}. Note that the outward unit normal on $e$ is $+\tilde{n}_i$ for $K^-$ and $-\tilde{n}_i$ for $K^+$, so that the element-boundary integrals arising from integration by parts carry signs consistent with~\cref{eq:traction_ref_normal}. On boundary faces $e \subset \partial\tilde\Omega$, $\tilde{n}_i$ denotes the outward unit normal.

% ------ Figure: Domain overview ------

We work in a broken setting: fields are assumed sufficiently regular on each element for one-sided traces to exist, but no inter-element continuity is imposed a priori. Integrating~\cref{eq:bulk} by parts over each element and summing yields: find $(p,\,u_i)$ such that for all suitable test functions $(v,\,w_i)$,
\begin{equation}
	\begin{aligned}
		0 ={}& \int_{\tilde\Omega} \bigl(\beta\,\dot p + \alpha\,\dot u_{k,k}\bigr)v\;d\Omega
		+ \int_{\tilde\Omega} \gamma p_{,i}v_{,i}\;d\Omega
		- \mathcal{I}^{\mathrm{flow}}_{\mathcal{E}^h_i}(p,v)
		\\
		&+ \int_{\partial\tilde\Omega} q_i\tilde{n}_iv\;d\tilde{s}
		- \int_{\tilde\Omega} fv\;d\tilde{\Omega}
		+ \int_{\tilde\Omega} \sigma_{ij}w_{i,j}\;d\tilde{\Omega}
		\\
		&+ \mathcal{I}^{\mathrm{mech}}_{\mathcal{E}^h_i}(u,p,w)
		- \int_{\partial\tilde\Omega} \sigma_{ij}\tilde{n}_jw_i\;d\tilde{s},
	\end{aligned}
	\label{eq:weak_surrogate}
\end{equation}
where the first line is the flow residual (tested by $v$) and the second is the momentum residual (tested by $w_i$), and $f:=\sum_{a=1}^{n}\mathcal{B}_a(x)Q_a(t)$ is the volumetric fluid source from~\cref{eq:bulk}. The interior-face functionals collect contributions from element boundaries:
\begin{equation}
	\begin{aligned}
		\mathcal{I}^{\mathrm{flow}}_{\mathcal{E}^h_i}(p,v)
		&:= \sum_{e \in \mathcal{E}^h_i} \int_e \bigl(
		\avg{q_i}\tilde{n}_i\jmp{v}
		+ \jmp{q_i}\tilde{n}_i\avg{v}
		\bigr)d\tilde{s}, \\[4pt]
		\mathcal{I}^{\mathrm{mech}}_{\mathcal{E}^h_i}(u,p,w)
		&:= \sum_{e \in \mathcal{E}^h_i} \int_e \bigl(
		\avg{\sigma_{ij}}\tilde{n}_j\jmp{w_i}
		+ \jmp{\sigma_{ij}}\tilde{n}_j\avg{w_i}
		\bigr)d\tilde{s}.
	\end{aligned}
	\label{eq:interface_functionals}
\end{equation}
The physical and numerical treatment of each face is encoded through face-wise constitutive or numerical closure relations that constrain the jumps and averages appearing in~\cref{eq:interface_functionals}.

\subsection{Face-type classification and closure}
\label{sec:face_closure}

The interior faces split naturally into two disjoint subsets,
\begin{equation}
	\mathcal{E}^h_i
	= \mathcal{E}^h_{i,\circ} \;\cup\; \mathcal{E}^h_c,
	\label{eq:face_split}
\end{equation}
where $\mathcal{E}^h_{i,\circ}$ collects ordinary bulk faces away from the interface and $\mathcal{E}^h_c$ collects surrogate-interface faces. The interface functionals in~\cref{eq:weak_surrogate} decompose accordingly:
\begin{equation}
	\mathcal{I}^{\mathrm{flow}}_{\mathcal{E}^h_i}
	= \mathcal{I}^{\mathrm{flow}}_{\mathcal{E}^h_{i,\circ}}
	+ \mathcal{I}^{\mathrm{flow}}_{\mathcal{E}^h_c},
	\qquad
	\mathcal{I}^{\mathrm{mech}}_{\mathcal{E}^h_i}
	= \mathcal{I}^{\mathrm{mech}}_{\mathcal{E}^h_{i,\circ}}
	+ \mathcal{I}^{\mathrm{mech}}_{\mathcal{E}^h_c}.
	\label{eq:I_split}
\end{equation}
The two subsets are closed by different mechanisms. If the discrete spaces enforce $C^0$ continuity across $\mathcal{E}^h_{i,\circ}$---that is, the mesh is split only along $\tilde\Gamma_c$ and the basis functions are continuous on $\tilde\Omega\setminus\tilde\Gamma_c$---then jumps vanish identically on ordinary interior faces and hence $\mathcal{I}^{\mathrm{flow}}_{\mathcal{E}^h_{i,\circ}} = \mathcal{I}^{\mathrm{mech}}_{\mathcal{E}^h_{i,\circ}} = 0$; only the surrogate-interface contributions persist. This is the continuous Galerkin (CG) setting adopted in the numerical examples of this work. Alternatively, if fully discontinuous polynomial spaces are used, inter-element continuity on $\mathcal{E}^h_{i,\circ}$ is enforced weakly through numerical fluxes and penalty terms, as in the SIPG, NIPG, and IIPG formulations~\citep{Riviere2008}; the particular variant affects the symmetry of the bilinear form and the required penalty magnitude but does not alter the structure of the surrogate-interface terms. When a DG bulk discretization is adopted, Dirichlet boundary conditions are also imposed weakly via Nitsche-type terms on $\mathcal{E}^h_\partial$, whereas for CG discretizations they are imposed strongly through the function space.

On the surrogate-interface faces $\mathcal{E}^h_c$, inter-element continuity is \emph{not} enforced. Instead, $\mathcal{I}^{\mathrm{flow}}_{\mathcal{E}^h_c}$ and $\mathcal{I}^{\mathrm{mech}}_{\mathcal{E}^h_c}$ are closed by transferring the physical interface conditions from $\Gamma_c$ to $\tilde\Gamma_c$ via the shifted interface method developed in \cref{sec:sbm}. This transfer is independent of the CG or DG choice made on $\mathcal{E}^h_{i,\circ}$.

\section{The shifted interface method}
\label{sec:sbm}

The key feature of the method is that the \textit{true} crack $\Gamma_c$ does not align with the computational mesh. The shifted interface method addresses this by transferring interface conditions from $\Gamma_c$ to the surrogate $\tilde\Gamma_c$ via local expansions. In the following, we present the derivation procedure of the shifted interface method (cf \cite{li2023simple}).

% ------ Figure: Domain overview ------
\begin{figure*}[t]
	\centering
	\pagefiguregraphics[width=0.8\textwidth]{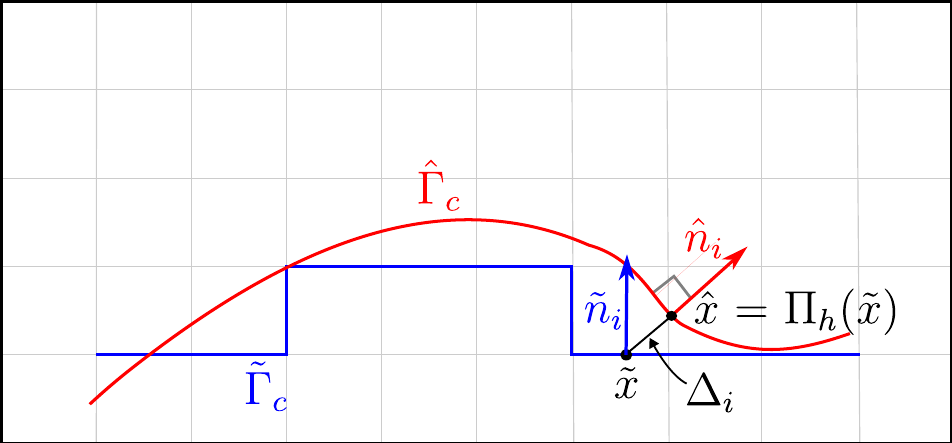}
	\caption{Shifted-interface geometry near the surrogate.
		For each point $\tilde x\in\tilde\Gamma_c$, the closest-point projection
		$\hat x=\Pi_h(\tilde x)\in\Gamma_c$ defines the gap vector
		$\Delta_i(\tilde x)=\hat x_i-\tilde x_i$.
		By the nearest-point property, $\Delta_i$ is normal to $\Gamma_c$ at $\hat x$
		(right-angle marker), hence $\Delta_i$ is parallel to the true normal $\hat n_i$.}
	\label{fig:shifted_interface}
\end{figure*}

\subsection{Shifts}
\label{sec:taylor_shifts}
We adopt here Taylor expansions as in~\cite{Li2020,Atallah2022,li2023simple,li2024complex,li2025shifted}. Other expansions are also possible in this framework (see~\cite{zorrilla2024shifted}), but our choice provides a good balance between accuracy and implementation simplicity. The surrogate interface is determined as follows.

For $\tilde{x} \in \tilde\Gamma_c$, define the mesh-dependent closest-point projection
\begin{equation}
	\Pi_h:\ \tilde\Gamma_c \to \Gamma_c, \qquad
	\tilde{x}\ \mapsto\ \hat{x} := \Pi_h(\tilde{x}),
	\label{eq:Pi_h}
\end{equation}
and the \emph{gap} (shift) vector
\begin{equation}
	\Delta_i(\tilde{x}) := \hat{x}_i - \tilde{x}_i,
	\label{eq:gap}
\end{equation}
with $\tilde{n}_i$ being the unit normal to $\tilde\Gamma_c$ at $\tilde{x}$ and $\hat{n}_i$ the unit normal to $\Gamma_c$ at $\hat{x} = \Pi_h(\tilde{x})$, as shown in~\cref{fig:shifted_interface}.

For any sufficiently smooth scalar field $\phi$, the first-order Taylor expansion gives
\begin{equation}
	\hat{\phi}
	= \tilde{\phi} + \Delta_j\tilde{\phi}_{,j}
	+ \mathcal{O}(\|\boldsymbol\Delta\|^2),
	\label{eq:SBM-value}
\end{equation}
and similarly for a vector (or tensor) quantity $\Phi_i$,
\begin{equation}
	\hat{\Phi}_i
	= \tilde{\Phi}_i + \Delta_\ell\tilde{\Phi}_{i,\ell}
	+ \mathcal{O}(\|\boldsymbol\Delta\|^2).
	\label{eq:vector-shift}
\end{equation}
Hereafter, $\mathcal{O}(\|\boldsymbol\Delta\|^2)$ remainders are dropped, and all shifted expressions are understood as first-order approximations.

\subsection{Derivation of the shifted interface forms}
\label{sec:derivation}

We now derive the shifted-interface contributions to the variational form by systematically applying the expansions of~\cref{sec:taylor_shifts} to the crack-face functionals.

The algebraic structure of the shift is identical for both flow and mechanics. To exploit this, we introduce a generic interface functional in which $\Phi_j$ denotes a vector-valued flux and $\psi$ a scalar test function on the surrogate. For flow, $\Phi_j = q_j$ and $\psi = v$. For mechanics, the generic framework is applied component-wise: for each displacement component $i = 1,\dots,d$, one sets $\Phi_j \to\sigma_{ij}$ (the $i$-th row of the Cauchy stress, a vector in~$j$) and $\psi \to w_i$; the full mechanics contribution is then obtained by summing over~$i$. The interface functional takes the general form
\begin{equation}
	\mathcal{I}_{\mathcal{E}^h_c}(\Phi, \psi)
	= \sum_{e \in \mathcal{E}^h_c} \int_e \bigl(
	\avg{\tilde{\Phi}_j}\tilde{n}_j\jmp{\psi}
	+ \jmp{\tilde{\Phi}_j}\tilde{n}_j\avg{\psi}
	\bigr)d\tilde{s},
	\label{eq:generic_interface}
\end{equation}
where tildes emphasize evaluation at the surrogate location $\tilde{x}$.

To relate surrogate and true interface quantities, we decompose the surrogate normal into its projection onto the true normal and its component orthogonal to the true normal (equivalently, tangential to $\Gamma_c$ at $\hat{x}$):
\begin{equation}
	\tilde{n}_j
	= (\tilde{n}_k\hat{n}_k)\hat{n}_j + \tilde{n}^{\perp}_j,
	\qquad
	 \tilde{n}^{\perp}_j := \tilde{n}_j - (\tilde{n}_k\hat{n}_k)\hat{n}_j.
	\label{eq:normal_decomp}
\end{equation}
Since $\tilde{n}_j$ and $\hat{n}_j$ are unit vectors, we define the mismatch angle $\varphi$ by
$\cos\varphi := \tilde{n}_k\hat{n}_k$. By construction $\tilde{n}^{\perp}_k\hat{n}_k = 0$, so $\tilde{n}^{\perp}_j$ lies in the tangent space of $\Gamma_c$ at $\hat{x} = \Pi_h(\tilde{x})$.

Applying the decomposition~\cref{eq:normal_decomp} to the average term in~\cref{eq:generic_interface} gives
\begin{equation}
	\avg{\tilde{\Phi}_j}\tilde{n}_j
	= \cos\varphi\avg{\tilde{\Phi}_j}\hat{n}_j
	+ \avg{\tilde{\Phi}_j}\tilde{n}^{\perp}_j.
	\label{eq:generic_decomp}
\end{equation}
The first term involves the normal component evaluated at the surrogate location. Rearranging the Taylor expansion~\cref{eq:vector-shift} to express surrogate values in terms of true-interface values, $\tilde{\Phi}_j= \hat{\Phi}_j - \Delta_\ell\tilde{\Phi}_{j,\ell}$ and substituting into the average gives
\begin{equation}
	\avg{\tilde{\Phi}_j}\hat{n}_j
	= \avg{\hat{\Phi}_j}\hat{n}_j
	- \avg{\Delta_\ell\tilde{\Phi}_{j,\ell}}\hat{n}_j.
	\label{eq:generic_shift_avg}
\end{equation}
Combining~\cref{eq:generic_decomp,eq:generic_shift_avg} yields the three-part structure:
\begin{equation}
	\avg{\tilde{\Phi}_j}\tilde{n}_j
	= \cos\varphi\avg{\hat{\Phi}_j}\hat{n}_j
	- \cos\varphi\avg{\Delta_\ell\tilde{\Phi}_{j,\ell}}\hat{n}_j
	+ \avg{\tilde{\Phi}_j}\tilde{n}^{\perp}_j.
	\label{eq:generic_complete}
\end{equation}
The same procedure applied to the jump term gives
\begin{equation}
	\jmp{\tilde{\Phi}_j}\tilde{n}_j
	= \cos\varphi\jmp{\hat{\Phi}_j}\hat{n}_j
	- \cos\varphi\jmp{\Delta_\ell\tilde{\Phi}_{j,\ell}}\hat{n}_j
	+ \jmp{\tilde{\Phi}_j}\tilde{n}^{\perp}_j.
	\label{eq:generic_jump_complete}
\end{equation}

Replacing back to the fields for the flow subproblem, $\Phi_j = q_j$ and $\psi = v$, substituting~\cref{eq:generic_complete,eq:generic_jump_complete}
becomes
\begin{equation}
	\begin{aligned}
		\mathcal{I}^{\mathrm{flow}}_{\mathcal{E}^h_c}(p,v)
		={}& \sum_{e \in \mathcal{E}^h_c} \int_e \Bigl\{
		\cos\varphi\,\avg{\hat{q}_j}\hat{n}_j\jmp{v}
		+ \cos\varphi\,\jmp{\hat{q}_j}\hat{n}_j\avg{v}
		\\
		&\qquad
		+ \avg{\tilde{q}_j}\tilde{n}^{\perp}_j\jmp{v}
		+ \jmp{\tilde{q}_j}\tilde{n}^{\perp}_j\avg{v}
		\\
		&\qquad
		- \cos\varphi\,\avg{\Delta_\ell\tilde{q}_{j,\ell}}\hat{n}_j\jmp{v}
		\\
		&\qquad
		- \cos\varphi\,\jmp{\Delta_\ell\tilde{q}_{j,\ell}}\hat{n}_j\avg{v}
		\Bigr\}d\tilde{s}.
	\end{aligned}
	\label{eq:flow_crack_general}
\end{equation}
Summing the generic result over displacement components $i = 1,\dots,d$ and identifying $\avg{\hat{\sigma}_{ij}}\hat{n}_j = \avg{\hat{t}_i}$, $\jmp{\hat{\sigma}_{ij}}\hat{n}_j = \jmp{\hat{t}_i}$, the mechanics interface functional reads
\begin{equation}
	\begin{aligned}
		\mathcal{I}^{\mathrm{mech}}_{\mathcal{E}^h_c}(u,p,w)
		={}& \sum_{e \in \mathcal{E}^h_c} \int_e \Big(
		\tfrac{1}{2}\cos\varphi\,\jmp{\hat{t}_i}\jmp{w_i}
		+ 2\cos\varphi\,\avg{\hat{t}_i}\avg{w_i}
		\Big)d\tilde{s}
		\\[2pt]
		&+ \mathcal{M}^{\perp}_{\mathcal{E}^h_c}(\tilde{\sigma},w)
		+ \mathcal{M}^{\Delta}_{\mathcal{E}^h_c}(\tilde{\sigma},w).
	\end{aligned}
	\label{eq:mech_crack_general}
\end{equation}
where the geometry-induced correction pieces, defined for succinctness, are
\begin{equation*}
	\begin{aligned}
		\mathcal{M}^{\perp}_{\mathcal{E}^h_c}(\tilde{\sigma},w)
		:={}& \sum_{e \in \mathcal{E}^h_c} \int_e
		\avg{\tilde{\sigma}_{ij}}\tilde{n}^{\perp}_j\jmp{w_i}\,d\tilde{s}
		\\
		&+ \sum_{e \in \mathcal{E}^h_c} \int_e
		\jmp{\tilde{\sigma}_{ij}}\tilde{n}^{\perp}_j\avg{w_i}\,d\tilde{s},
		\\[4pt]
		\mathcal{M}^{\Delta}_{\mathcal{E}^h_c}(\tilde{\sigma},w)
		:={}& -\sum_{e \in \mathcal{E}^h_c} \int_e
		\cos\varphi\,
		\avg{\Delta_\ell\tilde{\sigma}_{ij,\ell}}\hat{n}_j\jmp{w_i}\,d\tilde{s}
		\\
		&- \sum_{e \in \mathcal{E}^h_c} \int_e
		\cos\varphi\,
		\jmp{\Delta_\ell\tilde{\sigma}_{ij,\ell}}\hat{n}_j\avg{w_i}\,d\tilde{s}.
	\end{aligned}
\end{equation*}

\subsection{Interface constitutive closure on the true interface}
\label{sec:constitutive_closure}

The shifted forms~\cref{eq:flow_crack_general,eq:mech_crack_general} are closed once relations on the true interface $\Gamma_c$ are specified for the normal flux and the traction. These relations encode the interface physics---for example, a leaky hydraulic barrier, an impermeable membrane, a bonded interface, frictional contact, or cohesive behavior.

For flow, the interface model provides relations for $\avg{\hat{q}_j}\hat{n}_j$ and $\jmp{\hat{q}_j}\hat{n}_j$ in terms of the pressure traces and, if relevant, additional interface variables. For mechanics, the model provides relations for $\avg{\hat{t}_i}$ and $\jmp{\hat{t}_i}$, or equivalently, the traction $\hat{t}_i$ from~\cref{eq:avg_equals_transmitted}, which is a function of the displacement jump, its rate, and possibly pressure or other state variables.

The SIM is then applied to express the true-interface arguments (for example $\hat{p}^\pm$ and $\hat{u}_i^\pm$) in terms of surrogate quantities at $\tilde\Gamma_c$ using the Taylor expansions~\cref{eq:SBM-value,eq:vector-shift}. Substituting these expressions into the chosen interface law yields a closed formulation posed entirely on the computational mesh.

In the special case of a hydraulic barrier, conservation across the interface gives continuity of normal flux,
\begin{equation}
	\jmp{\hat{q}_j}\hat{n}_j = 0,
	\label{eq:flux_conservation}
\end{equation}
and the Robin transmission condition relates the average normal flux to the pressure jump,
\begin{equation}
	\avg{\hat{q}_j}\hat{n}_j = -T_n(\hat{x})\jmp{\hat{p}}.
	\label{eq:robin_flux}
\end{equation}
where $T_{n}(\hat{x})$ is the normal transmissivity. The true-interface pressure traces are expressed in terms of surrogate values by the Taylor shift~\cref{eq:SBM-value}.

For the mechanical response, the traction balance~\cref{eq:ic_traction_balance} gives $\avg{\hat{t}_i}=0$, so that by~\cref{eq:avg_equals_transmitted} the traction $\hat{t}_i$ satisfies $\jmp{\hat{t}_i}=2\hat{t}_i$. We prescribe $\hat{t}_i$ through the linear elastic-viscous spring law
\begin{equation}
	\hat{t}_i
	= K_{ij}(\hat{x})\jmp{\hat{u}_j}
	+ H_{ij}(\hat{x})\jmp{\dot{\hat{u}}_j},
	\label{eq:spring_law}
\end{equation}
where $K_{ij}(\hat{x})$ is the interface stiffness tensor and $H_{ij}(\hat{x})$ is the interface viscosity tensor, both assumed symmetric and positive semi-definite. We introduce a local orthonormal frame $\{R_{iA}\}_{A=0}^{d-1}$ at each point of $\Gamma_c$, where the composite index $A\in\{0,1,\ldots,d{-}1\}$ labels frame directions: $A=0$ corresponds to the normal, $A=\alpha$ ($\alpha=1,\ldots,d{-}1$) to the tangential directions.  Concretely, $R_{i0}=\hat{n}_i$ and $R_{i\alpha}=\hat{m}_i^{(\alpha)}$, where $\hat{m}_i^{(\alpha)}$ span the tangent plane of $\Gamma_c$ at $\hat{x}$.  Then
\begin{equation}
	K_{ij}=R_{iA}\bar{K}_{AB}R_{jB},
	\qquad
	H_{ij}=R_{iA}\bar{H}_{AB}R_{jB},
	\label{eq:KH_rotation_form}
\end{equation}
with local tensors $\bar{K}_{AB}$ and $\bar{H}_{AB}$ symmetric and positive semi-definite. In block form ($A$ partitioned as $0\mid\alpha$ and $B$ as $0\mid\beta$),
\begin{equation}
	\bar{K}_{AB}=
	\begin{bmatrix}
		K^{nn} & K^{nt}_{\beta} \\
		K^{nt}_{\alpha} & K^{tt}_{\alpha\beta}
	\end{bmatrix},
	\label{eq:stiffness_components}
\end{equation}
and analogously for $\bar{H}_{AB}$, where $K^{nn}$ governs normal opening stiffness, $K^{tt}_{\alpha\beta}$ governs tangential slip stiffness, and $K^{nt}_{\alpha}$ captures normal--tangential coupling.

Several physically relevant configurations are recovered by appropriate choices of these parameters and the transmissivity. In many applications, the normal--tangential coupling vanishes ($K^{n\!t} = 0$), so that~\cref{eq:spring_law} reduces to independent normal and tangential spring--dashpot pairs---this diagonal case is adopted in the numerical examples of this work. A traction-free crack corresponds to $K_{ij} = H_{ij} = 0$, so that $\hat{t}_i = 0$ and the interface offers no resistance to opening or sliding. At the opposite extreme, taking $K^{nn} \to \infty$ and $K^{tt}_{\alpha\beta} \to \infty$ enforces $\jmp{\hat{u}_j} \to \mathbf{0}$, recovering a fully bonded interface. An intermediate case of particular interest is frictionless contact, obtained by setting $K^{tt}_{\alpha\beta} = 0$ while retaining $K^{nn} > 0$: the interface resists normal opening but permits free tangential slip. For the hydraulic parameters, $T_n \to 0$ yields an impermeable interface with
$\avg{\hat{q}_j}\hat{n}_j \to 0$, while $T_n \to \infty$ enforces pressure continuity $\jmp{\hat{p}} \to 0$, provided the normal flux remains bounded. Note that for all cases considered within the current paper, the viscosity matrix $H_{ij}=0$ and also the cross-coupling between normal and tangential in the stiffness $K^{nt}=0$, but they are kept for the remaining derivations for generality. Finally, form~\cref{eq:stiffness_components} reuses the same form in the case of nonlinear  (e.g., elastoplasticity) incremental formulation, but here we will stick to the linear elastic constitutive laws.

\subsection{Enforcement of the constitutive relations}
\label{sec:enforcement}

\Cref{sec:derivation,sec:constitutive_closure} provide, respectively, the general shifted crack-face functionals~\cref{eq:flow_crack_general,eq:mech_crack_general} and the constitutive relations posed on the true interface $\Gamma_c$ together with their transfer to the surrogate $\tilde\Gamma_c$ via the SIM formulation. The remaining task is to incorporate these relations into the discrete
formulation.

Two different assembly viewpoints are useful. In the first, the transferred constitutive law is substituted directly into the shifted crack-face functionals, so that the interface physics enters only through face integrals during assembly. In the second, selected interface quantities are treated as additional unknowns supported on $\tilde\Gamma_c$, and the constitutive law is enforced as
explicit algebraic relations during the time integration.

A remark on terminology is in order. The labels ``weak'' and ``strong'' adopted here refer to the manner in which the interface constitutive law enters the discrete system, not to the Nitsche-type weak imposition of boundary conditions used in CutFEM. In particular, the strong enforcement strategy is \emph{not} a classical Lagrange multiplier approach: the auxiliary interface unknowns do not arise from a saddle-point variational principle, and the constitutive relations are enforced pointwise (i.e., collocated) at the interface degrees of freedom. Nevertheless, the method is not purely collocated either, since the feedback from the interface unknowns into the bulk equations retains a variational (integral) structure through the coupling blocks $\mathbf{G}_p$ and $\mathbf{G}_u$.

\subsubsection{Weak enforcement}
\label{sec:enforcement_substitution}

In the weak enforcement approach, the transferred closure relations
from~\cref{sec:constitutive_closure} are inserted
into~\cref{eq:flow_crack_general,eq:mech_crack_general}. For the
hydraulic barrier with linear elastic spring law
of~\cref{sec:constitutive_closure}, this produces closed crack-face
contributions posed entirely on the computational mesh.

For flow, substituting the Robin
condition~\cref{eq:robin_flux} and flux
conservation~\cref{eq:flux_conservation} into~\cref{eq:flow_crack_general},
with the true-interface pressure jump expressed via~\cref{eq:SBM-value}
as $\jmp{\hat{p}} = \jmp{\tilde{p} + \Delta_\ell\tilde{p}_{,\ell}}$,
gives
\begin{equation}
	\begin{aligned}
		\mathcal{I}^{\mathrm{flow}}_{\mathcal{E}^h_c}(p,v)
		={}& \sum_{e \in \mathcal{E}^h_c} \int_e \Bigl\{
		-\cos\varphi\,T_n(\hat{x})
		\jmp{\tilde{p} + \Delta_\ell\tilde{p}_{,\ell}}
		\jmp{v}
		\\
		&\qquad
		+ \avg{\tilde{q}_j}\tilde{n}^{\perp}_j\jmp{v}
		+ \jmp{\tilde{q}_j}\tilde{n}^{\perp}_j\avg{v}
		\\
		&\qquad
		- \cos\varphi\,
		\avg{\Delta_\ell\tilde{q}_{j,\ell}}\hat{n}_j\jmp{v}
		\\
		&\qquad
		- \cos\varphi\,
		\jmp{\Delta_\ell\tilde{q}_{j,\ell}}\hat{n}_j\avg{v}
		\Bigr\}d\tilde{s}.
	\end{aligned}
	\label{eq:flow_substituted}
\end{equation}
The first term is the Robin penalty transferred to the surrogate; the second term on the first line of~\cref{eq:flow_crack_general} vanishes by flux conservation. The second line collects the normal mismatch, and the third accounts for first-order Taylor corrections.

For mechanics, the traction balance~\cref{eq:ic_traction_balance} gives $\avg{\hat{t}_i} = 0$, eliminating the second term on the first line of~\cref{eq:mech_crack_general}, while $\tfrac{1}{2}\jmp{\hat{t}_i} = \hat{t}_i$ by~\cref{eq:avg_equals_transmitted}. Substituting the spring law~\cref{eq:spring_law} with the displacement jump expressed via~\cref{eq:vector-shift} as $\jmp{\hat{u}_j} = \jmp{\tilde{u}_j + \Delta_\ell\tilde{u}_{j,\ell}}$ (and analogously for the rate) yields
\begin{equation}
	\begin{aligned}
		\mathcal{I}^{\mathrm{mech}}_{\mathcal{E}_c^h}(u,p,w)
		={}& \sum_{e\in\mathcal{E}_c^h} \int_e \cos\varphi
		\Bigl(
		K_{ij}\jmp{\tilde{u}_j + \Delta_\ell \tilde{u}_{j,\ell}}
		+ H_{ij}\jmp{\dot{\tilde{u}}_j + \Delta_\ell \dot{\tilde{u}}_{j,\ell}}
		\Bigr)
		\jmp{w_i}\, d\tilde{s}
		\\
		&\quad + \mathcal{M}^{\perp}_{\mathcal{E}_c^h}(\tilde{\sigma},w)
		+ \mathcal{M}^{\Delta}_{\mathcal{E}_c^h}(\tilde{\sigma},w).
	\end{aligned}
	\label{eq:mech_substituted}
\end{equation}
where $K_{ij}$ and $H_{ij}$ are evaluated at $\hat{x} = \Pi_h(\tilde{x})$. The first line contains the full interface constitutive response transferred to the surrogate. The terms
$\mathcal{M}^{\perp}_{\mathcal{E}^h_c}$ and $\mathcal{M}^{\Delta}_{\mathcal{E}^h_c}$ collect the normal mismatch and first-order Taylor corrections, respectively.

\subsubsection{Strong enforcement}
\label{sec:enforcement_unknowns}

An alternative to weak enforcement is to treat the interface quantities, i.e., the one-sided tractions and normal fluxes, as additional independent unknowns defined on $\tilde\Gamma_c$ rather than expressing them directly in terms of the bulk fields. The shifted face functionals are then assembled in terms of these unknowns, and the constitutive law is enforced as separate algebraic equations at the interface degrees of freedom. This decoupling is particularly convenient for nonlinear interface laws (e.g., coulomb friction, elastoplasticity, damage), since the interface physics enters only through a pointwise update at the interface nodes and need not be embedded in the element-level variational assembly.

For flow, the one-sided normal fluxes
$\hat{q}^{n,\pm} := \hat{q}_j^\pm\hat{n}_j^\pm$ are introduced
as interface unknowns. By~\cref{eq:jump_flux,eq:avg_flux}, the
leading terms of~\cref{eq:flow_crack_general} become
\begin{equation}
	\begin{aligned}
		\mathcal{I}^{\mathrm{flow}}_{\mathcal{E}^h_c}(p,v)
		={}& \sum_{e \in \mathcal{E}^h_c} \int_e \Bigl\{
		\tfrac{1}{2}\cos\varphi\,\jmp{\hat{q}^n}\jmp{v}
		+ 2\cos\varphi\,\avg{\hat{q}^n}\avg{v}
		\\
		&\qquad
		+ \avg{\tilde{q}_j}\tilde{n}^{\perp}_j\jmp{v}
		+ \jmp{\tilde{q}_j}\tilde{n}^{\perp}_j\avg{v}
		\\
		&\qquad
		- \cos\varphi\,
		\avg{\Delta_\ell\tilde{q}_{j,\ell}}\hat{n}_j\jmp{v}
		\\
		&\qquad
		- \cos\varphi\,
		\jmp{\Delta_\ell\tilde{q}_{j,\ell}}\hat{n}_j\avg{v}
		\Bigr\}d\tilde{s}.
	\end{aligned}
	\label{eq:flow_unknown_form}
\end{equation}
At each interface degree of freedom on $\tilde\Gamma_c$, the constitutive law is enforced as the algebraic system
\begin{equation}
	\begin{aligned}
		\jmp{\hat{q}^n}
		&= -2T_n(\hat{x})
		\jmp{\tilde{p} + \Delta_\ell\tilde{p}_{,\ell}},
		\\
		\avg{\hat{q}^n} &= 0,
	\end{aligned}
	\label{eq:flow_algebraic_closure}
\end{equation}
where the first relation follows from the Robin condition~\cref{eq:robin_flux} via~\cref{eq:avg_flux}, and the second from flux conservation~\cref{eq:flux_conservation} via~\cref{eq:jump_flux}. True-interface pressure traces have been expressed in terms of surrogate quantities via~\cref{eq:SBM-value}. For the zero-thickness interface considered here, the second relation eliminates $\avg{\hat{q}^n}$; more general models with interfacial storage or fluid-conductive fractures prescribe a nontrivial relation for $\avg{\hat{q}^n}$ as well.

For mechanics, the one-sided tractions $\hat{t}_i^\pm = \hat{\sigma}_{ij}^\pm\hat{n}_j^\pm$ as defined
in~\cref{sec:bc_ic} are introduced as interface unknowns. By~\cref{eq:jump_traction,eq:avg_traction}, the leading terms of~\cref{eq:mech_crack_general} become
\begin{equation}
	\begin{aligned}
		\mathcal{I}^{\mathrm{mech}}_{\mathcal{E}_c^h}(u,p,w)
		={}& \sum_{e\in\mathcal{E}_c^h} \int_e
		\Bigl(
		\tfrac{1}{2}\cos\varphi\,\jmp{\hat{t}_i}\jmp{w_i}
		+ 2\cos\varphi\,\avg{\hat{t}_i}\avg{w_i}
		\Bigr)\, d\tilde{s}
		\\
		&\quad + \mathcal{M}^{\perp}_{\mathcal{E}_c^h}(\tilde{\sigma},w)
		+ \mathcal{M}^{\Delta}_{\mathcal{E}_c^h}(\tilde{\sigma},w).
	\end{aligned}
	\label{eq:mech_unknown_form}
\end{equation}
At each interface degree of freedom on $\tilde\Gamma_c$, the constitutive law is enforced as the algebraic system
\begin{equation}
	\begin{aligned}
		\avg{\hat{t}_i} &= 0,
		\\
		\jmp{\hat{t}_i}
		&= 2\Big(
		K_{ij}(\hat{x})
		\jmp{\tilde{u}_j + \Delta_\ell\tilde{u}_{j,\ell}}
		+ H_{ij}(\hat{x})
		\jmp{\dot{\tilde{u}}_j
			+ \Delta_\ell\dot{\tilde{u}}_{j,\ell}}
		\Big),
	\end{aligned}
	\label{eq:mech_algebraic_closure}
\end{equation}
where the first relation is the traction balance~\cref{eq:ic_traction_balance} and the second follows from the linear law~\cref{eq:spring_law} via~\cref{eq:avg_equals_transmitted}, with true-interface displacement traces expressed in terms of surrogate quantities via~\cref{eq:vector-shift}. For the zero-thickness interface considered here, the first relation eliminates $\avg{\hat{t}_i}$; more general models with surface stress or inertia~\citep{chambat2014jump,zemlyanova2023numerical} prescribe a nontrivial relation for $\avg{\hat{t}_i}$ as well.

In the discrete setting,~\cref{eq:flow_algebraic_closure,eq:mech_algebraic_closure} are evaluated pointwise at the interface nodes after each solution update (or nonlinear iteration), so that the element assembly of~\cref{eq:flow_unknown_form,eq:mech_unknown_form} involves only the interface unknowns and does not require knowledge of the constitutive law.

\section{Numerical implementation}
\label{sec:spaces_and_split}

We now specify the spatial finite-dimensional spaces used in this work and clarify that emph{semidiscrete} denotes spatial discretization only: after expansion in the finite element bases, the problem becomes a time-continuous system (here, a DAE) for the coefficient vectors. The bulk fields are approximated by continuous Galerkin finite elements. However, the surrogate interface $\tilde\Gamma_c$ must admit independent traces from each side in order to represent pressure and displacement jumps. This is achieved by splitting the mesh connectivity along $\tilde\Gamma_c$: vertices and facets on $\tilde\Gamma_c$ are duplicated so that elements on opposite sides do not share degrees of freedom. As a result, the computational domain remains geometrically identical, but the discrete spaces become two-sided on $\tilde\Gamma_c$, providing well-defined traces $(\cdot)^\pm$ and jumps and averages across $\tilde\Gamma_c$. We denote the resulting split mesh by $\mathcal{T}^{h,\mathrm{split}}$. In the split-mesh construction used throughout this work, connectivity is duplicated along surrogate crack faces to permit independent traces on the two sides of $\tilde\Gamma_c$. When a crack intersects the external boundary, this duplication is carried through to the boundary endpoint nodes so that two-sided traces are retained up to $\partial\Omega$. By contrast, for embedded cracks with interior tips, the split connectivity terminates at the surrogate tip vertex; these terminal vertices remain single-valued, creating a local split-to-continuous transition patch that is relevant for post-processing as discussed further in the numerical results.

\subsection{Bulk finite element spaces}
\label{sec:bulk_spaces}

For the displacement, we use a continuous bilinear space enriched by element bubbles, i.e., a MINI-type construction on quadrilateral elements motivated by the classical bubble-enriched MINI element~\cite{arnold1984stable}. Compared with alternatives that maintain primal form, such as Taylor--Hood ($\mathbb{Q}_2/\mathbb{Q}_1$), this keeps the bulk approximation low order while adding only local bubble degrees of freedom. Let $\mathbb{Q}_1(K)$ denote scalar bilinear polynomials on $K$, and let $b_K$ be an element bubble function on $K$ (vanishing on $\partial K$). The discrete displacement space is
\begin{equation}
	\begin{aligned}
		\bm{V}^h
		:={}& \bigl\{ \bm{w}^h \in [C^0(\mathcal{T}^{h,\mathrm{split}})]^d :\;
		\bm{w}^h|_K \in [\mathbb{Q}_1(K) \oplus \mathrm{span}\{b_K\}]^d,
		\\
		&\qquad \forall K \in \mathcal{T}^{h,\mathrm{split}} \bigr\},
	\end{aligned}
	\label{eq:Vh_def}
\end{equation}
with essential boundary conditions enforced by restricting to $\bm{V}^{h,0} \subset \bm{V}^h$.

For the pressure, we use continuous bilinear elements,
\begin{equation}
	\begin{aligned}
		Q^h
		:={}& \bigl\{ v^h \in C^0(\mathcal{T}^{h,\mathrm{split}}) :\;
		v^h|_K \in \mathbb{Q}_1(K),
		\\
		&\qquad \forall K \in \mathcal{T}^{h,\mathrm{split}} \bigr\},
	\end{aligned}
	\label{eq:Qh_def}
\end{equation}
with $Q^{h,0} \subset Q^h$ incorporating essential pressure boundary conditions when present. The pair $(\bm{V}^h, Q^h)$ is therefore a bubble-enriched low-order $\mathbb{Q}_1/\mathbb{Q}_1$ approximation. Such enrichments are widely used to improve stability in Stokes- and Biot-type discretizations; see, e.g.,~\cite{arnold1984stable,rodrigo2016stability}.

Because the mesh is split only along $\tilde\Gamma_c$, functions in $\bm{V}^h$ and $Q^h$ are continuous on $\tilde\Omega \setminus \tilde\Gamma_c$ and discontinuous across $\tilde\Gamma_c$. Consequently, for faces $e \in \mathcal{E}^h_{i,\circ}$ the jumps vanish identically,
\begin{equation}
	\jmp{v^h} = 0, \qquad \jmp{\bm{w}^h} = \bm{0}
	\qquad\text{on } \mathcal{E}^h_{i,\circ},
	\label{eq:jumps_vanish_bulk_faces}
\end{equation}
whereas for $e \in \mathcal{E}^h_c$ the two-sided traces are distinct and $\jmp{\cdot}$ is generally nonzero.

\subsection{Interface unknown spaces}
\label{sec:interface_spaces}

For the interface-unknown formulation of~\cref{sec:enforcement_unknowns}, the one-sided normal fluxes $\hat{q}^{n,\pm}$ and the one-sided tractions $\hat{t}_i^\pm$ are considered as additional unknowns supported on $\tilde\Gamma_c$. These variables enter the discrete interface functionals through facewise pairings with traces of the bulk test functions. To obtain a compatible discrete pairing, we choose interface spaces that are no richer than the corresponding trace spaces on $\tilde\Gamma_c$. Since the element bubble $b_K$ vanishes on $\partial K$, the trace of $\bm{V}^h$ on any face $e$ reduces to $[\mathbb{Q}_1(e)]^d$, and the trace of $Q^h$ is $\mathbb{Q}_1(e)$. We therefore define
\begin{align}
	\Lambda^h_q &:= \bigl\{ \mu^h \in L^2(\tilde\Gamma_c) :
	\mu^h|_e \in \mathbb{Q}_1(e)
	\;\;\forall e \in \mathcal{E}^h_c \bigr\},
	\label{eq:Lambda_q}
	\\[4pt]
	\bm{\Lambda}^h_t &:= \bigl\{ \bm{\lambda}^h
	\in [L^2(\tilde\Gamma_c)]^d :
	\bm{\lambda}^h|_e \in [\mathbb{Q}_1(e)]^d
	\;\;\forall e \in \mathcal{E}^h_c \bigr\}.
	\label{eq:Lambda_t}
\end{align}
The one-sided flux unknowns $\hat{q}^{n,\pm} \in \Lambda^h_q$ and the one-sided traction unknowns $\hat{t}_i^\pm \in \bm{\Lambda}^h_t$ are defined on each face of $\tilde\Gamma_c$; their jumps and averages then enter the interface functionals~\cref{eq:flow_unknown_form,eq:mech_unknown_form}.

\subsection{Semidiscrete problem: weak enforcement}
\label{sec:semidiscrete_direct}

In the weak enforcement approach of~\cref{sec:enforcement_substitution}, the constitutive law is embedded in the crack-face functionals. The semidiscrete problem reads: find $(p^h(t), \bm{u}^h(t)) \in Q^h \times \bm{V}^h$ such that for all $(v^h, \bm{w}^h) \in Q^{h,0} \times \bm{V}^{h,0}$,
\begin{equation}
	\begin{aligned}
		0 ={}& \int_{\tilde\Omega} \beta\dot{p}^hv^h\;d\tilde\Omega
		+ \int_{\tilde\Omega} \alpha\dot{u}^h_{k,k}v^h\;d\tilde\Omega
		+ \int_{\tilde\Omega} \gamma p^h_{,i}v^h_{,i}\;d\tilde\Omega
		\\
		&+ \int_{\partial\tilde\Omega} q_i\tilde{n}_iv^h\;d\tilde{s}
		- \mathcal{I}^{\mathrm{flow}}_{\mathcal{E}^h_c}(p^h,v^h)
		- \int_{\tilde\Omega} fv^h\;d\tilde\Omega
		\\
		&+ \int_{\tilde\Omega} \sigma_{ij}(\bm{u}^h,p^h)w^h_{i,j}\;d\tilde\Omega
		- \int_{\partial\tilde\Omega} \sigma_{ij}(\bm{u}^h,p^h)\tilde{n}_jw^h_i\;d\tilde{s}
		\\
		&+ \mathcal{I}^{\mathrm{mech}}_{\mathcal{E}^h_c}(\bm{u}^h,p^h,\bm{w}^h),
	\end{aligned}
	\label{eq:semidiscrete_direct}
\end{equation}
where $\mathcal{I}^{\mathrm{flow}}_{\mathcal{E}^h_c}$ and $\mathcal{I}^{\mathrm{mech}}_{\mathcal{E}^h_c}$ are the closed forms given by~\cref{eq:flow_substituted,eq:mech_substituted}. The only interior-face contributions that survive are those on $\mathcal{E}^h_c$, since jumps vanish on $\mathcal{E}^h_{i,\circ}$ by~\cref{eq:jumps_vanish_bulk_faces}.

Let $\mathbf{p}$ and $\mathbf{u}$ denote the vectors of nodal pressure and displacement degrees of freedom, with $N_I$ the scalar shape functions for pressure and $\bm{N}^k_i$ the displacement shape functions (including the bubble enrichment). Expanding~\cref{eq:semidiscrete_direct} in the finite element bases yields
\begin{equation}
	\begin{aligned}
		&\underbrace{
			\begin{bmatrix}
				\mathbf{M}_p & \mathbf{C} \\[3pt]
				\mathbf{0}   & \mathbf{D}^c_u
			\end{bmatrix}
		}_{\displaystyle\mathbf{E}}
		\begin{bmatrix}
			\dot{\mathbf{p}} \\[3pt] \dot{\mathbf{u}}
		\end{bmatrix}
		\\
		&\qquad
		+ \underbrace{
			\begin{bmatrix}
				\mathbf{K}_p + \mathbf{K}^c_p
				+ \mathbf{S}^c_p + \mathbf{R}^c_p
				& \mathbf{0} \\[3pt]
				-\mathbf{C}^T
				+ \mathbf{S}^c_{up} + \mathbf{R}^c_{up}
				& \mathbf{K}_u + \mathbf{K}^c_u
				+ \mathbf{S}^c_{uu} + \mathbf{R}^c_{uu}
			\end{bmatrix}
		}_{\displaystyle\mathbf{K}}
		\begin{bmatrix}
			\mathbf{p} \\[3pt] \mathbf{u}
		\end{bmatrix}
		\\
		&\qquad
		=
		\begin{bmatrix}
			\mathbf{f}_p \\[3pt] \mathbf{f}_u
		\end{bmatrix},
	\end{aligned}
	\label{eq:matrix_direct}
\end{equation}
or, equivalently, in compact form with $\mathbf{z}:=[\mathbf{p}, \mathbf{u}]^T$ and $\mathbf{f}:=[\mathbf{f}_p, \mathbf{f}_u]^T$,
\begin{equation}
	\mathbf{E}\dot{\mathbf{z}} + \mathbf{K}\mathbf{z} = \mathbf{f}.
	\label{eq:matrix_direct_compact}
\end{equation}
Here $\mathbf{M}_p$, $\mathbf{C}$, $\mathbf{K}_p$, and $\mathbf{K}_u$ are standard bulk volume-integral blocks (with $\mathbf{C}$ the Biot coupling and $-\mathbf{C}^T$ its transpose); $\mathbf{K}^c_p$, $\mathbf{K}^c_u$, and $\mathbf{D}^c_u$ are constitutive interface contributions; $\mathbf{S}^c_p$, $\mathbf{S}^c_{uu}$, $\mathbf{S}^c_{up}$ are Taylor correction terms; and $\mathbf{R}^c_p$, $\mathbf{R}^c_{uu}$, $\mathbf{R}^c_{up}$ are normal-mismatch corrections. All interface blocks are assembled over $\mathcal{E}^h_c$ from the three lines of the substituted forms~\cref{eq:flow_substituted,eq:mech_substituted}; the explicit element-level definitions of every block are collected in~\cref{app:matrix_blocks} for completeness. The pressure-coupling blocks $\mathbf{S}^c_{up}$ and $\mathbf{R}^c_{up}$ arise from the Biot pressure contribution to the effective stress and modify the $(2,1)$ block of~\cref{eq:matrix_direct} alongside $-\mathbf{C}^T$. The right-hand side $\mathbf{f}_p$ collects the source term and boundary flux contributions, and $\mathbf{f}_u$ the boundary traction contributions. Since the momentum equation is quasi-static, the lower-left block of $\mathbf{E}$ is zero. The block $\mathbf{D}^c_u$ is present only when the interface law contains rate dependence (interface viscosity); for purely rate-independent interface laws, $\mathbf{D}^c_u=\mathbf{0}$. The system~\cref{eq:matrix_direct} is a differential-algebraic equation (DAE) of index~1 even when $\mathbf{D}^c_u $ is nonzero because these quantities only exist on the crack and thus do not provide full diagonal terms.

\subsection{Semidiscrete problem: strong enforcement}
\label{sec:semidiscrete_unknown}

In the interface-unknown approach of~\cref{sec:enforcement_unknowns}, the one-sided fluxes and tractions are retained as independent unknowns and the constitutive law is enforced algebraically. The semidiscrete problem reads: find
\[
\begin{aligned}
	&(p^h(t), \bm{u}^h(t), \hat{q}^{n,+}, \hat{q}^{n,-},
	\hat{t}_i^+, \hat{t}_i^-)
	\\
	&\in Q^h \times \bm{V}^h \times \Lambda^h_q \times \Lambda^h_q
	\times \bm{\Lambda}^h_t \times \bm{\Lambda}^h_t
\end{aligned}
\]
such that for all
$(v^h, \bm{w}^h) \in Q^{h,0} \times \bm{V}^{h,0}$,
\begin{equation}
	\begin{aligned}
		0 ={}& \int_{\tilde\Omega} \beta\dot{p}^hv^h\;d\tilde{\Omega}
		+ \int_{\tilde\Omega} \alpha\dot{u}^h_{k,k}v^h\;d\tilde{\Omega}
		+ \int_{\tilde\Omega} \gamma p^h_{,i}v^h_{,i}\;d\tilde{\Omega}
		\\
		&+ \int_{\partial\tilde\Omega} q_i\tilde{n}_iv^h\;d\tilde{s}
		- \mathcal{I}^{\mathrm{flow}}_{\mathcal{E}^h_c}
		(p^h,\hat{q}^{n,\pm};v^h)
		- \int_{\tilde\Omega} fv^h\;d\tilde{\Omega}
		\\[4pt]
		&+\int_{\tilde\Omega}
		\sigma_{ij}(\bm{u}^h,p^h)w^h_{i,j}\;d\tilde{\Omega}
		- \int_{\partial\tilde\Omega}
		\sigma_{ij}(\bm{u}^h,p^h)\tilde{n}_jw^h_i\;d\tilde{s}
		\\
		&+ \mathcal{I}^{\mathrm{mech}}_{\mathcal{E}^h_c}
		(\bm{u}^h,p^h,\hat{t}_i^\pm;\bm{w}^h),
	\end{aligned}
	\label{eq:semidiscrete_unknown}
\end{equation}
where $\mathcal{I}^{\mathrm{flow}}_{\mathcal{E}^h_c}$ and $\mathcal{I}^{\mathrm{mech}}_{\mathcal{E}^h_c}$ are the forms given by~\cref{eq:flow_unknown_form,eq:mech_unknown_form}. The semicolon separates trial quantities (including the interface unknowns $\hat{q}^{n,\pm} \in \Lambda^h_q$ and $\hat{t}_i^\pm \in \bm{\Lambda}^h_t$) from test functions. The interface unknowns enter the first line of~\cref{eq:flow_unknown_form,eq:mech_unknown_form} through $\jmp{\hat{q}^n}$, $\avg{\hat{q}^n}$, $\jmp{\hat{t}_i}$, and $\avg{\hat{t}_i}$, while the bulk fields $p^h$ and $\bm{u}^h$ enter the remaining correction terms.

Let $\boldsymbol{\lambda}_q = [\hat{\mathbf{q}}^{n,+},\; \hat{\mathbf{q}}^{n,-}]^T$ and $\boldsymbol{\lambda}_t = [\hat{\mathbf{t}}^+,\; \hat{\mathbf{t}}^-]^T$ collect the interface degrees of freedom in $\Lambda^h_q$ and $\bm{\Lambda}^h_t$, and let $\boldsymbol{\lambda} = [\boldsymbol{\lambda}_q,\; \boldsymbol{\lambda}_t]^T$. To reemphasize, $\lambda$ is not a Lagrange multiplier as discussed in~\cref{sec:enforcement}. Expanding~\cref{eq:semidiscrete_unknown} in the bulk and interface bases yields a bulk semidiscrete system coupled to algebraic interface closures:
\begin{equation}
	\begin{aligned}
		&\underbrace{
			\begin{bmatrix}
				\mathbf{M}_p & \mathbf{C}\\[3pt]
				\mathbf{0} & \mathbf{0}
			\end{bmatrix}
		}_{\displaystyle \mathbf{E}_{\mathrm{bulk}}}
		\dot{\mathbf{z}}
		\\
		&\qquad
		+ \underbrace{
			\begin{bmatrix}
				\mathbf{K}_p + \mathbf{S}^c_p + \mathbf{R}^c_p
				& \mathbf{0} \\[3pt]
				-\mathbf{C}^T + \mathbf{S}^c_{up} + \mathbf{R}^c_{up}
				& \mathbf{K}_u + \mathbf{S}^c_{uu} + \mathbf{R}^c_{uu}
			\end{bmatrix}
		}_{\displaystyle \mathbf{K}_{\mathrm{bulk}}}
		\mathbf{z}
		\\
		&\qquad
		+ \underbrace{
			\begin{bmatrix}
				\mathbf{G}_p\\[3pt]
				\mathbf{G}_u
			\end{bmatrix}
		}_{\displaystyle \mathbf{G}}
		\boldsymbol{\lambda}
		\\
		&\qquad = \mathbf{f},
	\end{aligned}
	\label{eq:matrix_unknown}
\end{equation}
where the coupling blocks $\mathbf{G}_p$ and $\mathbf{G}_u$, which couple the bulk and interface unknowns through the first line of~\cref{eq:flow_unknown_form,eq:mech_unknown_form}, are defined in~\cref{app:matrix_blocks} for completeness. The interface variables enter the bulk equations only through the term $\mathbf{G}\boldsymbol{\lambda}$, which may equivalently be moved to the right-hand side and interpreted as an induced interface load vector.

The interface unknowns are not determined by an additional variational equation. Instead, they are obtained from the algebraic closures~\cref{eq:flow_algebraic_closure,eq:mech_algebraic_closure}, evaluated pointwise at each interface node. Writing these compactly,
\begin{equation}
	\boldsymbol{\lambda}
	= \mathcal{C}\!\big(\mathbf{p}, \mathbf{u}, 
	\dot{\mathbf{u}};\tilde\Gamma_c\big),
	\label{eq:algebraic_closure_compact}
\end{equation}
where $\mathcal{C}$ encodes the constitutive law transferred to the surrogate via~\cref{eq:SBM-value,eq:vector-shift}. $\boldsymbol{\lambda}$ is updated via~\cref{eq:algebraic_closure_compact} at each nonlinear iteration (reducing to a single evaluation for linear laws) during time integration, while the coupling blocks $\mathbf{G}_p$, $\mathbf{G}_u$ together with the correction blocks $\mathbf{S}^c$, $\mathbf{R}^c$ remain fixed geometry-dependent quantities that need not be reassembled. Equations~\cref{eq:matrix_unknown,eq:algebraic_closure_compact} therefore define a semi-explicit DAE system for the bulk state $\mathbf{z}$ and interface variables $\boldsymbol{\lambda}$.

\subsection{Post-processing on the true crack geometry}
\label{sec:postprocess_true}

The shifted interface method enforces the interface conditions on the mesh-fitted surrogate interface $\tilde\Gamma_c$, whereas the physical interface of interest is the true crack $\Gamma_c$. Consequently, the discrete solution is naturally represented on the surrogate geometry. For the purpose of reporting quantities on $\Gamma_c$ and producing visualizations in which the crack appears at its physical location, we perform a post-processing step that reuses the projection $\Pi_h$, gap vector $\Delta_\ell$, and expansion relations introduced in Section~\ref{sec:sbm}, as described in~\cite{li2024complex}. It is worth emphasizing that this step is not trivial but critical for the correct interpretation of the numerical results and the subsequent convergence analyses.

\subsubsection{Post-processing on the projected crack geometry}
\label{subsec:postprocess_projection}

The procedure to calculate the fields over the true crack consists of three stages: (i) reconstruction of smooth nodal gradient fields by $L^{2}$ projection, (ii) first-order Taylor transfer of the primary fields from the surrogate crack to the true crack, and (iii) construction of projected coordinates for visualization. We describe each in turn.

The primary unknowns, pore pressure~$p^h$ and displacement~$u^h_i$, reside in continuous finite element spaces ($Q^h$ and~$\bm{V}^h$, respectively; see~\cref{eq:Qh_def,eq:Vh_def}) and are therefore directly available at every mesh node. Many quantities of engineering interest, however, depend on spatial derivatives of these fields. The Darcy flux requires~$p^h_{,j}$, while the effective stress and total stress require~$u^h_{i,j}$. Within each element, these derivatives are well defined, but because $N_{I,j}$ is, in general, discontinuous across element boundaries, the element-wise gradient $f^h_{,j}\big|_K$ of any CG field~$f^h$ jumps from one element to the next. Direct evaluation at a shared node would therefore yield as many gradient values as there are elements meeting at that node, and these values are neither unique nor, for low-order elements, optimal in accuracy.

To obtain a single, smooth nodal representation of each gradient component, we employ an $L^{2}$ projection of the discontinuous element-wise derivative back into the continuous space~$Q^h$. Concretely, for a scalar CG field $f^h\in Q^h$ and spatial direction~$j$, the projected derivative~$g_{j}^h\in Q^h$ is the unique function satisfying
\begin{equation}
	\int_{\tilde\Omega} g_{j}^hN_I\;dV
	\;=\;
	\int_{\tilde\Omega}
	f^h_{,j}N_I\;d\tilde\Omega
	\qquad \forall\;N_I\in Q^h.
	\label{eq:post_L2proj}
\end{equation}
In matrix form this reads
\begin{equation}
	\begin{aligned}
		M\mathbf{g}_j &= G_j\mathbf{f},
		\\
		M_{IJ}&=\int_{\tilde\Omega}N_IN_J\;d\tilde\Omega,
		\qquad
		\bigl(G_j\bigr)_{IJ}
		=\int_{\tilde\Omega}N_I N_{J,j}\;d\tilde\Omega,
	\end{aligned}
	\label{eq:post_L2proj_matrix}
\end{equation}
where $M$ is the consistent CG mass matrix and $G_j$ is the weak gradient operator, both assembled on the surrogate mesh. Because $M$ is symmetric positive-definite, the system is solved once per spatial direction by a direct factorization that is reused for every field component. Applying \cref{eq:post_L2proj_matrix} component-wise to the displacement field yields the full displacement gradient~$u^h_{i,j}$, from which the strain tensor and, through the constitutive law, the effective and total stress tensors are evaluated pointwise at each node. Similarly, the reconstructed pressure gradient furnishes a nodal Darcy flux field. The resulting nodal gradient fields are globally continuous and, for structured meshes, often exhibit optimally convergent nodal behavior, making them
well-suited for visualization and for the residual diagnostics reported below.

Let $\tilde x\in\tilde\Gamma_c$ be a node on the surrogate crack.
Its closest-point projection onto the true crack is
$\hat x=\Pi_h(\tilde x)\in\Gamma_c$, with the gap vector
$\Delta_j=\hat x_j-\tilde x_j$ as defined
in~\cref{eq:Pi_h,eq:gap}. Because the CG solution is computed on the
surrogate mesh, the nodal values at~$\tilde x$ approximate the
solution at~$\tilde x$, not at~$\hat x$. To recover primary-field values at the physical crack
location we apply the first-order Taylor
expansion~\cref{eq:SBM-value}. For any primary scalar field~$f^h$ with reconstructed gradient
$f^h_{,j}$ from~\cref{eq:post_L2proj_matrix}, the transferred value is
\begin{equation}
	f^h(\hat x)
	\;\approx\;
	f^h(\tilde x)
	\;+\;
	\Delta_jf^h_{,j}(\tilde x),
	\label{eq:post_taylor_scalar}
\end{equation}
and analogously for each component of a vector field (cf.~\cref{eq:vector-shift}). In the numerical results of this paper, this first-order transfer is applied to the primary fields (pressure and displacement) when values are required on $\Gamma_c$. The transfer is performed only at degrees of freedom that lie on surrogate crack faces, while all interior nodes are left unchanged.

Derived quantities that depend on spatial gradients, such as Darcy flux and stress, are not Taylor-expanded in the results reported here. Instead, they are reconstructed on the surrogate mesh from the $L^2$-projected nodal gradients and then evaluated in the true-crack frame through projection onto the true normal and tangent directions during residual evaluation (see \cref{subsec:residual_conventions}). This choice is consistent with the omission of the gradient correction terms (third line of~\cref{eq:flow_substituted,eq:mech_substituted}) from the implementation (see~\cref{sec:numerical_results}). Since the Hessian-level shifted corrections were not used during assembly, applying a Taylor transfer to the derived fields in post-processing would introduce a correction that was absent from the solve and could distort the convergence diagnostics.

For visualization, a projected coordinate set is constructed by shifting every surrogate crack node to its closest-point projection on the true crack,
\begin{equation}
	\tilde x_j
	\;\mapsto\;
	\hat x_j
	= \tilde x_j + \Delta_j,
	\label{eq:post_proj_coords}
\end{equation}
while retaining the original element connectivity and leaving all non-crack nodes in place. The resulting mesh coincides with~$\Gamma_c$ along the crack, while the plotted field values are those obtained from the post-processing described above. As noted in~\cite{li2024complex}, this geometric projection may distort elements near corners or regions of high crack curvature, and can even produce inverted elements. For this reason, it is used solely for post-processing and does not modify the mesh employed in the solve.

\subsubsection{Residual evaluation conventions used in the numerical results}
\label{subsec:residual_conventions}

The interface residuals reported in Section~\ref{sec:numerical_results} are diagnostic, post-processed quantities. They are constructed from the reconstructed nodal gradient fields on the surrogate mesh and evaluated in the true-crack frame. Specifically, for the reported results, the flux and traction components on the true crack are computed by
\begin{equation}
	\hat{q}_n = \tilde{q}_i\,\hat{n}_i,
	\qquad
	\hat{t}_i = \tilde{\sigma}_{ij}\,\hat{n}_j,
	\label{eq:postprocess_projection}
\end{equation}
with tangential components obtained by projection onto the true tangent $\hat{m}_i$. No Taylor transfer is applied to the gradient-derived fields themselves.

For the strong (algebraic-constraint) enforcement, it is important to distinguish between (i) the algebraic closure residuals at interface nodes/DOFs and (ii) the post-processed diagnostic residuals reported in figures. The algebraic closure residuals are satisfied to solver tolerance, typically near machine precision, at each time step and nonlinear iteration, so they are not plotted. The reported residual profiles and $L^2$ norms are reconstructed diagnostic quantities computed from the primary fields and post-processed gradients. They therefore remain nonzero for both
enforcement strategies and converge according to the accuracy of the discretization and the post-processing pipeline.

% ======================================================================

\section{Numerical results}
\label{sec:numerical_results}

All numerical experiments are performed in 2D under the plane-strain assumption. The variational forms~\cref{eq:semidiscrete_direct,eq:semidiscrete_unknown} were implemented within \texttt{scikit-fem}~\citep{Gustafsson2020}. For time integration, we used \texttt{solve\_nivp}~\citep{rileysolve_nivp} with an adaptive SDIRK2 time integrator. The gradient correction blocks collected in~\cref{app:matrix_blocks} (third line), which involve second derivatives (Hessians) of the finite element shape functions, were omitted from the implementation. This choice was made for simplicity, since the standard \texttt{scikit-fem} assembly infrastructure does not readily provide element-level Hessian terms. For the $\mathbb{Q}_1$/$\mathbb{Q}_1$-bubble element discretization employed here, these corrections are small, and their omission is not expected to affect the quality of the results presented below. A full assessment of their influence is deferred to future work.

In all cases, the background mesh is a structured quadrilateral grid whose connectivity is split along the surrogate interface~$\tilde\Gamma_c$ as described in \cref{sec:spaces_and_split}: vertices and facets on~$\tilde\Gamma_c$ are duplicated so that elements on opposite sides no longer share degrees of freedom, while the mesh nodes themselves remain in place. For boundary-intersecting cracks, this duplication is carried to the boundary endpoints; for embedded cracks, the split connectivity terminates at the interior surrogate tips, where the displacement and pressure fields become single-valued by construction.

Unless otherwise stated, all interface residuals reported below are the post-processed diagnostic residuals defined using the conventions of \cref{subsec:residual_conventions}. In particular, for the strong enforcement strategy, the algebraic closure equations are satisfied to solver tolerance at the interface DOFs and thus are not presented.

The numerical investigation serves three purposes: (1) to verify the shifted interface method for poroelastic crack problems, (2) to compare the two enforcement strategies introduced in \cref{sec:enforcement}, i.e., weak enforcement (\cref{sec:enforcement_substitution}) and strong enforcement (\cref{sec:enforcement_unknowns}), and (3) to demonstrate how the method extends naturally to multiple cracks with heterogeneous interface properties.

To this end, we consider four scenarios of increasing complexity: (1) an offset mesh-aligned crack, (2) a boundary-intersecting angled crack, (3) an embedded angled crack, and (4) a multi-crack configuration with four simultaneously embedded cracks of distinct geometry and interface properties. The first three cases serve purposes~(1) and~(2) through systematic mesh-convergence studies, while the fourth addresses purpose~(3).

We note that, to the best of our knowledge, closed-form solutions for transient Biot in the presence of an embedded hydraulic barrier (impermeable crack with spring-type traction conditions) and a localized fluid source are not available in the literature, even in two dimensions. Verification is therefore carried out through self-convergence studies, in which each mesh is compared with the next finer mesh in the refinement sequence, supplemented by monitoring the interface residual norms.

The material parameters are taken as follows: shear modulus
$G = 22\times10^3$~MPa, Poisson's ratio $\nu = 0.25$, Biot coefficient
$\alpha = 0.25$, fluid viscosity $\eta = 2\times10^{-10}/3600$~MPa$\cdot$hr, fluid
compressibility $\beta = 8.5\times10^{-5}$~MPa$^{-1}$, and isotropic permeability
$k = 10^{-14}$~m$^2$ ($\approx 10$~mD, representative of fractured crystalline
rock~\citep{jaeger2007fundamentals}).
The fluid viscosity corresponds to water at approximately
120--150$^\circ$C~\citep{Huber2009}. The above parameters are consistent with fluid injection at depths
of 3--5~km under a typical geothermal gradient in a geothermal setting.

On the left and right boundaries, we impose homogeneous Dirichlet conditions for both displacement components ($u_x = u_y = 0$), while the top and bottom boundaries are traction-free. Homogeneous Dirichlet pressure conditions ($p = 0$) are prescribed on all four boundaries.

The domain is loaded by a localized fluid source placed at $\mathbf{x}_s = (-0.25, -0.25)$ and regularized with a compactly
supported Wendland $C^{2}$ radial basis function~\cite{Wendland1995} with support radius $R = 0.1$~km (details are given
in~\cref{app:wendland}). The injection rate is set to $Q = 10^{-5}$~km$^2$/hr ($\approx 2.8\times10^{-3}$~m$^2$/s per unit out-of-plane thickness);
this rate is elevated relative to typical field values, and the chosen boundary conditions are likewise simplified, for illustrative purposes.

Each simulation is integrated in time until a steady state is reached; for the single-crack cases, this corresponds to a final time of $t_{\mathrm{end}} = 354$~hr, and for the multi-crack case, $t_{\mathrm{end}} = 3000$~hr. All fields and residuals reported below are evaluated at this final time.

The crack is modeled as nearly welded, with high normal and tangential spring stiffnesses $k_n = k_t = 10^{8}$~MPa/km, and as hydraulically impermeable ($T_n = 0$). Structured quadrilateral meshes are employed with $n \in \{20, 40, 80, 160, 320\}$ elements per side, giving mesh sizes
$h = 1/n$. For each mesh, both enforcement strategies are applied: weak and strong.

To assess the quality of the interface solution quantitatively, we define six residual fields that measure how well the discrete solution
satisfies the interface conditions. These residuals are grouped by physical origin.
The \emph{balance residuals} measure conservation across the interface. The traction balance
residual~\cref{eq:ic_traction_balance} and the flux balance residual~\cref{eq:ic_flux_balance} are
\begin{equation}
	r_i^{\jmp{t}} := \jmp{\hat{\sigma}_{ij}}\,\hat{n}_j,
	\qquad
	r^{\jmp{q}} := \jmp{\hat{q}_j}\,\hat{n}_j,
	\label{eq:balance_residuals}
\end{equation}
both of which should vanish identically (see~\cref{sec:bc_ic}). The traction balance residual is decomposed into its normal and tangential
components $r_i^{\jmp{t}}\,\hat{n}_i$ and $r_i^{\jmp{t}}\,\hat{m}_i$. The \emph{constitutive residuals} measure how well the discrete solution satisfies the interface laws. For flow,~\cref{eq:robin_flux,eq:flux_conservation} give
\begin{equation}
	r^{\mathsf{J}}
	:= \avg{\hat{q}_j}\,\hat{n}_j
	- \mathsf{J}_\Gamma,
	\label{eq:constitutive_residual_flow}
\end{equation}
where $\mathsf{J}_\Gamma = -T_n\,\jmp{\hat{p}}$ is the prescribed normal flux from the Robin condition (which reduces to $\mathsf{J}_\Gamma = 0$ for the impermeable case $T_n = 0$). For mechanics,~\cref{eq:spring_law} gives
\begin{equation}
	r_i^{\mathsf{T}}
	:= \avg{\hat{\sigma}_{ij}}\,\hat{n}_j
	- \mathsf{T}_{\Gamma,i},
	\label{eq:constitutive_residual_mech}
\end{equation}
where
$\mathsf{T}_{\Gamma,i} = K_{ij}\,\jmp{\hat{u}_j}$
is the traction prescribed by the linear law. The constitutive residual is likewise decomposed into normal and
tangential components $r_i^{\mathsf{T}}\,\hat{n}_i$ and $r_i^{\mathsf{T}}\,\hat{m}_i$. In the discrete setting, the hatted quantities appearing in
\cref{eq:balance_residuals,eq:constitutive_residual_flow,eq:constitutive_residual_mech} are not evaluated on the true crack directly but are reconstructed from the surrogate solution as described in \cref{subsec:postprocess_projection,subsec:residual_conventions}. The primary fields (pressure and displacement) are transferred from surrogate nodes to the true crack via the first-order Taylor expansion~\cref{eq:post_taylor_scalar}, while the derived fields (Darcy flux $\tilde{q}_i$ and stress $\tilde{\sigma}_{ij}$) are obtained from the $L^2$-projected nodal gradients~\cref{eq:post_L2proj_matrix} on the surrogate mesh and projected onto the true-crack frame through $\hat{q}_n = \tilde{q}_i\,\hat{n}_i$ and $\hat{t}_i = \tilde{\sigma}_{ij}\,\hat{n}_j$ (cf.~\cref{eq:postprocess_projection}). The reported residuals, therefore, reflect the combined effect of the spatial discretization, nodal gradient reconstruction, and geometric frame transfer to the true crack.

% ======================================================================
% Aligned offset crack
% ======================================================================
\subsection{Offset crack}
\label{sec:aligned_offset}
\begin{figure*}[t]
  \centering
  \pagefiguregraphics{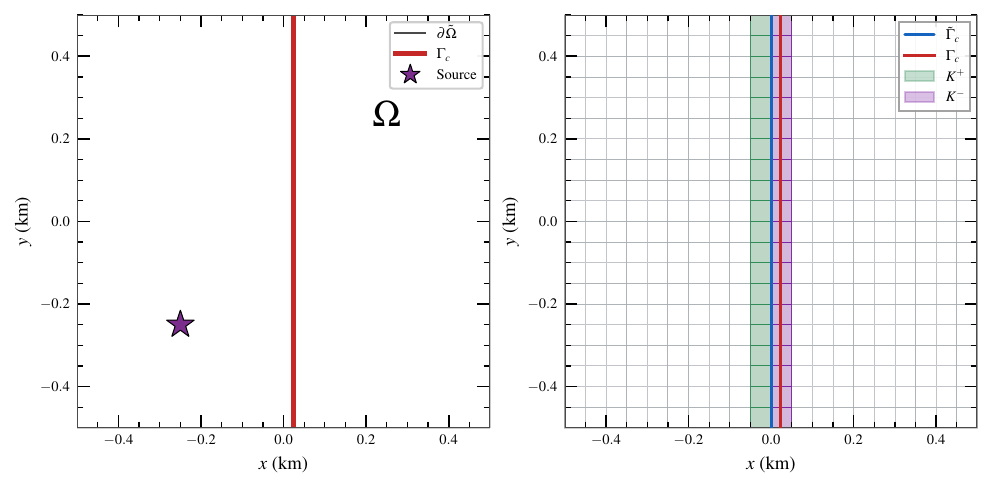}
  \caption{Mesh-aligned offset crack configuration. The
  crack is offset laterally by roughly half an element width, so that the
  surrogate interface lies on an interior column of element faces
  rather than on the domain midline.}
  \label{fig:domain_classification_offset}
\end{figure*}

In this test, the crack is perturbed laterally by roughly half an element width for the coarsest mesh, such that the surrogate interface $\tilde\Gamma_c$ still lies normal to vertical mesh faces but never overlaps with the true crack location for any refinement ( \cref{fig:domain_classification_offset}). Structured quadrilateral meshes with $n \in \{20,\,40,\,80,\,160,\,320\}$ elements per side are
used. Because the surrogate remains aligned with element faces, the gap vector ${\Delta}_i$ is purely horizontal and constant along the crack, and the normal mismatch vanishes ($\tilde{n}_i = \hat{n}_i$).

The purpose of this test is twofold: to verify that the formulation is insensitive to the lateral placement of the crack within the mesh, and to expose any differences between the two enforcement strategies when the crack no longer occurs at a position that ensures a mesh-fitted crack.

% ------ Figure: Domain classification (offset) ------

\subsubsection{Solution fields}
\label{sec:aligned_offset_fields}

\begin{figure*}[t]
  \centering
  \pagefiguregraphics{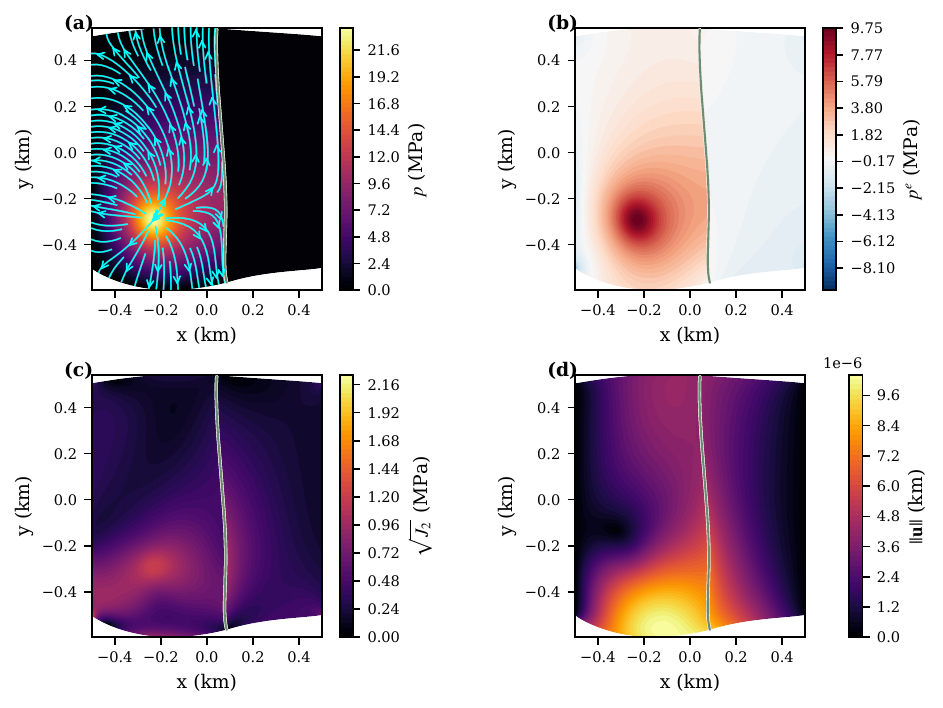}
  \caption{Computed solution fields on the finest mesh ($n = 320$) for
  the offset crack, displayed on the deformed configuration
  (displacement $\times 10^4$):
  (a)~pore pressure $p$ with Darcy flux vectors,
  (b)~mean effective stress
  $p^e = \tfrac{1}{3}\sigma^e_{kk}$,
  (c)~deviatoric stress invariant $\sqrt{J_2}$, and
  (d)~displacement magnitude $\|\mathbf{u}\|$.
  The crack is indicated by the white line.}
  \label{fig:fields_n160_offset}
\end{figure*}

\Cref{fig:fields_n160_offset} shows the computed fields on the finest mesh ($n = 320$), obtained using the weak (direct-substitution) enforcement strategy; the strong (algebraic-constraint) enforcement produces visually indistinguishable results. The lateral offset introduces a mild symmetry in the pressure and stress distributions, since the source is not equidistant from the two crack faces.

The pore pressure field (\cref{fig:fields_n160_offset}a) exhibits the expected localized high-pressure region around the source, with the impermeable crack ($T_n = 0$) acting as a hydraulic barrier that creates a clear pressure discontinuity across the interface. The overlaid Darcy flux vectors run parallel to the crack face, consistent with the zero-transmissivity condition, which permits no normal flux through the interface.
The mean effective stress $p^e = \tfrac{1}{3}\sigma^e_{kk}$ (\cref{fig:fields_n160_offset}b) displays a visible jump across the crack. This is expected: although the total traction is continuous (enforced by the high-stiffness spring law), the Biot pressure contribution $-\alpha p\delta_{ij}$ is discontinuous across the impermeable interface, and this discontinuity is inherited by the effective stress. In particular, the individual effective stress components $\sigma^e_{xx}$ and $\sigma^e_{yy}$ can each be discontinuous, even though the traction components $\sigma_{ij}\hat{n}_j$ are continuous. The deviatoric stress invariant $\sqrt{J_2}$ (\cref{fig:fields_n160_offset}c) shows stress concentrations near the source region, driven entirely by the poroelastic coupling since no external mechanical loads are applied. On the crack, $\sqrt{J_2}$ appears visually continuous; this is because the deviatoric part of the Cauchy stress is unaffected by the pore pressure ($\mathrm{dev}(\sigma_{ij}) = \mathrm{dev}(\sigma^e_{ij})$), and the near-zero displacement jump imposed by the high interface stiffness ensures that strains, and hence deviatoric stresses, are nearly continuous across the interface. The displacement magnitude (\cref{fig:fields_n160_offset}d) reflects the poroelastic swelling induced by the pore pressure, modulated by the clamped lateral boundaries. The field appears continuous across
the crack, consistent with the high interface stiffness ($k_n L / E \approx 2\times10^{3}$), which renders the displacement jump negligible.

% ------ Figure: Solution fields at n=320 (offset) ------

\subsubsection{Crack-face profiles}
\label{sec:aligned_offset_profiles}

\begin{figure*}[t]
  \centering
  \pagefiguregraphics{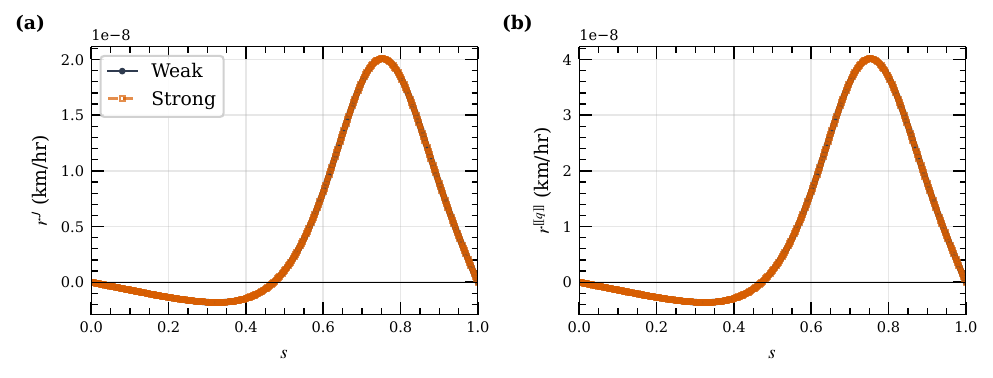}
  \caption{Flow interface residuals along the offset crack
  ($n = 320$):
  (a)~constitutive residual
  $r^{\mathsf{J}} = \avg{\hat{q}_j}\,\hat{n}_j
  - \mathsf{J}_\Gamma$ and
  (b)~flux balance residual
  $r^{\jmp{q}} = \jmp{\hat{q}_j}\,\hat{n}_j$.
  Both enforcement strategies are overlaid
  (indistinguishable).
 Residuals here are post-processed diagnostic quantities, while the residuals of the strong enforcement are zero (as shown in~\cref{fig:lambda_residual_flow_offset}).}
  \label{fig:profiles_flow_offset}
\end{figure*}

\begin{figure*}[t]
  \centering
  \pagefiguregraphics{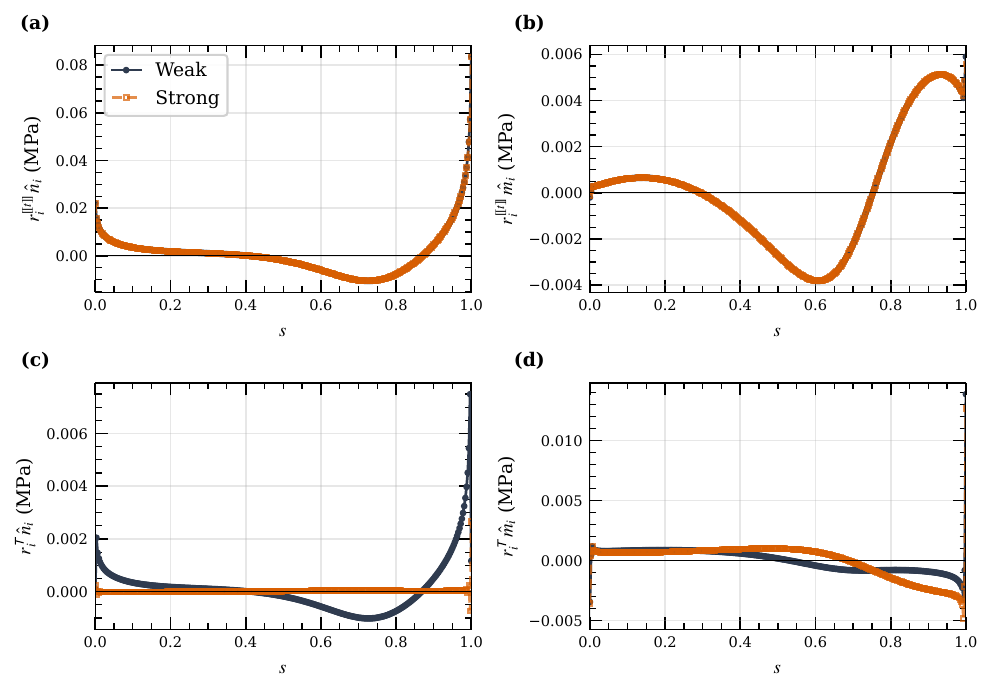}
  \caption{Mechanical interface residuals along the offset crack
  ($n = 320$):
  (a)~$r_i^{\jmp{t}}\,\hat{n}_i$,
  (b)~$r_i^{\jmp{t}}\,\hat{m}_i$,
  (c)~$r_i^{\mathsf{T}}\,\hat{n}_i$,
  (d)~$r_i^{\mathsf{T}}\,\hat{m}_i$.
 Residuals here are post-processed diagnostic quantities, while the residuals of the strong enforcement are zero (as shown in~\cref{fig:lambda_residual_mechanics_offset}).}
  \label{fig:profiles_mechanics_offset}
\end{figure*}

\Cref{fig:profiles_flow_offset,fig:profiles_mechanics_offset} display the interface residuals along the crack on the finest mesh ($n = 320$). The flow residuals (\cref{fig:profiles_flow_offset}) show that the two enforcement strategies are indistinguishable.

The mechanical balance residuals (\cref{fig:profiles_mechanics_offset}a,b) are likewise indistinguishable between the two strategies. The constitutive residuals (\cref{fig:profiles_mechanics_offset}c,d), however, reveal a clear difference. The strong enforcement produces residuals that are essentially zero along the crack interior, consistent with the exact algebraic satisfaction of the spring law at the interface degrees of freedom; the small, nonzero values near the crack tips reflect the post-processing error at nodes where the Taylor transfer is less accurate. The weak enforcement, by contrast, produces a structured oscillatory residual of order $10^{-2}$~MPa along the entire crack, reflecting the fact that the constitutive law is imposed only in an integral (weak) sense rather than pointwise. Both residuals remain small: relative to the Biot coupling stress scale $\alpha\,p_{\max} \approx 6$~MPa, the weak constitutive residual is less than~$0.2\%$.

% ------ Figure: Flow profiles (offset) ------

% ------ Figure: Mechanics profiles (offset) ------

\subsubsection{Strong enforcement}
\label{sec:aligned_offset_lambda}

\begin{figure*}[t]
  \centering
  \pagefiguregraphics{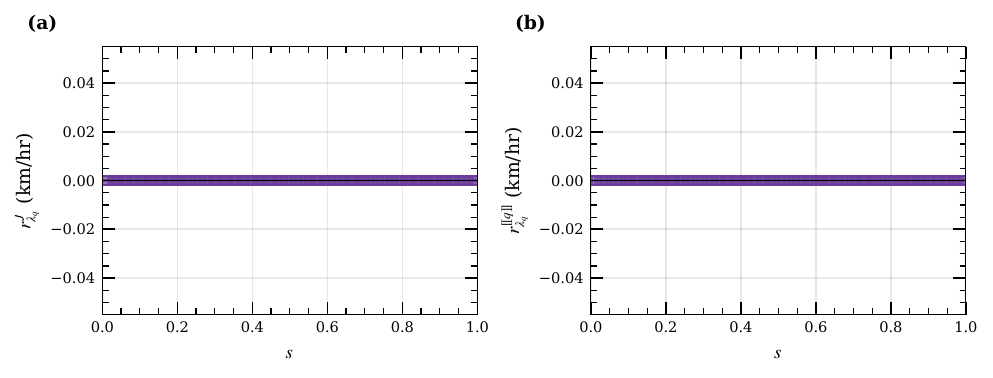}
  \caption{Flow constitutive residuals for the
  strong enforcement on the offset crack ($n = 320$):
  (a)~$r^{\mathsf{J}}_{\lambda_q}$ and
  (b)~$r^{\jmp{q}}_{\lambda_q}$.
  Both residuals are identically zero (to machine precision),
  confirming exact algebraic satisfaction of the Robin flux law.}
  \label{fig:lambda_residual_flow_offset}
\end{figure*}

\begin{figure*}[t]
  \centering
  \pagefiguregraphics{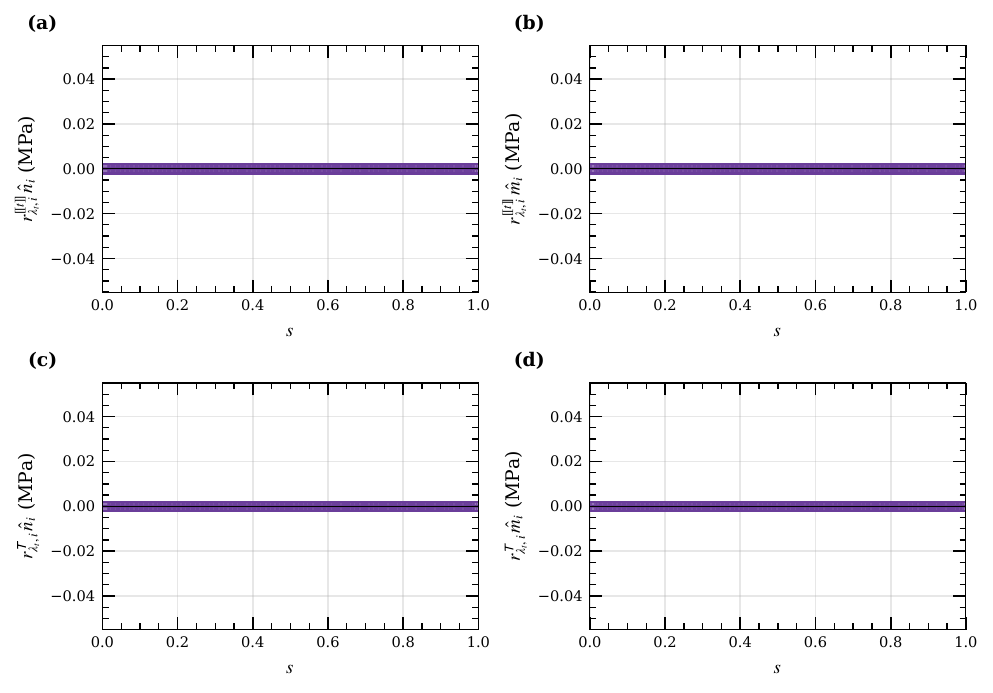}
  \caption{Mechanical constitutive residuals for
  the strong enforcement on the offset crack ($n = 320$):
  (a)~$r_{\lambda_t,i}^{\jmp{t}}\,\hat{n}_i$,
  (b)~$r_{\lambda_t,i}^{\jmp{t}}\,\hat{m}_i$,
  (c)~$r_{\lambda_t,i}^{\mathsf{T}}\,\hat{n}_i$,
  (d)~$r_{\lambda_t,i}^{\mathsf{T}}\,\hat{m}_i$.
  All components are identically zero, confirming exact
  algebraic satisfaction of the spring traction law.}
  \label{fig:lambda_residual_mechanics_offset}
\end{figure*}

The flow and mechanical profile residuals presented in
\cref{fig:profiles_flow_offset,fig:profiles_mechanics_offset} are
\emph{derived} quantities: they are computed by post-processing the
primary field solution, evaluating the one-sided flux and stress
traces, and substituting them into the constitutive and balance
laws. As such, their magnitudes reflect the combined effect of the
finite-element discretization error in the bulk fields and the
post-processing interpolation. In the strong (nodal) enforcement
strategy, however, the interface constitutive laws are imposed as
algebraic constraints within the semi-discrete system, so that the
time integrator---a stiffly accurate SDIRK scheme that solves the
resulting index-1 differential--algebraic system---enforces them
exactly (to solver tolerance) at every time step. To verify this
property independently,
\cref{fig:lambda_residual_flow_offset,fig:lambda_residual_mechanics_offset}
plot the constitutive residuals
$r^{\mathsf{J}}_{\lambda_q}$,
$r^{\jmp{q}}_{\lambda_q}$ (flow) and
$r_{\lambda_t,i}^{\jmp{t}}$,
$r_{\lambda_t,i}^{\mathsf{T}}$ (mechanics), evaluated directly
from the nodal degrees of freedom of
$\lambda_q$~and~$\lambda_t$ (see~\cref{eq:Lambda_q,eq:Lambda_t}) rather than from the derived
fields. All six residual components are identically zero (to
machine precision) along the entire crack, confirming the exact
algebraic satisfaction of the interface laws.

This result highlights a fundamental distinction between the two enforcement strategies. The strong enforcement satisfies the constitutive law \emph{pointwise} at each interface node: the values $\lambda_q$ and~$\lambda_t$ are constrained to match the Robin flux law and the spring traction law exactly by the algebraic structure of the semi-discrete system. The weak enforcement, by contrast, imposes these laws only in an
$L^2(\Gamma_c)$ integral sense via the bilinear form, so that they are satisfied on average across each interface element but not at individual nodes---hence the structured oscillatory residual visible in \cref{fig:profiles_mechanics_offset}c,d. The price of the pointwise satisfaction is the introduction of additional
unknowns into the system.

It is important to note that exact pointwise satisfaction of the algebraic constraint does not imply an exact solution of the continuous problem. The nodal values inherit the
finite-element discretization error present in the primal fields $(p^h, \mathbf{u}^h)$: because the interface is a codimension-one manifold embedded in the bulk mesh, its solution is controlled by the accuracy of the bulk approximation. In particular, the trace of the discrete fields on the surrogate facets is determined by the element-level interpolation in the adjacent elements, so that any bulk discretization error propagates directly to the interface. The constitutive law is therefore satisfied exactly \emph{for the discrete fields}, but the discrete fields themselves converge to the continuous solution only under mesh refinement.

% ------ Figure: λ-residual flow (offset) ------

% ------ Figure: λ-residual mechanics (offset) ------

\subsubsection{Convergence study}
\label{sec:aligned_offset_convergence}

\begin{figure*}[t]
  \centering
  \pagefiguregraphics{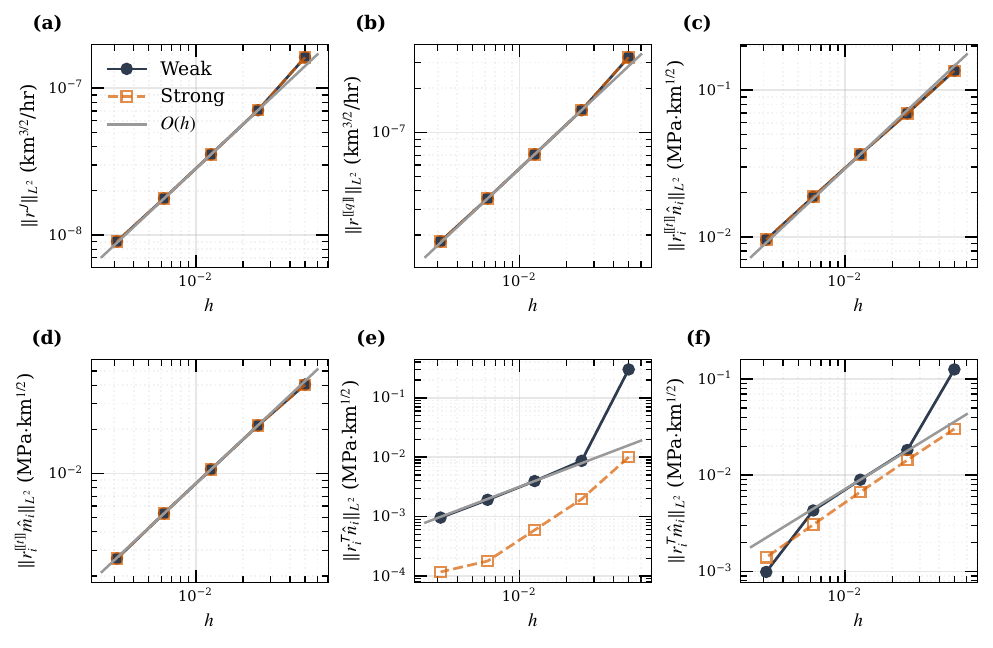}
  \caption{Convergence of interface residual norms for the
  mesh-aligned offset crack:
  (a)~$\|r^{\mathsf{J}}\|_{L^2}$,
  (b)~$\|r^{\jmp{q}}\|_{L^2}$,
  (c)~$\|r_i^{\jmp{t}}\,\hat{n}_i\|_{L^2}$,
  (d)~$\|r_i^{\jmp{t}}\,\hat{m}_i\|_{L^2}$,
  (e)~$\|r_i^{\mathsf{T}}\,\hat{n}_i\|_{L^2}$,
  (f)~$\|r_i^{\mathsf{T}}\,\hat{m}_i\|_{L^2}$.
  The solid gray line indicates $\mathcal{O}(h)$. The strong formulation satisfies the constitutive laws pointwise; errors stem from post-processing.}
  \label{fig:convergence_rates_offset}
\end{figure*}

\Cref{fig:convergence_rates_offset} displays the convergence
behavior. The flow and balance residuals are indistinguishable
between the two enforcement strategies and are listed once; the
constitutive residuals differ and are reported separately.

The flow constitutive and flux balance norms
(\cref{fig:convergence_rates_offset}a,b) converge at
$\mathcal{O}(h)$, and the traction balance norms
(\cref{fig:convergence_rates_offset}c,d) likewise converge at
$\mathcal{O}(h)$. These four residuals are
identical between the two enforcement strategies.

The constitutive residuals
(\cref{fig:convergence_rates_offset}e,f) reveal a meaningful
distinction. The strong enforcement converges consistently, with the
normal component $\|r_i^{\mathsf{T}}\,\hat{n}_i\|_{L^2}$
exhibiting an initially high rate prior to be being roughly $\mathcal{O}(h)$ and the
tangential component converging at $\mathcal{O}(h)$. The weak
enforcement shows pre-asymptotic behavior at the coarsest mesh
($n = 20$), where the constitutive residual is an order of
magnitude larger than at the next refinement level; from $n = 40$
onward, it settles into $\mathcal{O}(h)$ convergence. At the
finer mesh levels, the strong normal constitutive residual is approximately
ten times smaller than the weak counterpart, while the tangential
components are comparable. This advantage of the strong strategy is
expected: the algebraic closure satisfies the constitutive
law exactly at the interface degrees of freedom, so its constitutive
residual reflects only the post-processing error, whereas the weak
enforcement residual includes contributions from the integral
enforcement itself.

Despite this difference, all residual norms decrease monotonically
from $n = 40$ onward and both strategies converge to the same
solution under refinement. The convergence rates and magnitudes of
the balance and flux residuals are unaffected by the lateral offset,
confirming that the formulation is insensitive to the placement of
the crack within the mesh provided the surrogate remains aligned
with element faces.

%\input{figures/aligned_offset/convergence_table}

% ------ Figure: Convergence rates (offset) ------

% ======================================================================
% 30-degree angled crack
% ======================================================================
\subsection{Angled crack}
\label{sec:angled_crack}

\begin{figure*}[t]
  \centering
  \pagefiguregraphics{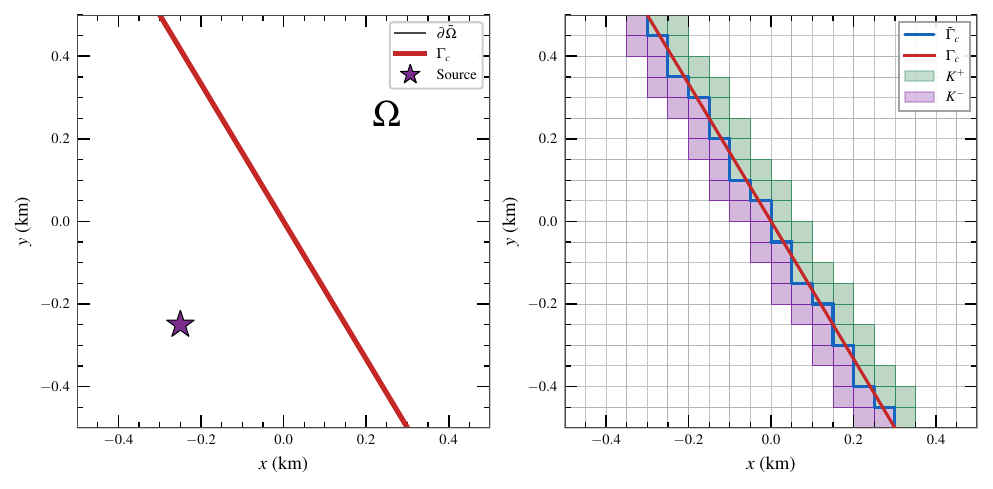}
  \caption{Angled crack configuration ($\theta = \arctan(0.6) \approx 30.96^\circ$).
  (a)~Domain geometry: the crack~$\Gamma_c$ (red) is rotated
  by $\arctan(0.6)$ from the horizontal. The point fluid source (purple
  star) is at $(-0.25,\,-0.25)$.
  (b)~Element classification on a coarse mesh: the surrogate
  interface~$\tilde\Gamma_c$ (blue staircase) approximates the
  true crack~$\Gamma_c$ (red) by splitting connectivity along
  the nearest column of element faces. Elements are classified as
  $K^+$ (green) and $K^-$ (purple).}
  \label{fig:domain_classification_30angle}
\end{figure*}

The preceding test verified the method on a mesh-aligned
configuration where the surrogate and true interface normals coincide. We now introduce a genuine geometric
mismatch by rotating the crack by $\theta = \arctan(0.6) \approx 30.96^\circ$ with respect to the mesh
lines, so that the surrogate interface~$\tilde\Gamma_c$ is a
staircase approximation of the true crack~$\Gamma_c$
(\cref{fig:domain_classification_30angle}). This test exercises the
full shifted interface machinery and reveals
differences between the two enforcement strategies that are absent in
the aligned cases.

\subsubsection{Problem setup}
\label{sec:angled_setup}

The domain, material parameters, boundary conditions, and source location are as described in~\cref{sec:numerical_results}. The
crack is rotated by $\theta = \arctan(0.6) \approx 30.96^\circ$ about the domain center,
giving a crack of length $L_c = \sqrt{1+0.6^2} \approx 1.166$~km that
extends diagonally across the domain. The surrogate
interface~$\tilde\Gamma_c$ is constructed by splitting the mesh connectivity along the column of element faces closest to the true crack, producing a staircase pattern. As a result, the surrogate normal~$\tilde{n}_i$ is piecewise constant and alternates between horizontal and vertical orientations and the gap vector~$\Delta_i$ is nonzero at every surrogate node. Structured quadrilateral meshes with
$n \in \{20,\,40,\,80,\,160,\,320\}$ elements per side are used, with the same interface constitutive model as before: high spring stiffnesses $k_n = k_t = 10^{8}$~MPa/km and zero transmissivity ($T_n = 0$).

An important geometric feature of this configuration is the
treatment of the crack tips. The crack extends to the domain
boundary, so the split connectivity is carried all the way to the boundary endpoint nodes. Consequently, this case does not contain the interior split-to-continuous transition that arises for embedded cracks. The dominant post-processing error near the crack endpoints instead comes from the localized staircase endpoint geometry, the projection onto the true-crack frame, and the interaction with the boundary Dirichlet condition. As will be seen below, these effects
primarily impact the post-processed flux residuals.

% ------ Figure: Domain classification (30 angle) ------

\subsubsection{Solution fields}
\label{sec:angled_fields}

\begin{figure*}[t]
  \centering
  \pagefiguregraphics{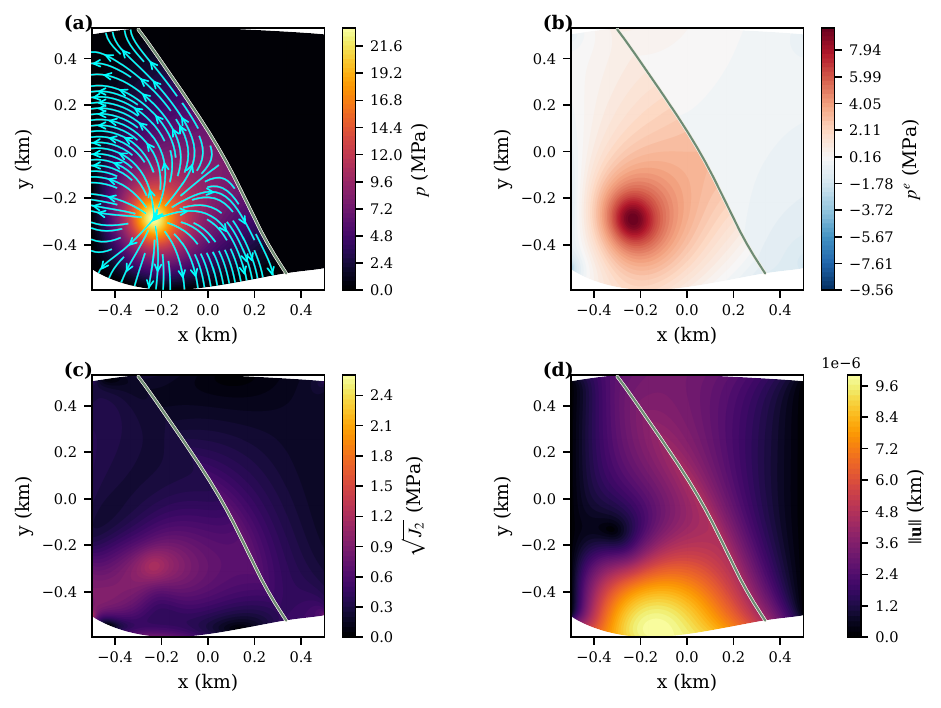}
  \caption{Computed solution fields on the finest mesh ($n = 320$)
  for the angled crack with $\theta = \arctan(0.6)$, displayed on the deformed
  configuration (displacement $\times 10^4$):
  (a)~pore pressure $p$ with Darcy flux vectors,
  (b)~mean effective stress
  $p^e = \tfrac{1}{3}\sigma^e_{kk}$,
  (c)~deviatoric stress invariant $\sqrt{J_2}$, and
  (d)~displacement magnitude $\|\mathbf{u}\|$.}
  \label{fig:fields_n160_30angle}
\end{figure*}

The computed fields on the finest mesh ($n = 320$) are shown
in~\cref{fig:fields_n160_30angle}; both enforcement strategies
produce visually indistinguishable results. The fields are
qualitatively consistent with the aligned cases: the pore pressure (\cref{fig:fields_n160_30angle}a) shows a localized high-pressure region around the source with a clear discontinuity across the impermeable crack, and the Darcy flux vectors run parallel to the interface. The angled crack orientation creates a more pronounced asymmetry in the pressure and stress distributions than in the preceding cases. The mean effective stress (\cref{fig:fields_n160_30angle}b) displays the expected jump across the crack, the deviatoric stress invariant (\cref{fig:fields_n160_30angle}c) shows concentrations near the source and crack tips, and the displacement magnitude
(\cref{fig:fields_n160_30angle}d) appears continuous across the crack. Notably, the post-processed crack geometry appears as a smooth diagonal line at the physical location of~$\Gamma_c$, confirming that the coordinate projection and Taylor transfer of~\cref{sec:postprocess_true} successfully recover the true crack geometry from the staircase surrogate.

% ------ Figure: Solution fields at n=320 (30 angle) ------

\subsubsection{Crack-face profiles}
\label{sec:angled_profiles}

\begin{figure*}[t]
	\centering
	\pagefiguregraphics{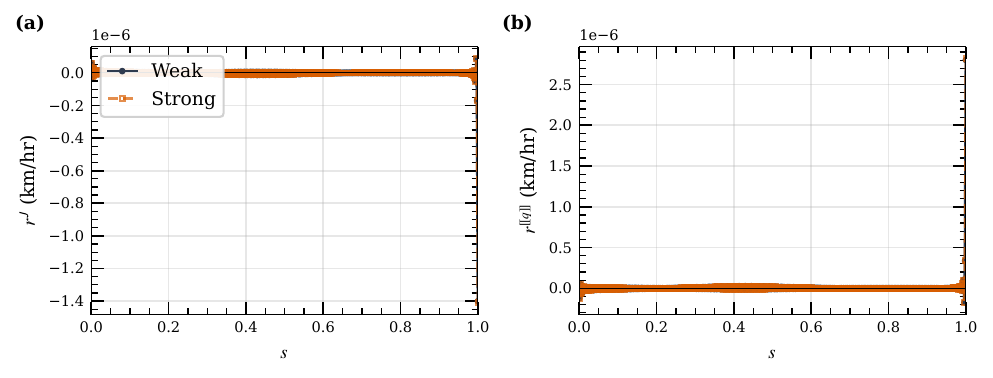}
	\caption{Flow interface residuals along the angled
		crack with $\theta = \arctan(0.6)$ ($n = 320$):
		(a)~constitutive residual
		$r^{\mathsf{J}} = \avg{\hat{q}_j}\,\hat{n}_j
		- \mathsf{J}_\Gamma$, and
		(b)~flux balance residual
		$r^{\jmp{q}} = \jmp{\hat{q}_j}\,\hat{n}_j$.
		Both enforcement strategies are overlaid. The strong formulation satisfies the constitutive laws pointwise; errors stem from post-processing.}
	\label{fig:profiles_flow_30angle}
\end{figure*}

\begin{figure*}[t]
	\centering
	\pagefiguregraphics{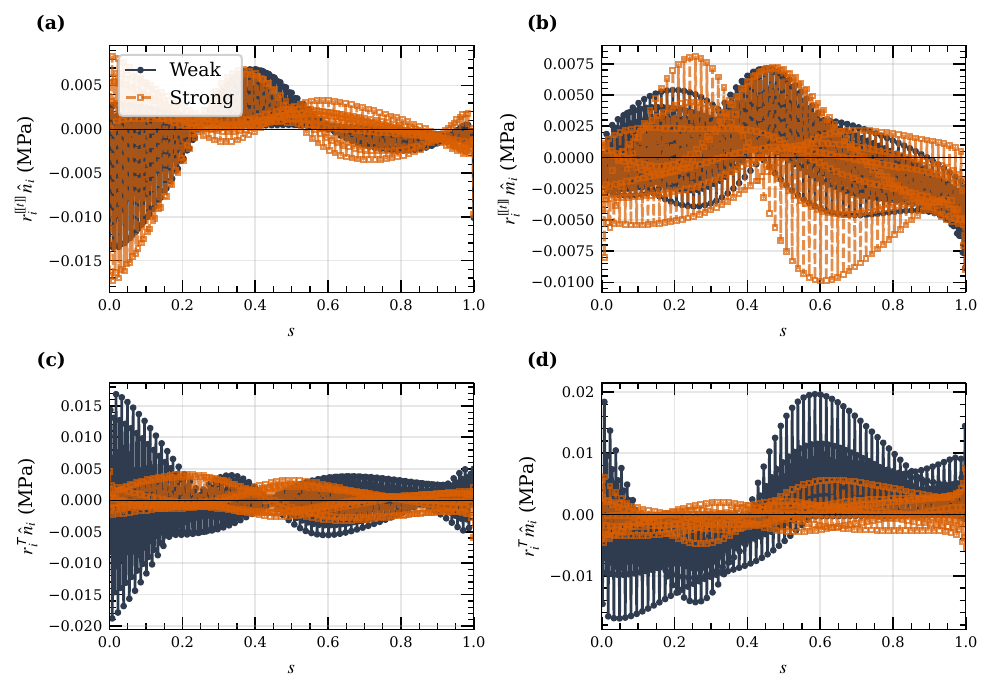}
	\caption{Mechanical interface residuals along the
		angled crack with $\theta = \arctan(0.6)$ ($n = 320$):
		(a)~$r_i^{\jmp{t}}\,\hat{n}_i$ (normal traction balance),
		(b)~$r_i^{\jmp{t}}\,\hat{m}_i$ (tangential traction balance),
		(c)~$r_i^{\mathsf{T}}\,\hat{n}_i$ (normal constitutive residual),
		(d)~$r_i^{\mathsf{T}}\,\hat{m}_i$ (tangential constitutive
		residual).
		Both enforcement strategies are overlaid. The strong formulation satisfies the constitutive laws pointwise; errors stem from post-processing.}
	\label{fig:profiles_mechanics_30angle}
\end{figure*}

\Cref{fig:profiles_flow_30angle,fig:profiles_mechanics_30angle} display the interface residuals along the crack on the finest mesh ($n = 320$). The flow residuals
(\cref{fig:profiles_flow_30angle}) are of order~$10^{-6}$~km/hr along the crack interior. For reference, the characteristic Darcy velocity is $q_{\mathrm{ref}} = (k/\eta)\,p_{\max}/L \approx 4 \times 10^{-6}$~km/hr, so the pointwise residuals amount to roughly~$25\%$ of this scale. This is approximately two orders of magnitude larger than in the aligned cases, reflecting the additional error introduced by the geometric mismatch between the staircase surrogate and the true crack. Both enforcement strategies produce nearly indistinguishable flow residuals. A prominent feature is the sharp spike near the crack tips ($s \approx 0$ and $s \approx 1$), where both the constitutive residual
$r^{\mathsf{J}}$ and the balance residual $r^{\jmp{q}}$ increase by roughly an order of magnitude relative to the interior values. The origin of this spike and its impact on the global convergence rate are discussed in~\cref{sec:angled_convergence}.

% ------ Figure: Flow profiles (30 angle) ------

% ------ Figure: Mechanics profiles (30 angle) ------

\subsubsection{Convergence study}
\label{sec:angled_convergence}

\begin{figure*}[t]
	\centering
	\pagefiguregraphics{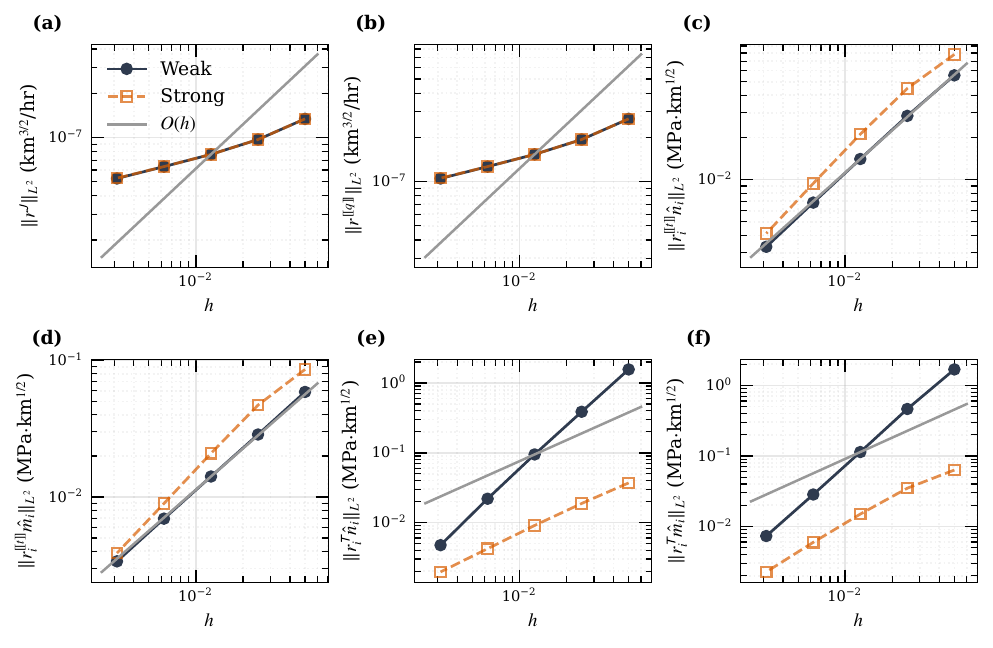}
	\caption{Convergence of interface residual norms for the
		angled crack with $\theta = \arctan(0.6)$:
		(a)~$\|r^{\mathsf{J}}\|_{L^2}$,
		(b)~$\|r^{\jmp{q}}\|_{L^2}$,
		(c)~$\|r_i^{\jmp{t}}\,\hat{n}_i\|_{L^2}$,
		(d)~$\|r_i^{\jmp{t}}\,\hat{m}_i\|_{L^2}$,
		(e)~$\|r_i^{\mathsf{T}}\,\hat{n}_i\|_{L^2}$,
		(f)~$\|r_i^{\mathsf{T}}\,\hat{m}_i\|_{L^2}$.
		The solid gray line indicates $\mathcal{O}(h)$. The strong formulation satisfies the constitutive laws pointwise; errors stem from post-processing. }
	\label{fig:convergence_rates_30angle}
\end{figure*}

\Cref{fig:convergence_rates_30angle}
displays the convergence behavior. All residual norms reported in
this section are computed from the \emph{derived} one-sided field
traces (flux and stress), not from the additional nodal degrees of freedom in the case of the strong enforcement. For
both enforcement strategies, the same post-processing pipeline is
used. Consequently, for the strong enforcement the constitutive
residuals do not measure convergence of the constitutive law
itself---which is already satisfied exactly at the interface nodes
(cf.\ \cref{sec:aligned_offset_lambda})---but rather the accuracy
of the post-processing reconstruction. The flow residuals are
nearly indistinguishable between the two enforcement strategies
and are listed once; all mechanical residuals differ and are
reported separately. The convergence behavior separates into three
distinct regimes depending on the residual type.

%\input{figures/30_angle/convergence_table}

% ------ Figure: Convergence rates (30 angle) ------

The flux norms $\|r^{\mathsf{J}}\|_{L^2}$ and $\|r^{\jmp{q}}\|_{L^2}$
(\cref{fig:convergence_rates_30angle}a,b) converge at approximately $\mathcal{O}(h^{0.3})$, substantially slower than the $\mathcal{O}(h)$ rate observed in the aligned cases. We
attribute this degradation to a localized post-processing error near the surrogate crack tips, where three factors interact: (i) $L^2$~gradient reconstruction is least accurate in the staircase
endpoint patch where the surrogate crack terminates on the boundary; (ii) the projection onto the true-crack normal $\hat{n}_i$ (at angle $\theta = \arctan(0.6)$ to the mesh lines) mixes both gradient components, contaminating the normal flux with the tangential fluxes; and (iii) the tips lie on the domain boundary~$\partial \Omega$, where the Dirichlet condition constrains the pressure value but not its gradient, compounding the mismatch. This hypothesis is confirmed by the tip-trimming study discussed at the end of this section.

The traction balance norms (\cref{fig:convergence_rates_30angle}c,d), by contrast, converge at $\mathcal{O}(h)$ for both enforcement strategies and are not subject to the same tip-dominated degradation. The key difference is the magnitude of the field jump at the interface: the displacement jump is negligibly small ($k_n L / E \approx 2 \times 10^3$, giving $\jmp{u_i} \sim 10^{-8}$~km), so the endpoint-induced stress-projection error remains mild and converges at the same rate as the
interior. The strong enforcement produces balance residuals approximately $1.5$--$2$ times larger than the weak enforcement across all mesh levels, which is the counterpart of the constitutive residual behavior discussed next: the strong strategy pins $\boldsymbol{\lambda}$ to satisfy the spring law exactly, constraining the bulk solution, so the traction balance is satisfied only through the bulk equations; the weak strategy distributes both balance and constitutive errors through the variational formulation, which appears to yield a slightly more accurate traction balance, as evidenced by the post-processed gradient reconstruction.

The constitutive residual norms (\cref{fig:convergence_rates_30angle}e,f) show the most striking
distinction between the two strategies. For the weak enforcement, the constitutive residual genuinely measures how well the constitutive law is satisfied, since the constitutive relation is imposed only
in an integral sense; this residual converges at approximately $\mathcal{O}(h^2)$, starting from values exceeding $1$~MPa$\cdot$km$^{1/2}$ at $n = 20$ and decreasing to
approximately $2$--$3 \times10^{-2}$~MPa$\cdot$km$^{1/2}$ at $n = 320$. For the strong enforcement, by contrast, the constitutive law is already satisfied exactly at the interface nodes
(cf.\ \cref{sec:aligned_offset_lambda}), so the derived-field residual shown here measures only the post-processing error---the mismatch between the enforced values and the one-sided
stress traces reconstructed from the bulk fields via the $L^2$~gradient projection. This post-processing residual converges at approximately $\mathcal{O}(h)$, starting from values roughly
two orders of magnitude smaller than the weak counterpart, so that at the finer mesh levels the strong residuals remain approximately five times smaller. The faster $\mathcal{O}(h^2)$ rate of the weak enforcement reflects the convergence of the integral enforcement error, which dominates its residual; the $\mathcal{O}(h)$ rate of the strong enforcement is set by the first-order accuracy of the gradient reconstruction and the projection onto~$\hat{n}_i$.

\begin{figure}[!htbp]
		\centering
		\pagefiguregraphics{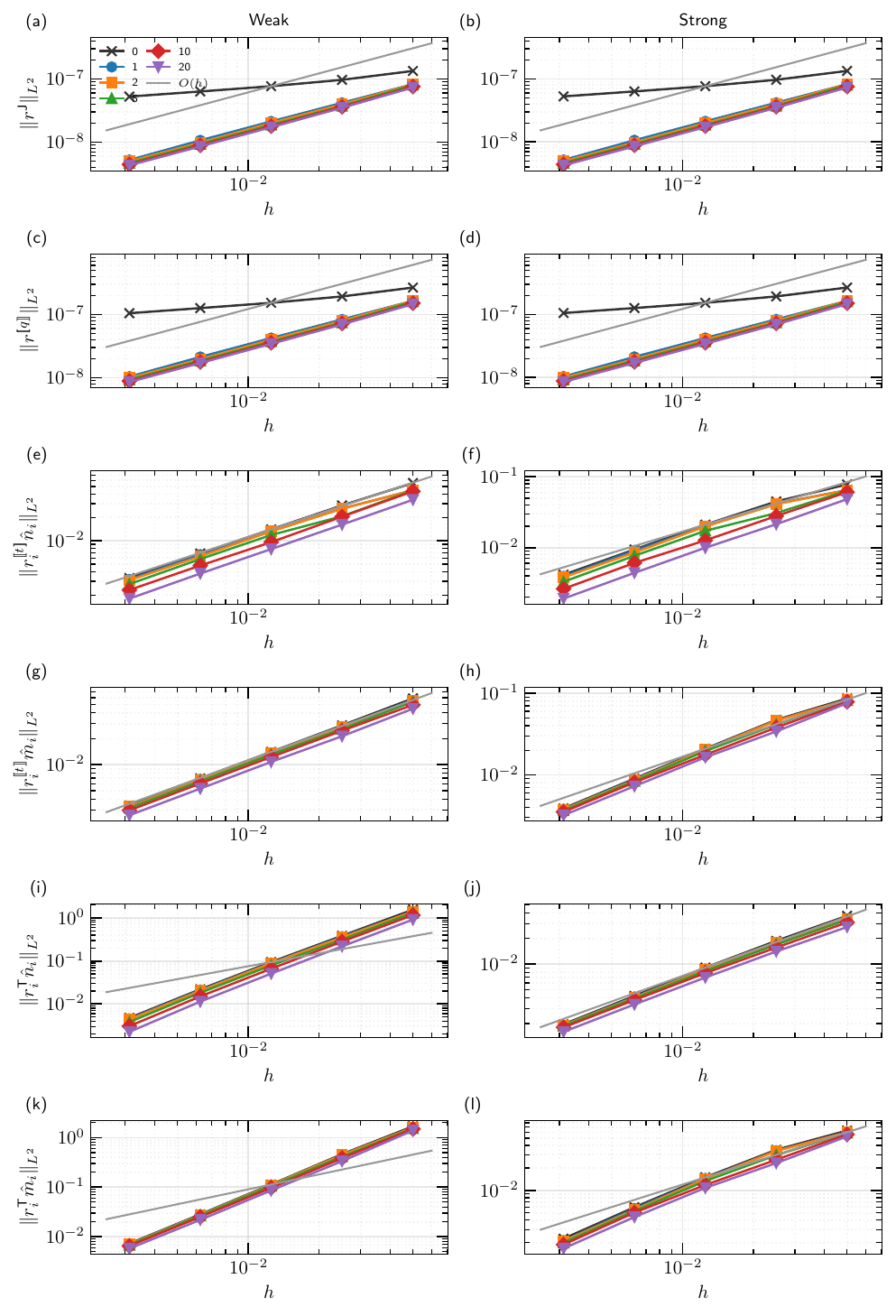}
	\caption{Effect of tip trimming on the convergence of interface
		residual norms for the angled crack with
		$\theta = \arctan(0.6)$. Each curve
		corresponds to excluding a percentage~$\epsilon$ of the crack
		length at each tip before computing the $L^2$~norm. Left
		column: weak enforcement; right column: strong enforcement.
		Rows from top to bottom:
		(a,b)~$\|r^{\mathsf{J}}\|_{L^2}$,
		(c,d)~$\|r^{\jmp{q}}\|_{L^2}$,
		(e,f)~$\|r_i^{\jmp{t}}\,\hat{n}_i\|_{L^2}$,
		(g,h)~$\|r_i^{\jmp{t}}\,\hat{m}_i\|_{L^2}$,
		(i,j)~$\|r_i^{\mathsf{T}}\,\hat{n}_i\|_{L^2}$,
		(k,l)~$\|r_i^{\mathsf{T}}\,\hat{m}_i\|_{L^2}$.
		The solid gray line indicates $\mathcal{O}(h)$.}
	\label{fig:tip_trimming}
	\end{figure}

To verify the tip-error hypothesis and to assess the tip contribution to all residual components, we exploit the fact that, in classical fracture mechanics, crack-tip singular fields decay rapidly with distance from the tip; the post-processing error concentrated near the crack endpoints is expected
to be similarly localized. We therefore repeat the $L^2$~norm computation after excluding a percentage~$\epsilon$ of the crack length at each tip. \Cref{fig:tip_trimming} shows the results for
$\epsilon \in \{0,\,1,\,2,\,5,\,10,\,20\}\%$. The effect on the flux residuals is dramatic (\cref{fig:tip_trimming}a--d): the untrimmed norms ($\epsilon = 0$) exhibit the sub-first-order rate
discussed above, while even a $1\%$ trim substantially improves the convergence. For $\epsilon \geq 5\%$, the flux residuals recover the $\mathcal{O}(h)$ rate observed in the aligned cases, confirming that the degraded global rate is entirely attributable to the localized tip patches. The traction and constitutive residuals (\cref{fig:tip_trimming}e--l) also benefit from tip trimming, with improved convergence rates and tighter curves as $\epsilon$ increases, confirming that tip effects contribute to all residual components even when they do not dominate the global rate.

% ------ Figure: Tip trimming ------

Despite the differences in individual residual components, both enforcement strategies produce solutions that converge under refinement. At $n = 160$, the flux constitutive
residual $\|r^{\mathsf{J}}\|_{L^2} \approx 6 \times 10^{-8}$~km$^{3/2}$/hr, which relative to the injection rate $Q = 10^{-5}$~km$^2$/hr is of order~$10^{-3}$, the same relative accuracy as in the aligned cases. All mechanical residual norms at $n = 160$ are below $3 \times 10^{-2}$~MPa$\cdot$km$^{1/2}$, less than $0.5\%$ of the Biot coupling stress $\alpha\,p_{\max} \approx 6$~MPa. These results confirm that the shifted interface formulation yields convergent, physically accurate solutions even when the surrogate interface does not align with the true crack geometry.

\begin{figure}[!htbp]
		\centering
		\pagefiguregraphics{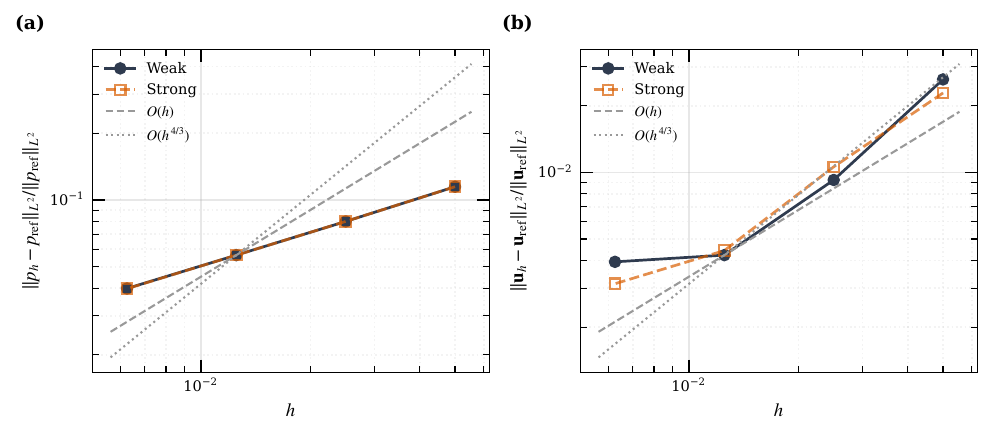}
	\caption{Bulk-field $L^2(\Omega)$ self-convergence for the
		angled crack with $\theta = \arctan(0.6)$. Each plotted point shows the
		relative error~\cref{eq:self_convergence} for one
		adjacent-level comparison:
		$n=20$ against $40$, $40$ against $80$, $80$ against
		$160$, and $160$ against $320$.
		(a)~Pressure, (b)~displacement.
		The gray $O(h)$ and $O(h^{4/3})$ lines are shown only as visual references. In particular, the $O(h^{4/3})$ guide is motivated by reduced-regularity corner benchmarks and is not claimed here as a sharp theoretical rate for the present fracture problem.}
	\label{fig:self_convergence_30angle}
	\end{figure}

The interface residual convergence study above measures how well the discrete solution satisfies the interface conditions, but does not directly assess the accuracy of the bulk fields (pressure and displacement) over the entire domain. Since no closed-form solution is available for this problem, we assess bulk-field convergence through a self-convergence study based on the shifted nodal fields
stored on the split surrogate mesh. For each mesh level $h$, the coarse solution is compared with the next finer level $h_{\mathrm{ref}}$ using the relative discrete $L^2$ error
\begin{equation}
	e_f
	= \frac{\|f_h - \Pi_h f_{h_{\mathrm{ref}}}\|_{M_h}}
	       {\|\Pi_h f_{h_{\mathrm{ref}}}\|_{M_h}},
	\label{eq:self_convergence}
\end{equation}
where $f$ denotes either $p$ or~$\mathbf{u}$, $\Pi_h$ transfers the finer shifted nodal field into the coarse nodal space, and $\|\mathbf{v}\|_{M_h}^2 := \mathbf{v}^\mathsf{T} M_h \mathbf{v}$ with $M_h$ the consistent scalar mass matrix on the coarse split mesh (applied componentwise for displacement). When the split meshes are nested, the finer shifted nodal values are taken directly at matching split nodes; otherwise, they are transferred by a consistent coarse-space $L^2$ projection. In the present computations, the mesh sequence is $n \in \{20,\,40,\,80,\,160,\,320\}$, and the plotted self-convergence points correspond to the adjacent comparisons $20\!\to\!40$, $40\!\to\!80$, $80\!\to\!160$, and $160\!\to\!320$. The $n=320$ level is therefore used only as the terminal reference and is not plotted. The gray $\mathcal{O}(h)$ and $\mathcal{O}(h^2)$ lines in \cref{fig:self_convergence_30angle} are visual reference slopes only.

% \Cref{fig:self_convergence_30angle} shows the results. The pressure error (\cref{fig:self_convergence_30angle}a) converges slower than $\mathcal{O}(h)$ for both enforcement strategies, with the weak and strong curves essentially coinciding. The displacement error, on the other hand, (\cref{fig:self_convergence_30angle}b) decays initially more rapidly prior to an apparent reduction of convergence lower than $\mathcal{O}(h)$. Rates less than optimal are consistent with expectations for bilinear elements on a non-smooth domain: solutions of elliptic problems near re-entrant corners or crack-type singularities possess reduced Sobolev regularity~\citep{Grisvard2011,Dauge2008,Demlow2016}, and the replacement of the true interface by a staircase surrogate may itself limit the achievable convergence order~\citep{DupontGuzmanScott2020}. Critically, the bulk fields converge with mesh refinement, even though some post-processed interface residuals exhibit degraded convergence rates due to the localized tip artifacts discussed above.

\Cref{fig:self_convergence_30angle} shows the bulk-field $L^2(\Omega)$ self-convergence results for the angled crack. In \cref{fig:self_convergence_30angle}a, the pressure error decreases under mesh refinement for both enforcement strategies, with the weak and strong curves nearly coincident, but with a rate below the smooth-solution $O(h^2)$ benchmark. In \cref{fig:self_convergence_30angle}b, the displacement error decreases more rapidly on the coarser meshes before transitioning to a slower asymptotic trend, again below the smooth-solution rate. This behavior is consistent with reduced regularity near non-smooth geometric features. For a related Poisson benchmark on a domain with a $3\pi/2$ re-entrant corner, \citet{atallah2021analysis} report $u\in H^{1+r}(\Omega)$ for any $r<2/3$, with observed $L^2$-convergence close to $O(h^{4/3})$ rather than $O(h^2)$. We therefore include the $O(h^{4/3})$ line only as a qualitative reference for bulk $L^2$ behavior in a reduced-regularity regime, not as a claimed sharp asymptotic rate for the present coupled fracture problem. In crack problems, the limiting exponent may depend on the local crack-tip singularity and on the conditions imposed on the two crack faces \citep{costabel2002crack}. A precise theoretical derivation of the asymptotic convergence rate for the present problem exceeds the scope of this work and will be addressed in a future study. The main point here is, therefore, that the bulk fields converge robustly under refinement for both enforcement strategies, even though some post-processed interface residuals remain more sensitive to localized tip artifacts.

% ------ Figure: Self-convergence bulk (30 angle) ------

\FloatBarrier

	\subsection{Embedded angled crack}
\label{sec:embedded_crack}

The previous test considered a crack that extends to the domain boundary, so that each crack tip coincides with a point on~$\partial \Omega$. We now examine the more challenging configuration of an \emph{embedded} crack whose tips lie entirely within the domain interior. This arrangement creates two localized singularities of a qualitatively different character from the boundary-intersecting case: here, the split connectivity terminates in the bulk rather than continuing to a boundary endpoint, so the element patches surrounding each tip must reconcile two-sided duplicated fields with an unsplit interior neighborhood. As we shall see, these interior tips produce severe localized post-processing artifacts that contaminate the global convergence rates. In classical fracture mechanics, crack-tip singular fields are known to decay rapidly with distance from the tip. The same spatial localization is expected for the post-processing error here. This motivates a tip-trimming convergence
analysis, introduced at the end of this section, in which a small fraction of the crack length near each tip is excluded from the $L^2$~norm computation, thereby isolating the convergence behavior
of the interior where the shifted interface approximation is well-resolved.

\subsubsection{Problem setup}
 \label{sec:embedded_setup}
\begin{figure}[!htbp]
 \centering
 \pagefiguregraphics{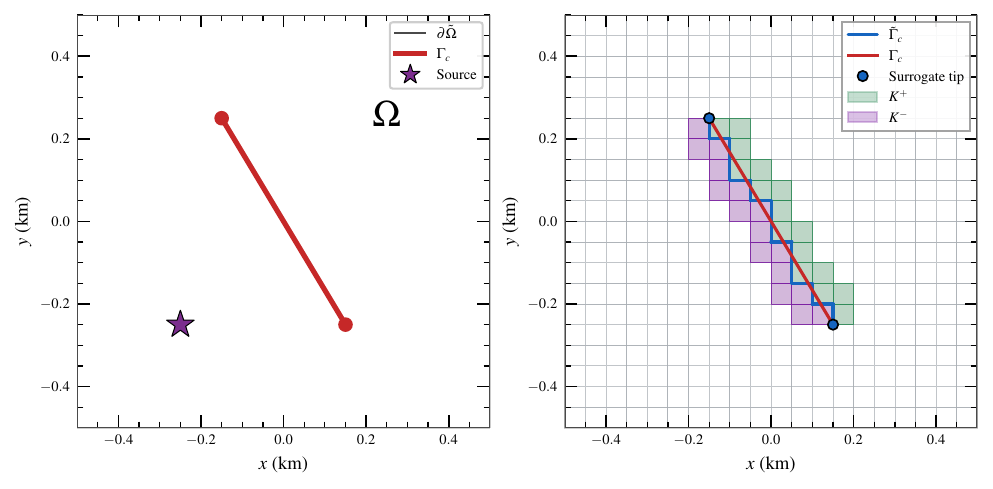}
 \caption{Embedded angled crack configuration ($\theta = \arctan(0.6) \approx 30.96^\circ$, $L_c = \tfrac12\sqrt{1+0.6^2} \approx 0.583$~km). (a)~Domain geometry: the crack~$\Gamma_c$ (red) is rotated by $\arctan(0.6)$ from the horizontal with both tips in the domain interior. The point fluid source (purple star) is at $(-0.25,\,-0.25)$. (b)~Element classification on a coarse mesh: the surrogate interface~$\tilde\Gamma_c$ (blue staircase) approximates the true crack~$\Gamma_c$ (red). Surrogate tip locations (black circles) lie in the interior; elements are classified as $K^+$ (green) and $K^-$ (purple).}
\label{fig:domain_classification_embedded}
\end{figure}

  The domain, material parameters, boundary conditions, and source location are as described in~\cref{sec:numerical_results}. The crack orientation is the same $\theta = \arctan(0.6) \approx 30.96^\circ$ as in the preceding test, but the crack length is reduced to $L_c = \tfrac12\sqrt{1+0.6^2} \approx 0.583$~km so that both tips remain well inside~$\tilde{\Omega}$ (\cref{fig:domain_classification_embedded}a). The surrogate interface~$\tilde\Gamma_c$ is once again a staircase approximation of the true crack~$\Gamma_c$ (\cref{fig:domain_classification_embedded}b), constructed identically to the boundary-intersecting case. The same interface constitutive model is used: spring stiffnesses $k_n = k_t = 10^{8}$~MPa/km and zero transmissivity ($T_n = 0$). Structured quadrilateral meshes with $n \in \{20,\,40,\,80,\,160,\,320\}$ elements per side are used.
  
   An important geometric distinction from the boundary-intersecting case is the character of the surrogate crack tips. When the crack extends to~$\partial \Omega$, the endpoint nodes remain duplicated on the boundary, and the two-sided traces persist up to~$\partial \Omega$. For the embedded crack, by contrast, the tip vertex is an interior node whose connectivity transitions from duplicated (two independent traces $p^\pm$, $u_i^\pm$) to standard (a single shared degree of freedom) within the element patch (\cref{fig:domain_classification_embedded}b). This interior split-to-continuous transition is more severe: the gradient reconstruction must reconcile element contributions with incompatible numbers of active degrees of freedom, and no boundary condition constrains the solution to reduce the mismatch. The resulting tip artifact affects \emph{all} post-processed residual components, not only the flux residuals as in the boundary-intersecting case.
   
 % ------ Figure: Domain classification (embedded)

\FloatBarrier
 \subsubsection{Solution fields} 
 \label{sec:embedded_fields}
\begin{figure}[!htbp]
 \centering
 \pagefiguregraphics{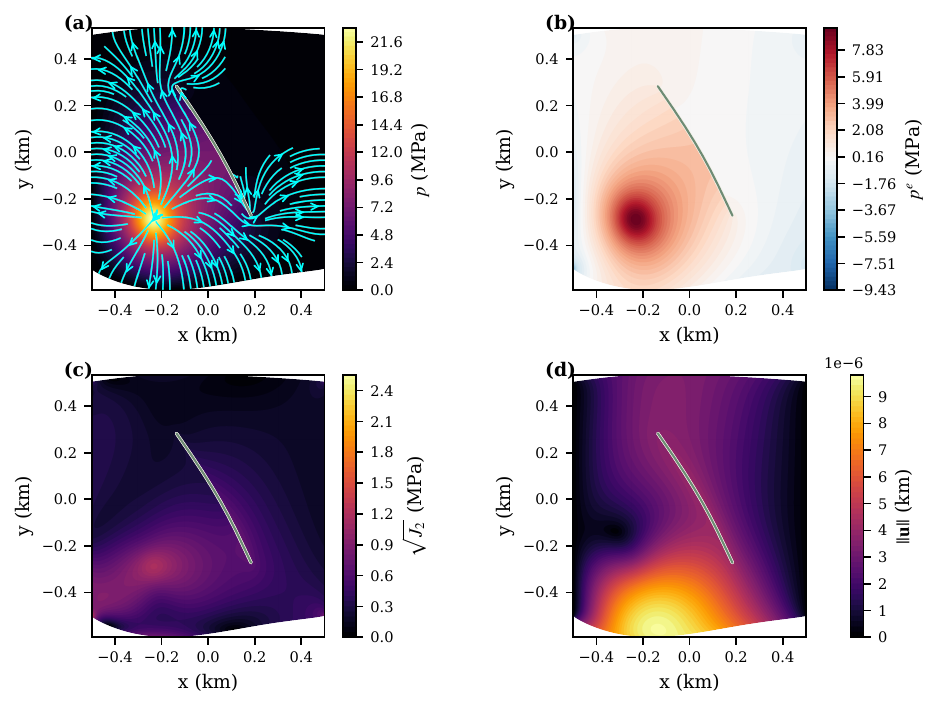}
 \caption{Computed solution fields on the finest mesh ($n = 320$) for the embedded angled crack with $\theta = \arctan(0.6)$, displayed on the deformed configuration (displacement $\times 10^4$): (a)~pore pressure $p$ with Darcy flux vectors, (b)~mean effective stress $p^e = \tfrac{1}{3}\sigma^e_{kk}$, (c)~deviatoric stress invariant $\sqrt{J_2}$, and (d)~displacement magnitude $\|\mathbf{u}\|$.}
 \label{fig:fields_n160_embedded}
\end{figure}

  The computed fields on the finest mesh ($n = 320$) are shown in~\cref{fig:fields_n160_embedded}; both enforcement strategies produce visually indistinguishable results. The fields in this scenario produce different visual results from the boundary-intersecting case. The pore pressure (\cref{fig:fields_n160_embedded}a) exhibits a high-pressure lobe around the source with a sharp discontinuity across the impermeable crack, and the Darcy flux vectors are deflected to flow parallel to the interface and then around the tip, as expected. The shorter crack confines the pressure jump to the central portion of the domain, creating a more compact perturbation pattern. The mean effective stress (\cref{fig:fields_n160_embedded}b) and deviatoric invariant (\cref{fig:fields_n160_embedded}c) show little to no stress concentrations near both embedded crack tips. The displacement magnitude (\cref{fig:fields_n160_embedded}d) appears continuous across the crack, consistent with the high spring stiffness.

  % ------ Figure: Solution fields at n=320 (embedded)

\FloatBarrier

\subsubsection{Crack-face profiles}
\label{sec:embedded_profiles}

\begin{figure}[!htbp]
		\centering
		\pagefiguregraphics{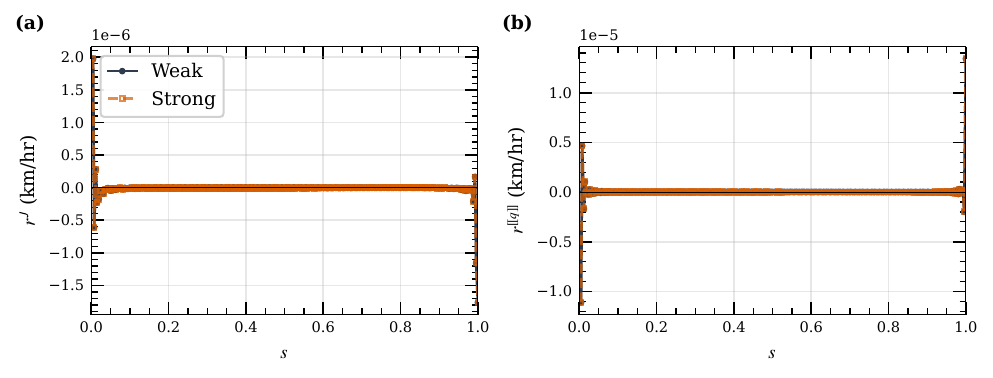}
	\caption{Flow interface residuals along the embedded
		angled crack with $\theta = \arctan(0.6)$ ($n = 320$):
		(a)~constitutive residual
		$r^{\mathsf{J}} = \avg{\hat{q}_j}\,\hat{n}_j
		- \mathsf{J}_\Gamma$, and
		(b)~flux balance residual
		$r^{\jmp{q}} = \jmp{\hat{q}_j}\,\hat{n}_j$.
		Sharp spikes appear at both crack tips, reflecting the
		interior tip artifact.
		Both enforcement strategies are overlaid. The strong formulation satisfies the constitutive laws pointwise; errors stem from post-processing.}
	\label{fig:profiles_flow_embedded}
	\end{figure}

\begin{figure}[!htbp]
		\centering
		\pagefiguregraphics{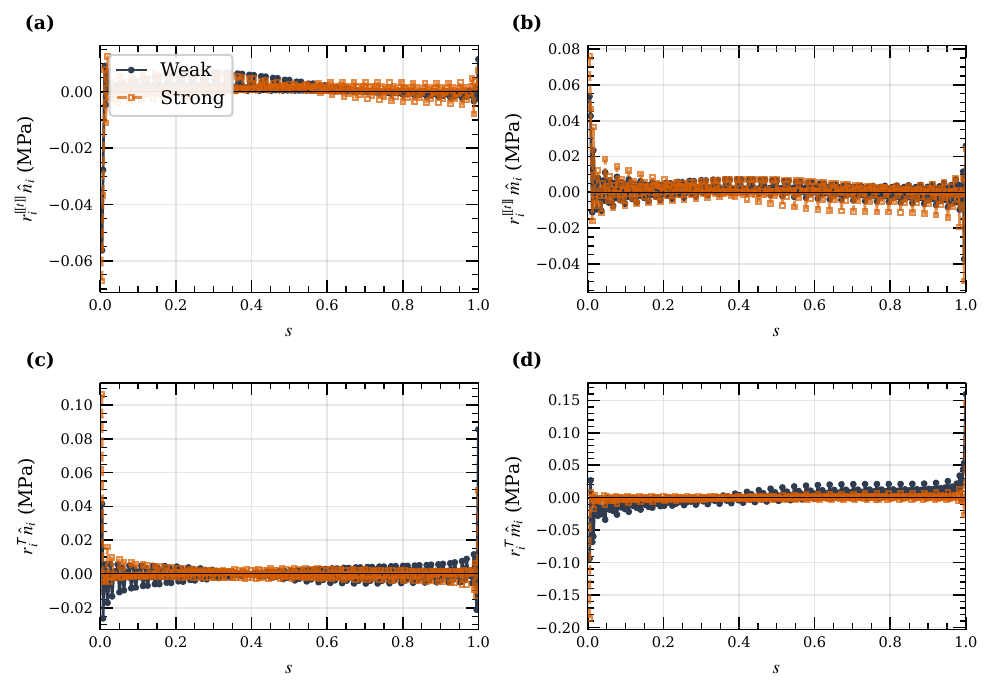}
	\caption{Mechanical interface residuals along the embedded
		angled crack with $\theta = \arctan(0.6)$ ($n = 320$):
		(a)~$r_i^{\jmp{t}}\,\hat{n}_i$ (normal traction balance),
		(b)~$r_i^{\jmp{t}}\,\hat{m}_i$ (tangential traction balance),
		(c)~$r_i^{\mathsf{T}}\,\hat{n}_i$ (normal constitutive residual),
		(d)~$r_i^{\mathsf{T}}\,\hat{m}_i$ (tangential constitutive
		residual). The strong formulation satisfies the constitutive laws pointwise; errors stem from post-processing.}
	\label{fig:profiles_mechanics_embedded}
	\end{figure}

\Cref{fig:profiles_flow_embedded,fig:profiles_mechanics_embedded}
display the interface residuals along the crack on the finest mesh ($n = 320$). The flow residuals
(\cref{fig:profiles_flow_embedded}) are of order~$10^{-6}$~km/hr along the crack interior, comparable to the boundary-intersecting case. Both enforcement strategies produce nearly indistinguishable profiles, and the staircase-induced oscillations from the previous test are again present. A key difference from the boundary-intersecting case is the character of the tip spikes: both ends of the crack now show pronounced peaks in the constitutive residual $r^{\mathsf{J}}$ and in the balance residual $r^{\jmp{q}}$, and these peaks are more severe than in the boundary-intersecting case. The peaks arise because the interior split-to-continuous transition produces a larger gradient reconstruction error than the boundary-intersecting variant, where the duplicated endpoint traces persist to~$\partial \Omega$ and the dominant error is confined to a smaller boundary endpoint patch.

The mechanical residuals (\cref{fig:profiles_mechanics_embedded}) exhibit the same qualitative features as in the boundary-intersecting case, namely staircase oscillations throughout the interior and a separation between the two enforcement strategies in the constitutive components, but
with sharp spikes appearing at both crack tips in all four mechanical residual components. In the boundary-intersecting case, the tip error was confined largely to the flux residuals, while the mechanical residuals converged cleanly at~$\mathcal{O}(h)$. Here, the interior tips contaminate
the traction balance (\cref{fig:profiles_mechanics_embedded}a,b) and the constitutive residuals (\cref{fig:profiles_mechanics_embedded}c,d) alike. The normal traction balance $r_i^{\jmp{t}}\,\hat{n}_i$ reaches peak values of approximately $0.04$~MPa for both strategies, while the constitutive residuals for the weak enforcement (\cref{fig:profiles_mechanics_embedded}c,d, dark grey) oscillate up to $0.05$--$0.08$~MPa in the interior, with much larger tip spikes. The strong enforcement (orange) maintains smaller constitutive residuals (within~$\approx 0.01$~MPa in the interior), consistent with the trade-off identified in~\cref{sec:angled_profiles}.

% ------ Figure: Flow profiles (embedded) ------

% ------ Figure: Mechanics profiles (embedded) ------

\FloatBarrier

\subsubsection{Convergence study}
\label{sec:embedded_convergence}

\begin{figure}[!htbp]
		\centering
		\pagefiguregraphics{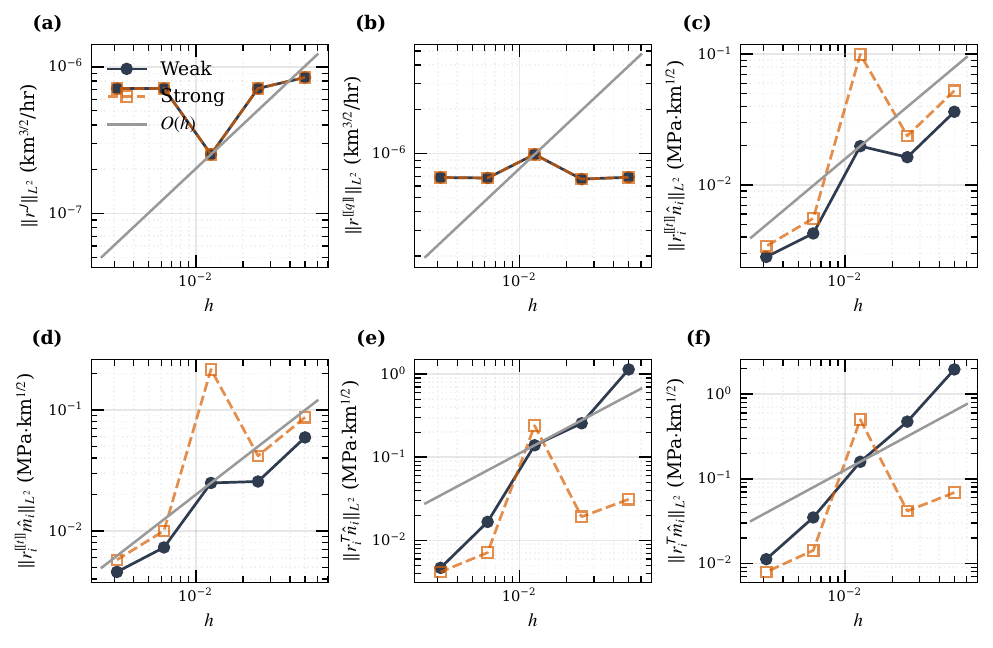}
	\caption{Convergence of interface residual norms for the
		embedded angled crack with $\theta = \arctan(0.6)$:
		(a)~$\|r^{\mathsf{J}}\|_{L^2}$,
		(b)~$\|r^{\jmp{q}}\|_{L^2}$,
		(c)~$\|r_i^{\jmp{t}}\,\hat{n}_i\|_{L^2}$,
		(d)~$\|r_i^{\jmp{t}}\,\hat{m}_i\|_{L^2}$,
		(e)~$\|r_i^{\mathsf{T}}\,\hat{n}_i\|_{L^2}$,
		(f)~$\|r_i^{\mathsf{T}}\,\hat{m}_i\|_{L^2}$.
		The solid gray line indicates $\mathcal{O}(h)$. The strong formulation satisfies the constitutive laws pointwise; errors stem from post-processing.}
	\label{fig:convergence_rates_embedded}
	\end{figure}

\Cref{fig:convergence_rates_embedded} displays the convergence behavior. The results reveal a significantly different convergence landscape compared with the boundary-intersecting case. The flux norms $\|r^{\mathsf{J}}\|_{L^2}$ and $\|r^{\jmp{q}}\|_{L^2}$ (\cref{fig:convergence_rates_embedded}a,b) exhibit essentially no convergence. The overall rate is approximately $\mathcal{O}(h^{0.03})$, and the finest-mesh values ($\|r^{\mathsf{J}}\| \approx 8.7 \times 10^{-7}$, $\|r^{\jmp{q}}\| \approx 6.7 \times 10^{-7}$) are comparable to the coarsest-mesh values. In the boundary-intersecting case
(\cref{sec:angled_convergence}), the same residuals converged at approximately~$\mathcal{O}(h^{0.3})$. The further degradation here is attributable to the more severe interior tip artifact.
Whereas the boundary-intersecting crack remains split up to the Dirichlet-constrained boundary endpoints, the embedded tip transitions from split to unsplit interior nodes where no external constraint modifies the gradient reconstruction. The resulting $\mathcal{O}(1)$ pointwise error at each tip dominates the global $L^2$~norm and prevents convergence over the mesh
range considered.

%\input{figures/30_angle_sub/convergence_table}

% ------ Figure: Convergence rates (embedded) ------

The traction balance norms (\cref{fig:convergence_rates_embedded}c,d) converge at approximately~$\mathcal{O}(h)$ for the strong enforcement and at a comparable rate for the weak enforcement when the anomalous $n = 40$ data point is excluded. At $n = 40$, weak enforcement yields traction balance norms two orders of magnitude larger than those at neighboring mesh levels. The normal
component spikes to $\|r_i^{\jmp{t}}\,\hat{n}_i\|_{L^2} \approx 2.24$~MPa$\cdot$km$^{1/2}$, while at $n = 20$ and $n = 80$ the values are $3.70 \times 10^{-2}$ and $1.69 \times 10^{-2}$~MPa$\cdot$km$^{1/2}$, respectively. The $n = 40$ surrogate tip happens to coincide with a particularly
unfavorable element configuration in which the closest-point projection maps a large cluster of surrogate quadrature points to the same small segment of the true crack near the tip, producing
a localized but intense artifact that contaminates the global norm. The strong enforcement escapes this anomaly because its algebraic closures pin the interface unknowns, preventing the variational
form from amplifying the tip error into the bulk solution.

The constitutive residuals (\cref{fig:convergence_rates_embedded}e,f) show non-monotone behavior for both strategies, although on different mesh levels. The weak enforcement shows the same $n = 40$ spike discussed above, with $\|r_i^{\mathsf{T}}\,\hat{n}_i\|_{L^2}$ reaching $1.17 \times 10^1$~MPa$\cdot$km$^{1/2}$, two orders of magnitude above the values at other mesh levels. The strong enforcement instead shows a non-monotone increase at $n = 80$, where the normal constitutive residual rises to $8.50 \times 10^{-2}$~MPa$\cdot$km$^{1/2}$ from $1.82 \times 10^{-2}$ at $n = 40$ before decreasing to $7.01 \times 10^{-3}$ at $n = 160$. These non-monotone entries
are indicative of tip-dominated convergence. 

\begin{figure}[!htbp]
		\centering
		\pagefiguregraphics{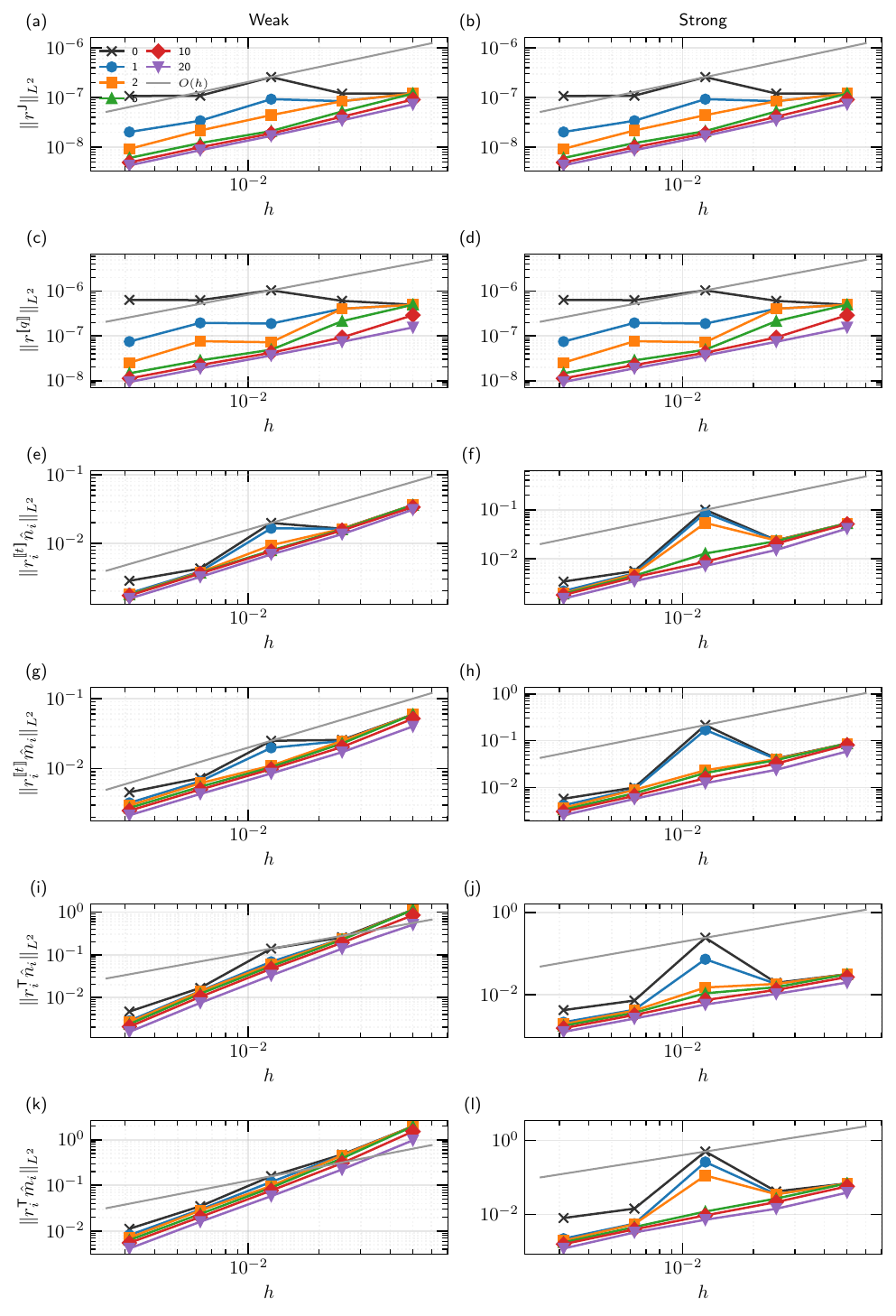}
	\caption{Effect of tip trimming on the convergence of interface
		residual norms for the embedded angled crack with
		$\theta = \arctan(0.6)$.
		Each curve corresponds to excluding a
		percentage~$\epsilon$ of the crack length at each tip before
		computing the $L^2$~norm. Left column: weak enforcement;
		right column: strong enforcement.
		Rows from top to bottom:
		(a,b)~$\|r^{\mathsf{J}}\|_{L^2}$,
		(c,d)~$\|r^{\jmp{q}}\|_{L^2}$,
		(e,f)~$\|r_i^{\jmp{t}}\,\hat{n}_i\|_{L^2}$,
		(g,h)~$\|r_i^{\jmp{t}}\,\hat{m}_i\|_{L^2}$,
		(i,j)~$\|r_i^{\mathsf{T}}\,\hat{n}_i\|_{L^2}$,
		(k,l)~$\|r_i^{\mathsf{T}}\,\hat{m}_i\|_{L^2}$.
		The solid gray line indicates $\mathcal{O}(h)$.}
	\label{fig:tip_trimming_embedded}
	\end{figure}

To confirm that the degraded convergence is caused exclusively by the localized tip patches, we repeat the tip-fraction sensitivity analysis from~\cref{sec:angled_convergence}. The $L^2$~norms are
recomputed after excluding a percentage~$\epsilon$ of the crack length at each tip, for
$\epsilon \in \{0,\,1,\,2,\,5,\,10,\,20\}\%$. \Cref{fig:tip_trimming_embedded} displays the results.

% ------ Figure: Tip trimming (embedded) ------

The effect on the flux residuals is dramatic (\cref{fig:tip_trimming_embedded}a--d). The untrimmed curves ($\epsilon = 0$) are flat, confirming the near-zero convergence rate seen in~\cref{fig:convergence_rates_embedded}. A $1\%$ trim immediately recovers a rate close to~$\mathcal{O}(h)$, and for $\epsilon \geq 2\%$ the rate stabilizes at approximately~$1.03$. This recovery is sharper than in the boundary-intersecting case, where $5\%$ trimming was needed to reach~$\mathcal{O}(h)$, and reflects the fact that the embedded tip artifact, while more intense, is
also more tightly localized.

The traction balance residuals (\cref{fig:tip_trimming_embedded}e--h) converge at approximately~$\mathcal{O}(h)$ for all trimming levels, excluding $\epsilon = 0$, confirming that the tip patches do not dominate these components globally. This is, of course, not unexpected, as we should not expect ideal convergence behavior of the reconstructed fields over the true crack due to the high (theoretically infinite) gradients close to the tip. The constitutive residuals (\cref{fig:tip_trimming_embedded}i--l) show the clearest distinction between the two strategies. The strong enforcement converges at approximately~$\mathcal{O}(h)$ for all trimming percentages~$\epsilon \geq 1\%$, consistent with the boundary-intersecting case. The weak enforcement converges at approximately~$\mathcal{O}(h)$, starting from much smaller absolute values. At $n = 160$ with $5\%$
trimming, the strong constitutive residuals are $\|r_i^{\mathsf{T}}\,\hat{n}_i\| \approx 3.5 \times 10^{-3}$ and $\|r_i^{\mathsf{T}}\,\hat{m}_i\| \approx 3.3 \times 10^{-3}$~MPa$\cdot$km$^{1/2}$, roughly three times smaller than the weak counterparts, reproducing the same trade-off observed in the
boundary-intersecting case.

Despite the challenging interior-tip geometry, both enforcement strategies yield residuals that tend to zero once the localized tip artifacts are accounted for via trimming. All mechanical residual
norms at $n = 160$ are below $5 \times 10^{-2}$~MPa$\cdot$km$^{1/2}$, less than~$1\%$ of the
Biot coupling stress $\alpha\,p_{\max} \approx 6$~MPa. The non-convergence of the untrimmed flux norms does not indicate a failure of the method but rather a limitation of the post-processing. The
discrete solution itself converges, and the residuals are small once the tip patches are excluded. Specialized tip-enrichment strategies (for example, crack-tip singular-function enrichments or adaptive local refinement) could, in principle, eliminate these artifacts but lie outside the scope of the present work. We note that the degraded rates near crack tips are consistent
with classical elliptic regularity theory for domains with geometric singularities~\citep{Grisvard2011,Dauge2008,Demlow2016}. However, these results apply to scalar elliptic problems; determining the precise singular exponents for the coupled poroelastic system considered here would require a dedicated regularity analysis and is left to future work.

\begin{figure}[!htbp]
		\centering
		\pagefiguregraphics{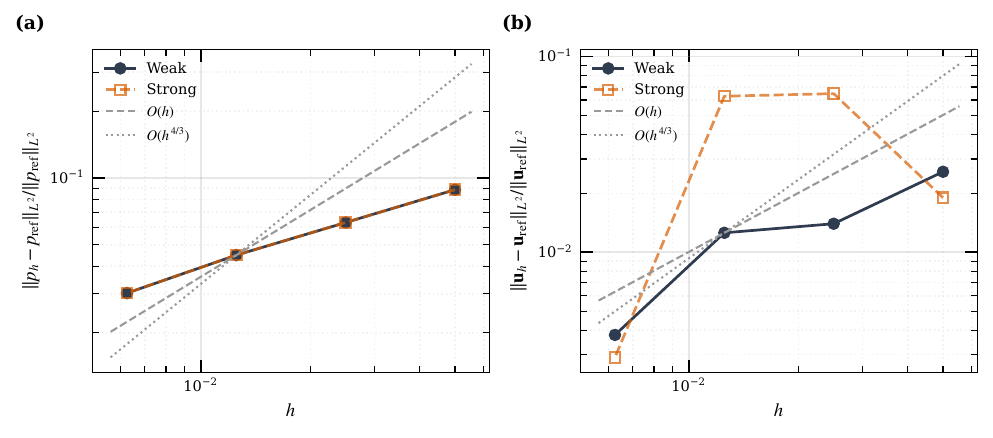}
	\caption{Bulk-field $L^2(\Omega)$ self-convergence for the
		embedded angled crack with $\theta = \arctan(0.6)$. Format as in
		\cref{fig:self_convergence_30angle}.}
	\label{fig:self_convergence_embedded}
	\end{figure}

The bulk-field self-convergence study (\cref{fig:self_convergence_embedded}), computed using the same procedure as in \cref{sec:angled_convergence} (\cref{eq:self_convergence}), reinforces this conclusion. The pressure error (\cref{fig:self_convergence_embedded}a) converges at a rate slightly lower than~$\mathcal{O}(h)$. The displacement error (\cref{fig:self_convergence_embedded}b) does not yet exhibit a clean asymptotic rate for either enforcement strategy over the present mesh range. In particular, the strong-enforcement curve is non-monotone, so it should not be interpreted as a clean $\mathcal{O}(h)$ regime, while the weak curve decreases more steadily. These trends are consistent with the boundary-intersecting case and confirm that the bulk solution quality is not degraded by the embedded-tip configuration, even though the post-processed interface residuals are more severely affected.

% ------ Figure: Self-convergence bulk (embedded) ------

\begin{figure}[!htbp]
		\centering
		\pagefiguregraphics{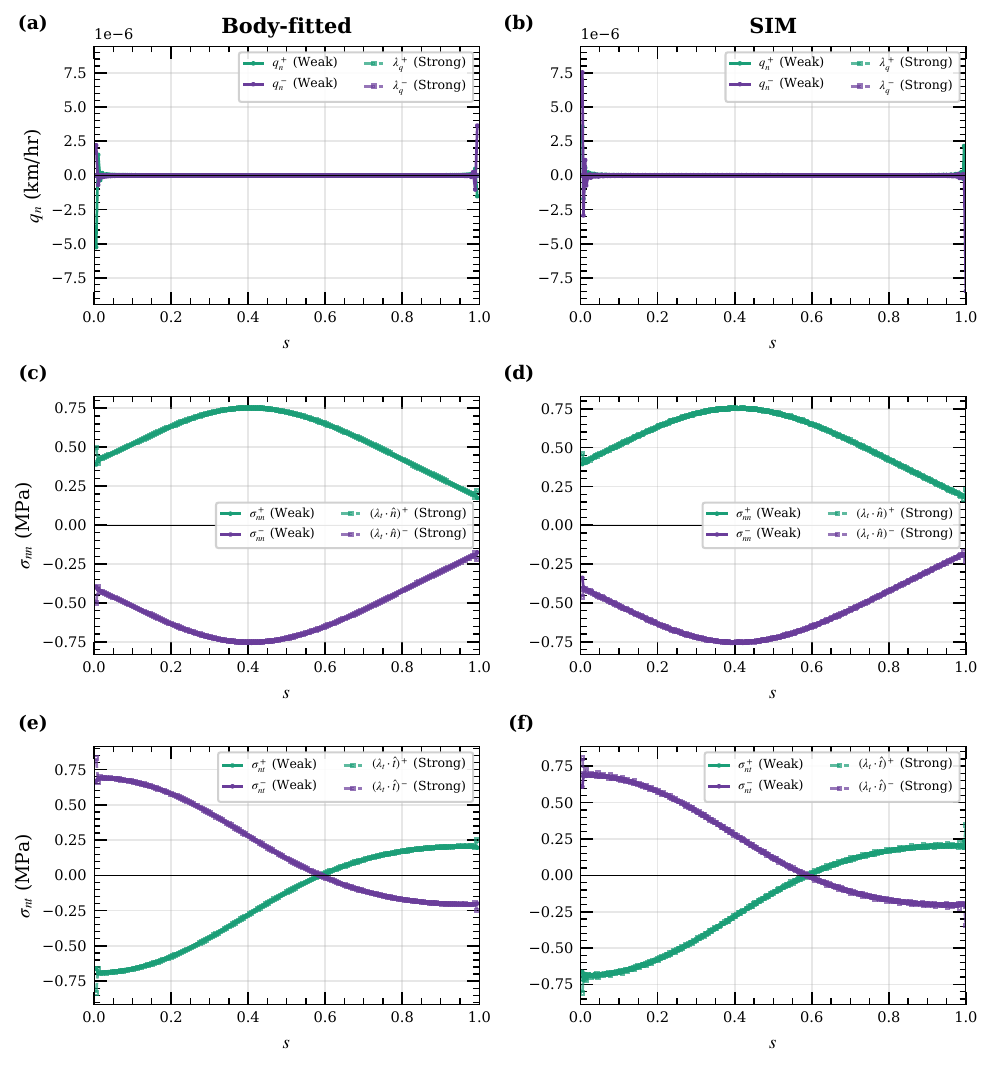}
	\caption{One-sided interface quantities on the finest mesh
		($n = 320$) for the embedded crack with
		$\theta = \arctan(0.6)$:
		body-fitted mesh (left column) versus
		shifted interface method (right column).
		Rows: (a,b)~normal flux $q_n$,
		(c,d)~normal traction $\sigma_{nn}$,
		(e,f)~tangential traction $\sigma_{nt}$.
		Filled circles show the weak-enforcement post-processed
		traces ($(+)$, green; $(-)$, purple); open squares show
		the strong-enforcement interface unknowns.}
	\label{fig:traction_perside_embedded}
	\end{figure}

As a final verification, \cref{fig:traction_perside_embedded} compares the one-sided interface tractions and fluxes on the finest mesh ($n = 320$) between a body-fitted (conformal) mesh and the shifted interface method. In the conformal case, the mesh conforms exactly to the crack, so the surrogate and true interfaces coincide, and no shifting is required; this solution serves as a reference to isolate the effect of the unfitted interface treatment.
The normal flux $q_n$ (\cref{fig:traction_perside_embedded}a,b) is essentially zero on both sides in both cases, consistent with the zero transmissivity imposed on this crack.
The normal traction $\sigma_{nn}$ (\cref{fig:traction_perside_embedded}c,d) and the tangential traction $\sigma_{nt}$ (\cref{fig:traction_perside_embedded}e,f) show excellent agreement between the body fitted and SIM solutions along the entire crack interior. The one-sided traces from weak enforcement (derived via the post-processing pipeline) closely match the strong-enforcement values ($\lambda_t$), and both mirror the body-fitted reference. The characteristic staircase-induced oscillations visible in the SIM tangential traction (\cref{fig:traction_perside_embedded}f) are absent from the body fitted solution (\cref{fig:traction_perside_embedded}e), implying that these oscillations arise from the normal mismatch in combination with the local expansion between the surrogate and true interfaces rather than from the underlying solution accuracy. Discrepancies appear only at the crack tips, where the body-fitted solution also exhibits large gradients due to the split-to-continuous transition, albeit without the additional normal-projection error of the SIM case.
This comparison demonstrates that the loss of convergence in the interface residuals reported above is not attributable to a deficiency of the SIM itself, but rather to localized post-processing singularities at the crack tips—an artifact that is present, in milder form, even on the body-fitted mesh.

% ------ Figure: Traction per-side conformal vs SBM (embedded) ------

\FloatBarrier

% ======================================================================
% Multi-crack
% ======================================================================
\subsection{Multi-crack configuration}
\label{sec:multi_crack}

The preceding test cases each considered a single crack in order to isolate the effects of interface orientation, mesh offset, and tip treatment. In practice, however, subsurface formations contain multiple fractures with varying geometries and hydraulic characteristics. The shifted interface formulation accommodates this naturally: each crack is defined by its own parametric curve, assigned its own interface constitutive model, and the mesh connectivity is split independently along the surrogate of each crack. No special treatment is required beyond supplying the list of crack descriptors to the assembler, as the framework is nonintrusive. Our implementation automatically constructs the surrogate interfaces, classifies elements, and assembles the coupled system for an arbitrary number of non-intersecting cracks on a single structured mesh.

\begin{figure}[!htbp]
	  \centering
	  \pagefiguregraphics{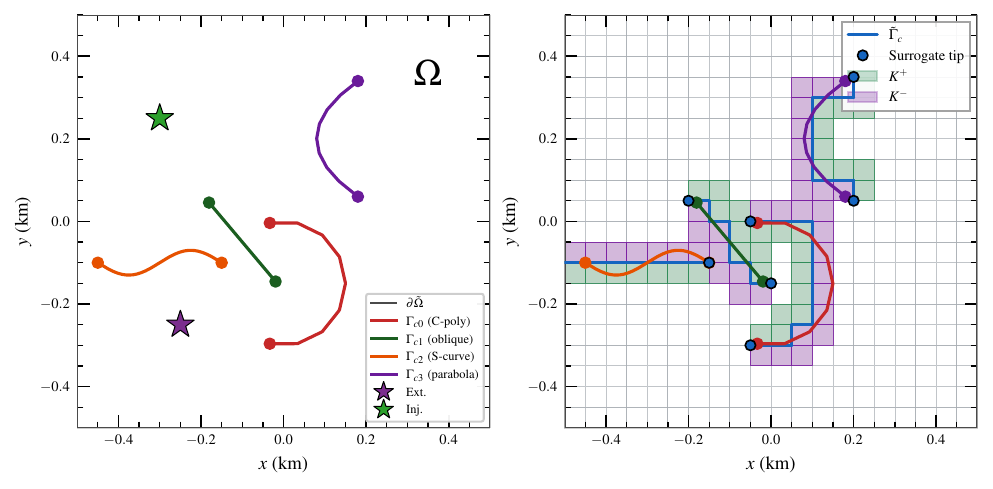}
  \caption{Multi-crack configuration.
  (a)~Domain geometry showing the four cracks: a C-shaped
  polyline~$\Gamma_{c0}$ (red), an oblique straight
  line~$\Gamma_{c1}$ (green), a smooth
  S-curve~$\Gamma_{c2}$ (orange), and a parabolic
  arc~$\Gamma_{c3}$ (purple).  The injection (green star) and
  extraction (purple star) wells are indicated.
  (b)~Element classification on a coarse mesh: the surrogate
  interfaces~$\tilde\Gamma_c$ (blue staircase segments)
  approximate the true cracks. Elements are classified as
  $K^+$ (green) and~$K^-$ (purple) independently for each crack.}
  \label{fig:domain_classification_multicrack}
	\end{figure}

To demonstrate this capability we place four embedded cracks in the
same $1\times1$~km domain used throughout this section
(\cref{fig:domain_classification_multicrack}a). The four crack
geometries are deliberately chosen to sample a range of shapes:
\begin{enumerate}
  \item A \emph{C-shaped polyline} ($\Gamma_{c0}$) centered at
        $(0.0,\,-0.15)$, consisting of eight linear segments
        inscribed on a circular arc of radius $R = 0.15$~km that
        subtends roughly $200^\circ$.  This crack is modelled as
        an impermeable, welded barrier ($T_n = 0$,
        $k_n = k_t = 10^{8}$~MPa/km).
  \item An \emph{oblique straight line} ($\Gamma_{c1}$) centered
        at $(-0.10,\,-0.05)$ with length $0.25$~km and inclination
        $\theta = 0.7$~rad ($\approx 40^\circ$).  This crack is
        modelled as compliant ($k_n = k_t = 10^{4}$~MPa/km) and
        impermeable ($T_n = 0$).
  \item A \emph{smooth S-curve} ($\Gamma_{c2}$) centered at
        $(-0.3,\,-0.10)$ with horizontal extent $L = 0.30$~km and
        a sinusoidal transverse amplitude $A = 0.03$~km.  This
        crack is semi-permeable ($T_n = 5$~km/(MPa$\cdot$hr)) and nearly welded ($k_n = k_t = 10^{8}$~MPa/km).
  \item A \emph{parabolic arc} ($\Gamma_{c3}$) centered at
        $(0.18,\,0.20)$ with half-width $0.14$~km and height
        $0.10$~km, discretized as an eight-segment polyline.
        This crack is modelled as a nearly welded, impermeable barrier
        ($T_n = 0$, $k_n = k_t = 10^{8}$~MPa/km).
\end{enumerate}
The combination of curved, straight, and polyline interfaces, each with a different set of interface parameters, exercises the generality of the implementation across geometries and constitutive
regimes.

Boundary conditions are the same as in the single-crack tests: homogeneous Dirichlet displacement on the left and right boundaries, traction-free top and bottom, and homogeneous Dirichlet pressure on all four sides.  In place of the single injection source used previously, the domain is now loaded by two localized sources (again as Wendland~$C^{2}$ with support radius $R = 0.1$~km): an extraction well at $\mathbf{x}_{\mathrm{ext}} = (-0.25,\,-0.25)$ and an injection well at $\mathbf{x}_{\mathrm{inj}} = (-0.30,\,0.25)$. This injection--extraction pair drives fluid through the fractured region, producing richer interactions between the cracks and the flow field. The simulation is run on a single structured quadrilateral mesh with $n = 160$ elements per side ($h = 1/160$~km) using the weak (direct-substitution) enforcement strategy; no convergence study is performed, as the focus here is on demonstrating the multi-crack capability rather than on formal rate verification, which has already been established in the preceding single-crack tests.

% ------ Figure: Domain classification (multi-crack) ------

\FloatBarrier

\subsubsection{Solution fields}
\label{sec:multi_crack_fields}

\begin{figure}[!htbp]
	  \centering
	  \pagefiguregraphics{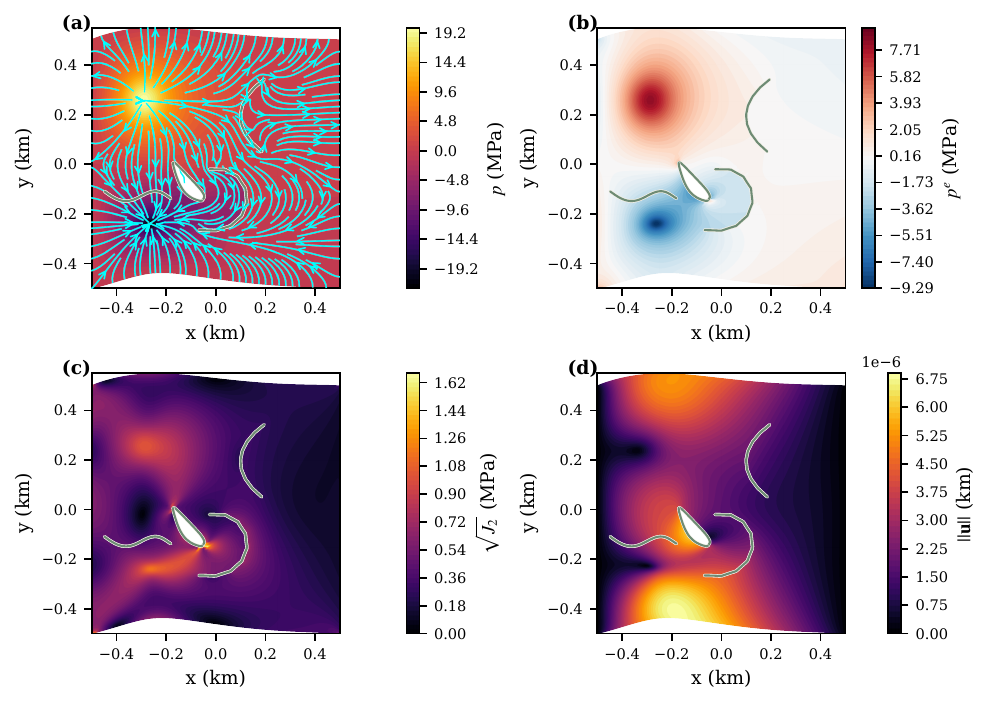}
  \caption{Computed solution fields on the $n = 160$ mesh for the
  multi-crack configuration, displayed on the deformed configuration
  (displacement $\times 10^4$):
  (a)~pore pressure $p$ with Darcy flux streamlines,
  (b)~mean effective stress $p^e = \tfrac{1}{3}\sigma^e_{kk}$,
  (c)~deviatoric stress invariant $\sqrt{J_2}$, and
  (d)~displacement magnitude $\|\mathbf{u}\|$.
  The four cracks are visible as olive colored traces.
  The interaction between the injection--extraction doublet and
  the heterogeneous crack network produces a rich spatial pattern
  in all field variables and the opening of one of the cracks.}
  \label{fig:fields_multicrack}
	\end{figure}

\Cref{fig:fields_multicrack} shows the computed fields on the $n = 160$ mesh, displayed on the deformed configuration (displacement magnified by a factor of~$10^4$).

The pore pressure field (\cref{fig:fields_multicrack}a) reveals a complex interaction between the four cracks and the injection--extraction doublet. The injection well generates a high-pressure lobe in the upper-left quadrant, while the extraction well produces a low-pressure region in the lower-left. The Darcy flux streamlines (cyan) show how each crack modifies the flow pattern according to its hydraulic character. The impermeable cracks ($\Gamma_{c0}$, $\Gamma_{c1}$, and~$\Gamma_{c3}$) act as hydraulic barriers, deflecting streamlines around their tips and forcing flow to take longer paths. The semi-permeable S-curve~$\Gamma_{c2}$ allows partial through-flow, producing a less abrupt deflection. The overall streamline pattern is smooth away from the cracks and exhibits the expected source--sink topology of the doublet.

The mean effective stress $p^e = \tfrac{1}{3}\sigma^e_{kk}$ (\cref{fig:fields_multicrack}b) displays localized stress concentrations at the crack tips and visible jumps across the impermeable interfaces, consistent with the discontinuity in the Biot pressure contribution observed in the single-crack tests. The stress pattern around the C-shaped crack is particularly noteworthy: the enclosed concave region between the crack tips develops an intensified effective stress, reflecting the combined influence of the hydraulic barrier and the curvature.

The deviatoric stress invariant $\sqrt{J_2}$ (\cref{fig:fields_multicrack}c) shows stress concentrations near the crack tips and around the source and sink locations, driven by the poroelastic coupling. The compliant oblique crack~$\Gamma_{c1}$ produces a pronounced deviatoric stress concentration consistent with its lower spring stiffness, which allows larger displacement jumps and correspondingly sharper strain gradients. The high-stiffness cracks show less intense but still visible tip concentrations.

The displacement magnitude (\cref{fig:fields_multicrack}d) is dominated by the Biot-induced swelling from the injection well and the contraction from the extraction well, modulated by the clamped lateral boundaries. The displacement field is continuous across the high-stiffness cracks, as expected. In contrast, the compliant oblique crack~$\Gamma_{c1}$ ($k_n = k_t = 10^{4}$~MPa/km) exhibits a clearly visible opening in the accentuated deformed configuration: the nearby extraction well draws fluid from the surrounding matrix, lowering the pore pressure and inducing local contraction that pulls the two crack faces apart. This crack opening is readily apparent as a widened opening (gap) in the deformed-mesh visualization and illustrates how heterogeneous interface stiffness interacts with the poroelastic loading to produce qualitatively different mechanical responses along different fractures within the same domain.

From a practical standpoint, the multi-crack simulation requires no additional algorithmic intervention beyond the single-crack case. The only additional computational cost is the assembly of the interface operators for each crack and the enlargement of the split-connectivity mesh, both of which scale linearly with the number of cracks. This example illustrates that the shifted interface framework is well-suited to application-driven problems involving multiple fractures with arbitrary geometries and heterogeneous interface properties.

% ------ Figure: Solution fields (multi-crack) ------

\FloatBarrier

\section{Conclusions}

This work adapted the shifted interface method to coupled transient poroelasticity with embedded interfaces. Starting from the standard Biot equations describing flow and deformation in a porous medium traversed by an internal discontinuity, we derived a unified framework that transfers both hydraulic transmission (Robin-type flux-pressure coupling) and mechanical response (spring-type traction-displacement coupling) from the true crack to a surrogate interface fitted to the background mesh. The derivation decomposes a generic interface flux into true-normal and tangential-mismatch components, yielding shifted bilinear forms whose structure is independent of the specific constitutive law.

Two enforcement strategies were formulated and compared: (i) weak enforcement, in which the interface
constitutive laws are inserted directly into the variational form, and (ii) strong enforcement, in
which auxiliary interface unknowns mediate the coupling and the constitutive closure is imposed pointwise at interface degrees of freedom. Both strategies operate on the same split-connectivity mesh and share the same bulk and interface operators; they differ only in whether the constitutive relation is enforced variationally or algebraically.

The method was validated on three single-crack test cases of increasing geometric complexity: an offset mesh-aligned crack, a boundary-intersecting angled crack with
$\theta = \arctan(0.6)$, and an embedded angled crack with the same orientation and both tips in the domain interior. Across these cases, the numerical results show that the method remains reliable over a broad range of interface geometries and clarify the role of post-processing in the observed residual convergence. For the offset mesh-aligned crack, the interface residuals converge at first order, confirming the correctness of the shifted poroelastic coupling and the equivalence of the two enforcement strategies in geometrically favorable settings. For angled cracks, the formulation remains accurate, but the convergence of post-processed flux residuals is degraded by highly localized artifacts near surrogate crack tips. For embedded cracks, these are further amplified by the interior split-to-continuous transition. The tip-trimming study shows that excluding a small fraction of crack length near each tip recovers first-order behavior, indicating that the degradation is a local effect consistent with the reduced regularity near crack-tip singularities known from elliptic theory~\citep{Grisvard2011,Dauge2008}, rather than a deficiency of the underlying discretization. Characterizing the precise singular exponents for the coupled poroelastic system is deferred to future work. A fourth test demonstrated the multi-crack capability of the
framework by embedding four cracks of distinct geometry and interface properties, including curved, straight, and polyline interfaces with varying permeability and stiffness, driven by an injection--extraction doublet. The resulting fields confirm that the formulation handles multiple non-intersecting cracks on a single structured mesh with no additional algorithmic complexity.

A second key conclusion concerns the trade-off between enforcement strategies. The strong enforcement gives smaller constitutive residuals in absolute magnitude because the interface constitutive
laws are satisfied to solver tolerance at interface degrees of freedom. The weak enforcement, in contrast, permits larger constitutive residuals on coarse meshes but can exhibit faster asymptotic decay of those residuals. In practice, neither strategy is uniformly preferable. The strong approach is attractive when pointwise satisfaction of interface laws is important, while the weak approach remains competitive and, in some regimes, appears asymptotically more favorable in constitutive residual convergence.

Taken together, the numerical results support the shifted interface method as a practical framework for poroelastic crack modeling on non-body-fitted meshes, and in particular as a flexible tool for
simulating geometrically complex embedded cracks without requiring mesh conformity to the interface. This makes the approach especially attractive for application-driven poromechanics problems in which crack geometry is irregular, evolving, or difficult to mesh directly, including subsurface energy systems, fractured reservoirs, geothermal injection, and coupled hydro-mechanical processes in heterogeneous media. A natural next step is to extend the present formulation to more advanced interface constitutive models, including nonlinear hydraulic transmissivity, frictional and damage-based mechanical laws, and rate- or state-dependent responses. Further improvements may also come from incorporating the Hessian-based correction terms omitted here, especially for curved interfaces and higher-order discretizations. Beyond this, extending to three dimensions, accounting for intersecting fractures, and modeling fracture propagation would significantly broaden the method's applicability to realistic poromechanical simulations.

\printcredits

\section*{Declaration of competing interest}
The authors declare that they have no known competing financial interests or personal relationships that could have influenced the work reported in this paper.

\section*{Acknowledgments}
Funded by the European Union. Views and opinions expressed are, however, those of the author(s) only and do not necessarily reflect those of the European Union or the European Research Council Executive Agency.  Neither the European Union nor the granting authority can be held responsible for them. This work is supported by the ERC grant INJECT, no. 101087771, doi: 10.3030/101087771.

% %% The Appendices part is started with the command \appendix;
% %% appendix sections are then done as normal sections
\appendix
\numberwithin{equation}{section}
% \section{Example Appendix Section}

% ============================================================
\section{Explicit matrix block definitions}
\label{app:matrix_blocks}
% ============================================================

This appendix collects the element-level definitions of all matrix
blocks appearing in the semidiscrete
systems~\cref{eq:matrix_direct,eq:matrix_unknown}. Throughout,
$N_I$ denotes the scalar shape functions for pressure and
$\bm{N}^k_i$ the displacement shape functions (including the
bubble enrichment).

The bulk blocks are
\begin{equation}
	(\mathbf{M}_p)_{IJ}
	= \int_{\tilde\Omega} \beta\,N_I\,N_J\;d\tilde{\Omega},
\end{equation}
\begin{equation}
	(\mathbf{C})_{Ij}
	= \int_{\tilde\Omega} \alpha\,N_I\,\bm{N}_{j,k}^k\;d\tilde{\Omega},
\end{equation}
\begin{equation}
	(\mathbf{K}_p)_{IJ}
	= \int_{\tilde\Omega} \gamma\,N_{I,i}\,N_{J,i}\;d\tilde{\Omega},
\end{equation}
\begin{equation}
	(\mathbf{K}_u)_{ij}
	= \int_{\tilde\Omega}
	\mathbb{C}_{klmn}\,\bm{N}^k_{i,l}\,\bm{N}^m_{j,n}\;d\tilde{\Omega}.
\end{equation}
with $\mathbb{C}_{klmn}$ the drained elasticity tensor. The Biot
coupling appears as $\mathbf{C}$ and its transpose $-\mathbf{C}^T$.

The interface blocks are assembled over $\mathcal{E}^h_c$ from the
three lines of the substituted
forms~\cref{eq:flow_substituted,eq:mech_substituted}. The
constitutive contributions (line one) give
\begin{equation}
	\begin{split}
		(\mathbf{K}^c_p)_{IJ}
		&= \sum_{e \in \mathcal{E}^h_c} \int_e
		\cos\varphi\,T_n\,
		\jmp{N_J + \Delta_\ell\,N_{J,\ell}}\,
		\jmp{N_I}\;d\tilde{s}.
	\end{split}
\end{equation}
\begin{equation}
	\begin{split}
			(\mathbf{K}^c_u)_{ij}
			&= \sum_{e \in \mathcal{E}^h_c} \int_e
			\cos\varphi\,K_{km}\,
			\jmp{\bm{N}^m_j + \Delta_\ell\,\bm{N}^m_{j,\ell}}\,
			\jmp{\bm{N}^k_i}\;d\tilde{s}.
	\end{split}
\end{equation}
\begin{equation}
	\begin{split}
		(\mathbf{D}^c_u)_{ij}
		&= \sum_{e \in \mathcal{E}^h_c} \int_e
		\cos\varphi\,H_{km}\,
		\jmp{\bm{N}^m_j + \Delta_\ell\,\bm{N}^m_{j,\ell}}\,
		\jmp{\bm{N}^k_i}\;d\tilde{s}.
	\end{split}
\end{equation}

The Taylor correction contributions (line three
of~\cref{eq:mech_substituted}) involve the stress gradient
$\tilde{\sigma}_{ij,\ell}
= \mathbb{C}_{ijkl}\,u_{k,l\ell} - \alpha\,p_{,\ell}\,\delta_{ij}$,
which couples to both displacement and pressure. Splitting
accordingly,
\begin{equation}
	\begin{split}
		(\mathbf{S}^c_p)_{IJ}
		&= \sum_{e \in \mathcal{E}^h_c} \int_e
		\cos\varphi\,\Big(
		\avg{\Delta_\ell\,(-\gamma\,N_{J,j\ell})}\,
		\hat{n}_j\,\jmp{N_I}
		\\
		&\qquad\qquad
		+ \jmp{\Delta_\ell\,(-\gamma\,N_{J,j\ell})}\,
		\hat{n}_j\,\avg{N_I}
		\Big)\;d\tilde{s}.
	\end{split}
\end{equation}
\begin{equation}
	\begin{split}
		(\mathbf{S}^c_{uu})_{ij}
		&= \sum_{e \in \mathcal{E}^h_c} \int_e
		\cos\varphi\,\Big(
		\avg{\Delta_\ell\,\mathbb{C}_{kmpq}\,\bm{N}^q_{j,p\ell}}\,
		\hat{n}_m\,\jmp{\bm{N}^k_i}
		\\
		&\qquad\qquad
		+ \jmp{\Delta_\ell\,\mathbb{C}_{kmpq}\,\bm{N}^q_{j,p\ell}}\,
		\hat{n}_m\,\avg{\bm{N}^k_i}
		\Big)\;d\tilde{s}.
	\end{split}
\end{equation}
\begin{equation}
	\begin{split}
		(\mathbf{S}^c_{up})_{iJ}
		&= \sum_{e \in \mathcal{E}^h_c} \int_e
		\cos\varphi\,\Big(
		\avg{\Delta_\ell\,(-\alpha\,N_{J,\ell}\,\delta_{km})}\,
		\hat{n}_m\,\jmp{\bm{N}^k_i}
		\\
		&\qquad\qquad
		+ \jmp{\Delta_\ell\,(-\alpha\,N_{J,\ell}\,\delta_{km})}\,
		\hat{n}_m\,\avg{\bm{N}^k_i}
		\Big)\;d\tilde{s}.
	\end{split}
\end{equation}

The normal-mismatch contributions (line two) involve the stress
$\tilde{\sigma}_{ij}
= \mathbb{C}_{ijkl}\,u_{k,l} - \alpha\,p\,\delta_{ij}$, which
splits analogously:
\begin{equation}
	\begin{split}
		(\mathbf{R}^c_p)_{IJ}
		&= -\sum_{e \in \mathcal{E}^h_c} \int_e
		\Big(
		\avg{(-\gamma\,N_{J,j})}\,\tilde{n}^{\perp}_j\,\jmp{N_I}
		\\
		&\qquad\qquad
		+ \jmp{(-\gamma\,N_{J,j})}\,\tilde{n}^{\perp}_j\,\avg{N_I}
		\Big)\;d\tilde{s}.
	\end{split}
\end{equation}
\begin{equation}
	\begin{split}
		(\mathbf{R}^c_{uu})_{ij}
		&= -\sum_{e \in \mathcal{E}^h_c} \int_e
		\Big(
		\avg{\mathbb{C}_{kmpq}\,\bm{N}^q_{j,p}}\,
		\tilde{n}^{\perp}_m\,\jmp{\bm{N}^k_i}
		\\
		&\qquad\qquad
		+ \jmp{\mathbb{C}_{kmpq}\,\bm{N}^q_{j,p}}\,
		\tilde{n}^{\perp}_m\,\avg{\bm{N}^k_i}
		\Big)\;d\tilde{s}.
	\end{split}
\end{equation}
\begin{equation}
	\begin{split}
		(\mathbf{R}^c_{up})_{iJ}
		&= -\sum_{e \in \mathcal{E}^h_c} \int_e
		\Big(
		\avg{(-\alpha\,N_J\,\delta_{km})}\,
		\tilde{n}^{\perp}_m\,\jmp{\bm{N}^k_i}
		\\
		&\qquad\qquad
		+ \jmp{(-\alpha\,N_J\,\delta_{km})}\,
		\tilde{n}^{\perp}_m\,\avg{\bm{N}^k_i}
		\Big)\;d\tilde{s}.
	\end{split}
\end{equation}

The coupling blocks $\mathbf{G}_p$ and $\mathbf{G}_u$ arise from the
first line
of~\cref{eq:flow_unknown_form,eq:mech_unknown_form}:
\begin{equation}
	\begin{split}
		(\mathbf{G}_p)_{I\alpha}
		&= -\sum_{e \in \mathcal{E}^h_c} \int_e
		\cos\varphi\,\Big(
		\tfrac{1}{2}\,\jmp{N_I}\,\jmp{L_\alpha}
		\\
		&\qquad\qquad
		+ 2\,\avg{N_I}\,\avg{L_\alpha}
		\Big)\;d\tilde{s}.
	\end{split}
\end{equation}
\begin{equation}
	\begin{split}
		(\mathbf{G}_u)_{i\alpha}
		&= \sum_{e \in \mathcal{E}^h_c} \int_e
		\cos\varphi\,\Big(
		\tfrac{1}{2}\,\jmp{\bm{N}^k_i}\,\jmp{L^k_\alpha}
		\\
		&\qquad\qquad
		+ 2\,\avg{\bm{N}^k_i}\,\avg{L^k_\alpha}
		\Big)\;d\tilde{s}.
	\end{split}
\end{equation}
with $L_\alpha$ and $L^k_\alpha$ the basis functions of
$\Lambda^h_q$ and $\bm{\Lambda}^h_t$, respectively.

% ============================================================
\section{Wendland source regularization}
\label{app:wendland}
% ============================================================

The point fluid source at $\mathbf{x}_s$ is replaced by the normalized
Wendland $C^{2}$ radial basis function~\cite{Wendland1995}
\begin{equation}\label{eq:wendland}
	\begin{aligned}
		W(\mathbf{x};\,\mathbf{x}_s,R)
		&= \frac{7}{\pi R^{2}}\,\psi_{3,1}\!\bigl(r\bigr),
		\\
		\psi_{3,1}(r) &= (1 - r)_{+}^{4}(4r + 1),
		\qquad
		r = \frac{\lVert \mathbf{x} - \mathbf{x}_s \rVert}{R}\,,
	\end{aligned}
\end{equation}
where $R = 2\sigma$ is the support radius, $\sigma = 0.05$~km is the
characteristic width, and $(\cdot)_{+} = \max(\cdot,\,0)$ denotes the
positive part.  The function $\psi_{3,1}$ is the minimal-degree
polynomial in Wendland's family that is positive definite on
$\mathbb{R}^{d}$ for $d \le 3$ and belongs to $C^{2}(\mathbb{R})$.
The prefactor $7/(\pi R^{2})$ ensures unit mass in two dimensions,
\begin{equation}\label{eq:wendland-norm}
	\int_{\mathbb{R}^{2}} W(\mathbf{x};\,\mathbf{x}_s,R)\,
	\mathrm{d}\mathbf{x} = 1\,,
\end{equation}
so that $W \to \delta$ as $R \to 0$.

\bibliographystyle{cas-model2-names}

\bibliography{biblio}

\end{document}